\documentclass[a4paper,10pt,twocolumn]{article}

\usepackage[utf8]{inputenc}
\usepackage[T1]{fontenc}
\usepackage[english]{babel}
\usepackage{mathptmx}
\usepackage{amsmath,amssymb}
\usepackage{graphicx}
\usepackage{subcaption}
\usepackage{xcolor}
\usepackage{hyperref}
\usepackage{booktabs}
\usepackage{geometry}
\usepackage{makecell}
\usepackage{multirow}
\graphicspath{{./}{Images/}}

\begin{document}
	
	% Titre et auteurs
	\title{\textbf{
			Impact of mechanical constraints on tokamak design\\
			and implications for high field power plants}}
	\author{
		\small Timothé Auclair$^{1}$, Baptiste Boudes$^{1}$, Jean-Luc Duchateau$^{1}$, Eric Nardon$^{1}$, Laura Pittaluga$^{1}$, Yanick Sarazin$^{1}$, Finn Sutcliffe$^{1}$, Alexandre Torre$^{1}$ \\
		\small $^1$ CEA, IRFM, F-13108 Saint-Paul-lez-Durance, France\\
		\small \texttt{timothe.auclair@cea.fr}
	}
	\date{} % Pas de date
	
	% Titre
	\twocolumn[
	\begin{@twocolumnfalse}
		\maketitle
		
		% Keywords
		
		% Abstract
		\begin{abstract}
			Two analytical models for sizing the toroidal field coils and central solenoid of a tokamak are developed within the D0FUS system code: a pedagogical thin-cylinder model and a refined
			thick-cylinder and winding packs model. The refined model shows good agreement with six reference machines and the MADE magnet design code. When the high-field design space is explored for DEMO-class power plants (2~GW of fusion power, $Q=40$, $t_\mathrm{plateau}=2$~h), pushing the peak field at the Toroidal Field (TF) coil conductor up to $B_\mathrm{max} = 20$~T, the radial build emerges as the dominant constraint: in fact, in the baseline wedging/316L configuration, no viable design can be found beyond 20~T, making alternative strategies necessary. The primary levers identified are high-strength steels (e.g CHSN01), alternative mechanical architectures (bucking, plug), and reductions of the effective Central Solenoid (CS) flux demand (for example through auxiliary heating during ramp-up), each carrying an impact of the same order of magnitude on the minimum feasible major radius. Secondary optimisations (conductor shape, radial grading) are shown to provide additional but more modest gains. When all favourable levers are combined (CHSN01, bucking, etc.), compact machines ($R_0 < 4$~m) appear feasible. This suggests that, provided one accepts the associated risks of combining new approaches (CHSN01, bucking, etc.), high-temperature superconductors could unlock the path to compact electricity generating tokamaks.
		\end{abstract}
		\vspace{1cm}
	\end{@twocolumnfalse}
	]
	
	\section{Introduction}
	
\begin{figure}
		\centering
		\includegraphics[width=1\linewidth]{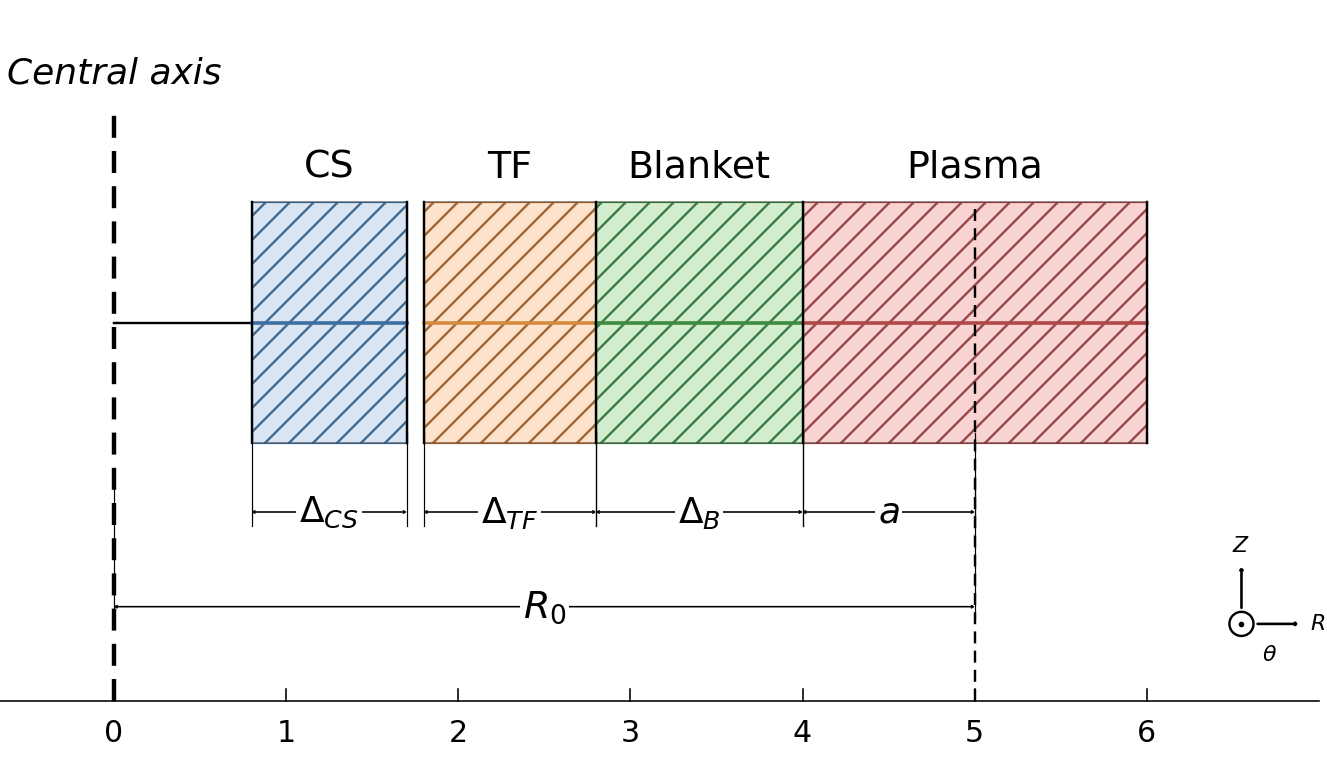}
		\caption{Typical radial build in D0FUS (wedging illustration)}
		\label{fig:radialbuildclean}
	\end{figure}

	A tokamak system code like D0FUS \cite{auclair2025tokamak} is designed to assess the performance, sizing, and cost of a given tokamak power plant. Therefore, its capability to accurately predict the dimensions of the main components of a tokamak is essential. The most critical thicknesses are those on the high-field side, where space is most constrained. A typical internal leg cross-section view, also referred to as the radial build, is shown in Fig. \ref{fig:radialbuildclean}. In D0FUS, the plasma radius (\(a\)) and the combined thickness of the first wall, breeding blanket, neutron shield, and associated gaps (\(\Delta_{B}\)) are input parameters. In contrast, the toroidal field (TF) coil thickness $\Delta_{TF}$ and the central solenoid (CS) thickness $\Delta_{CS}$ are key outputs of the code since they highly depend on the design point.
	
	To determine these coil thicknesses, two models have been developed. The first, named "Academic model", employs simplified assumptions (thin-cylinder stress theory, no steel in the winding pack) that prioritize pedagogical clarity. This level of description is typical of the magnet sizing modules found in most system codes \cite{panin2017mechanical, kovari2016process, duchateau2014conceptual}. It is described in Section~\ref{Academic}.
	
	The second, named "Refined model", introduces thick-cylinder stress theory, a composite Cable-In-Conduit Conductor (CICC) winding pack, and self-consistent CS axial stress. Other system codes have pursued similar refinements of their magnet modules \cite{swanson2022validation, morris2015implications}. The present model follows this trend while remaining mostly analytical. It does not aim at the level of multiphysics detail of dedicated magnet design codes such as MADE \cite{giannini2023magnet} or MADMACS \cite{zani2019parametric}, but approaches their predictive capability for radial build sizing. This model is presented in Section~\ref{D0FUS}.
	
	These models are then compared and benchmarked against existing designs and more detailed codes in Section~\ref{Benchmark}.
	
	The ability of high magnetic fields to reduce the size and cost of a tokamak power plant has recently been the subject of a scientific debate. Federici \emph{et al.}~\cite{federici2024relationship} argue that, under conventional design assumptions (wedging, proven materials, thick shielding), higher fields demand proportionally thicker structural support, offsetting the expected size reduction. In a Comment, Creely \emph{et al.}~\cite{creely2024comment} counter that these conclusions rest on specific design choices rather than fundamental limits, and that alternative assumptions restore the compactness benefit of high fields. Section~\ref{Results} contributes to this discussion. Anchored on the EU-DEMO1 2017 baseline and reproducing the Federici \emph{et al.}\ $R_0(B_0)$ constraint curves, D0FUS first confirms the structural limitation inherent to conventional wedging, and then quantifies which design levers, and which combinations of them, are required to overcome it.
	
	Consistent with the D0FUS philosophy of physically transparent models, all formulations presented here are two-dimensional (in-plane) and do not account for out-of-plane forces (OOP). Their evaluation requires dedicated 3D structural analyses, which will need to be addressed in future work.
	
	\section{Academic Model}
	\label{Academic}
	
	The radial build is here determined by three criteria: generating the required magnetic field $B_{max}$ on the inner leg of the TF coils, generating the magnetic flux $\Psi_{CS}$ required from the CS to initiate, ramp-up, and maintain the current, and finally, withstanding the mechanical stresses induced by the associated Lorentz forces.
	
	In this model, we adopt a simple two-layer coil model for both the CS and TF coils: the first layer consists of a pure conductor (superconductor, copper, insulation and cooling) designed to generate \(B_{\text{max}}\) or $\Psi_{CS}$. This layer is assumed not to handle any mechanical stress. The second layer is solely composed of steel to withstand all the mechanical stresses. This approach, although simplistic, makes it very easy to understand the underlying trends.
	
	Several limitations must be noted: firstly, any realistic winding pack in reality includes a significant portion of steel for force handling. Secondly, the steel stress is considered to be homogeneous and equal to the allowable limit $\sigma_{\text{lim}}$, neglecting stress concentration. That is why this model has to be viewed as a preliminary, pedagogical model to observe trends in coil thickness and the overall radial build. Finally, we choose for this model the thin-cylinder approximation, allowing simple mechanical expressions. These three limitations are alleviated in the Refined model presented in Section~\ref{D0FUS}.
	
	Throughout this work, tensile stresses are positive and compressive stresses are negative. When comparing to the allowable limit $\sigma_{\text{lim}}$, absolute values are used.
	
	\subsection{Toroidal field coil}
	
\begin{figure}[tbph]
		\centering
		\includegraphics[width=0.35\textwidth]{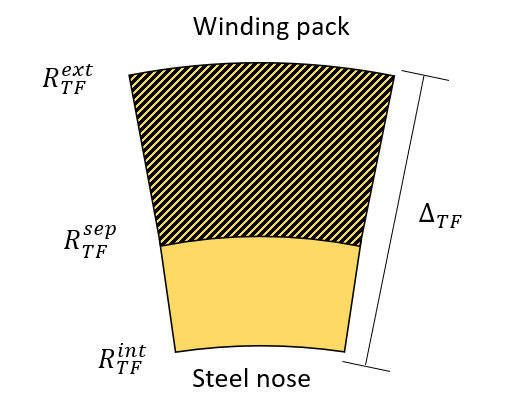}
		\caption{Schematic horizontal cross-section of the inner leg (high-field side) TF coil in the two layer approximation}
		\label{fig:2layerstf}
	\end{figure}

	The inner legs of the TF coils form a cylinder of outer radius \(R_{\text{TF}}^{\text{ext}}\) and inner radius \(R_{\text{TF}}^{\text{int}}\). We aim to determine \(R_{\text{TF}}^{\text{int}}\) and the associated TF coil thickness $\Delta_{TF}$ shown in Fig. \ref{fig:2layerstf}.
	
	\subsubsection{Magnetic field generation}
	
	The first requirement is to carry enough current to generate the maximum magnetic field on the inner leg \( B_{\text{max}} \). We can directly compute the internal radius of the conductor layer $R_{\text{TF}}^{\text{sep}}$ illustrated in Fig. \ref{fig:2layerstf}, in the thin cylinder approximation (\(\Delta R=R_{\rm TF}^{\rm ext}-R_{\rm TF}^{\rm sep}\ll R_{\rm TF}^{\rm ext}\)) as detailed in Appendix~\ref{Annexe_B_thinlayer}:
	\[
	B_{\rm max}
	\;=\;\mu_0\,J_{\rm TF}^{\rm wost}\bigl(R_{\rm TF}^{\rm ext}-R_{\rm TF}^{\rm sep}\bigr)
	\]
	
	\begin{equation}
		\quad\Longrightarrow\quad
		R_{\text{TF}}^{\text{sep}} = R_{\text{TF}}^{\text{ext}} - \frac{B_{\text{max}}}{J_{\rm TF}^{\rm wost} \mu_0}
		\label{equation_B_thin}
	\end{equation}
	
	Here, $J^{\rm wost}$ (for "without steel") is the engineering current density (in $A/m^2$) defined on the non-steel cross-section of the conductor. It accounts for the superconductor, copper stabilizer, cooling channels and insulation, but excludes the steel jacket which is sized independently for mechanical purposes. $J^{\rm wost}$ depends on the cooling temperature (a free parameter in the code, typically 4.2 K) and the magnetic field in which the conductor is immersed. Details of its determination are given in Appendix~\ref{JBappendix}.
	
	\subsubsection{Mechanical stress}
	
	The electromagnetic forces acting on the TF coil case can be decomposed into two components: a tensile force along the $z$-axis and a centering force along the $r$-axis directed toward the machine axis.
	
\begin{figure}
		\centering
		\begin{subfigure}[b]{1\linewidth}
			\centering
			\includegraphics[width=0.9\linewidth]{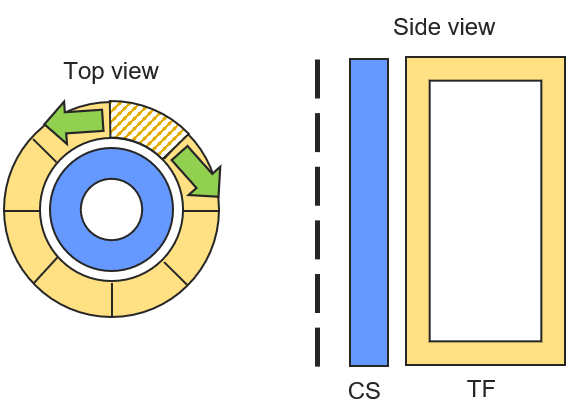}
			\caption{Illustration of the wedging configuration}
			\label{fig:topviewtf_a}
		\end{subfigure}
		
		\vspace{1em}
		
		\begin{subfigure}[b]{1\linewidth}
			\centering
			\includegraphics[width=0.9\linewidth]{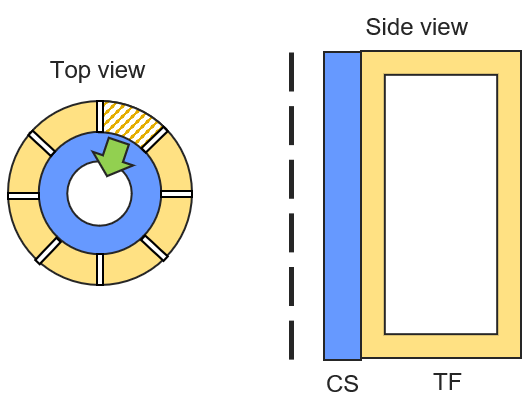}
			\caption{Illustration of the bucking configuration}
			\label{fig:topviewtf_b}
		\end{subfigure}
		
		\vspace{1em}
		
		\begin{subfigure}[b]{1\linewidth}
			\centering
			\includegraphics[width=0.9\linewidth]{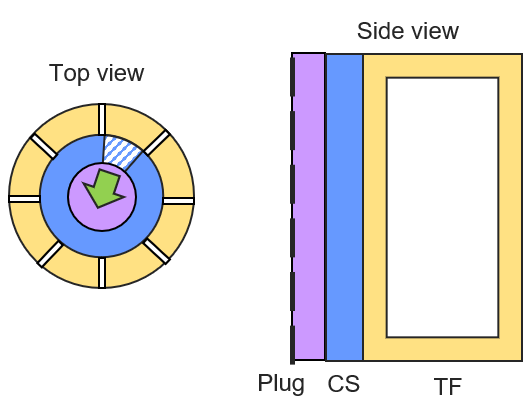}
			\caption{Illustration of the central plug configuration}
			\label{fig:topviewtf_c}
		\end{subfigure}
		\caption{Top and side view illustrating the three possible mechanical configurations}
		\label{fig:topviewtf}
	\end{figure}

	An important aspect of the resulting stress distribution is the so-called \emph{vault effect}, illustrated in Fig.~\ref{fig:topviewtf} which shows a sector of the tokamak with an inner TF coil leg (yellow) and an associated section of the CS (blue). This effect arises from the mechanical response of cylindrical shells to radial pressure. The coil structure must then be designed to withstand the resulting stress state. A common failure criterion is based on the Tresca stress, defined as the maximum absolute difference between principal stresses:
	
	\[
	\sigma_{\text{Tresca}} = \max(| \sigma_r - \sigma_\theta |, | \sigma_r - \sigma_z |, | \sigma_\theta - \sigma_z |)
	\]
	
	The Tresca criterion is adopted as the stress acceptability condition: it requires that the largest difference between any two principal stresses remains below a prescribed allowable stress, here taken as $\sigma_{\text{lim}} = \frac{2}{3}\sigma_{\text{yield}}$, providing a conservative bound for components subjected to multiaxial loading. In practice, the dominant stresses are the tensile $\sigma_z$ and centering $\sigma_r$ stress in the bucking configuration and the tensile $\sigma_z$ and vault $\sigma_\theta$ stress in the wedging configuration:
	
	\begin{equation}
		\text{Bucking case: }\sigma_{\text{Tresca}} = |\sigma_r - \sigma_z| \leqq \sigma_{\text{lim}}
	\end{equation}
	
	\begin{equation}
		\text{Wedging case: }\sigma_{\text{Tresca}} = |\sigma_\theta - \sigma_z| \leqq \sigma_{\text{lim}}
	\end{equation}
	
	Three mechanical architectures are considered throughout this work, illustrated in Fig.~\ref{fig:topviewtf}: the \emph{wedging} configuration, where the inboard legs of the TF coils are in toroidal contact (Fig.~\ref{fig:topviewtf_a}), the \emph{bucking} configuration, where they are separated and rest radially onto the central solenoid (Fig.~\ref{fig:topviewtf_b}), and the \emph{plug} configuration, a bucking variant complemented by a stiff central insert inside the CS bore (Fig.~\ref{fig:topviewtf_c}). For brevity, the labels \emph{wedging}, \emph{bucking} and \emph{plug} are used throughout the paper to refer to these three architectures.
	
	The wedging and bucking architectures are detailed in the next two subsections. The plug configuration involves the CS specifically and is therefore deferred to Section~\ref{Pluging}.
	
	\textbf{a - Bucking}
	\label{sec:bucking}
	
	The bucking configuration, shown in Fig. \ref{fig:topviewtf_b}, consists in supporting the inner leg of the TF coils on the CS to transfer the centering forces to it: the TF coils are assumed to have no contact between them.
	
	At the inner leg, where the toroidal field peaks at $B_\mathrm{max}$, the centering stress $\sigma_r$ can be approximated by the magnetic pressure:
	
	\begin{equation}
		\sigma_r = P_{\text{TF}} =  \frac{B_{max}^2}{2 \mu_0 }
		\label{equation0}
	\end{equation}
	
	The vertical force $F_z$, defined as the upward force acting on the upper half of a coil assumed to be vertically symmetric, is independent of the coil shape \cite{freidberg2015designing}. Its expression per coil for a circular geometry (also valid for D-shaped coils) is derived in Appendix~\ref{appendixB}.
	
	\begin{equation}
		F_z = \frac{\pi}{\mu_0 N_{coil}} B_0^2 R_0^2 \ln \left( \frac{R_0 + a + \Delta_B}{R_0 - a - \Delta_B} \right)
		\label{equationFz}
	\end{equation}
	
	with $B_0$ the central magnetic field and $N_{coil}$ the number of TF coils (the total vertical force being $N_{coil} F_z$).
	We can then express the tensile stress $\sigma_z$ by distributing a fraction $f_{z,\mathrm{WP}}$ of the total tension $N_{coil} F_z$ over the inboard leg cross-section. This assumption would be exact for a Princeton D-shaped coil (neglecting inter-coil structures). With the default $f_{z,\mathrm{WP}} = 1/2$ (equal sharing between inboard and outboard legs) and no external clamping ($F_{\rm clamp} = 0$), this leads to:
	
	\begin{equation}
		\sigma_z = \frac{N_{coil} F_z}{2 \pi ((R_{\text{TF}}^{\text{sep}})^2 - (R_{\text{TF}}^{\text{ int}})^2)}
		\label{equation1}
	\end{equation}
	
	In D0FUS, both $f_{z,\mathrm{WP}}$ and $F_{\rm clamp}$ (subtracted from $F_z$) are configurable parameters.\\
	
	\textbf{b - Wedging}
	\label{sec:wedging}
	
	For the wedging case, $\sigma_z$ is still given by Eq.~\ref{equation1}. Conversely, the centering force is now transmitted to the steel noses located at the inner radius of each TF coil (see Fig.~\ref{fig:2layerstf}). In contact with each other, they form a vault that transforms the radial stress into an azimuthal one $\sigma_\theta$ as shown in Fig.~\ref{fig:topviewtf}. In the limit of a thin cylinder shell subject to a sole external pressure exerted on it, the Lamé-Clapeyron theory \cite{LameClapeyron1833} provides a simple expression of this stress detailed in Appendix~\ref{app:thin_wall_stress}:
	
	\begin{equation}
		\sigma_{\theta} = \frac{P_{\text{TF}} R_{\text{TF}}^{\text{sep}}}{R_{\text{TF}}^{\text{sep}}-R_{\text{TF}}^{\text{int}}}
	\end{equation}
	
	\textbf{c - Solution}
	
	So far, the engineering current density $J_{\rm TF}^{\rm wost}$ is undetermined. As detailed in Appendix~\ref{JBappendix}, one can consider several scalings for different superconducting materials, both functions of temperature and magnetic field. This allows one to determine the winding pack thickness and associated radius $R_{\text{TF}}^{\text{sep}}$ from Section 2.1.2.a. Subsequently, setting $\sigma_{\text{Tresca}}$ to its limit $\sigma_{\text{lim}}$ (with a typical value of 660 MPa, if one considers 316L steel, or 1000 MPa, if one considers CHSN01 steel) allows $R_{\text{TF}}^{\text{int}}$ to be deduced. Solving for it for wedging in the thin cylinder approximation leads to (full calculus in Appendix~\ref{1rst_order_wedging}):
	
	\[
	R_{\text{TF}}^{\text{int}} 
	= R_{\text{TF}}^{\text{sep}}
	- \frac{B_{\max}^2 R_{\text{TF}}^{\text{sep}}}{2\mu_0 \sigma_{\text{lim}}}
	\left(1+\frac{1}{2}\ln \left( \frac{R_0 + a + \Delta_{B}}{R_0 - a - \Delta_{B}}\right)\right)
	\]
	
	and similarly in the bucking configuration (full calculus in Appendix~\ref{1rst_order_bucking}):
	
	\[
	\,R_{\text{TF}}^{\text{int}}
	= R_{\text{TF}}^{\text{sep}}
	- \frac{B_{\max}^2 R_{\text{TF}}^{\text{sep}}}{4\mu_0\left(\sigma_{\text{lim}}-\tfrac{B_{\max}^2}{2\mu_0}\right)} 
	\ln\left(\frac{R_0+a+\Delta_{B}}{R_0-a-\Delta_{B}}\right)
	\]
	
	\subsection{Central Solenoid}
	
	The CS is also treated making use of the two layer model as illustrated in Fig. \ref{fig:2layerscs}:
	
	\begin{figure}[tbph]
		\centering
		\includegraphics[width=0.33\textwidth]{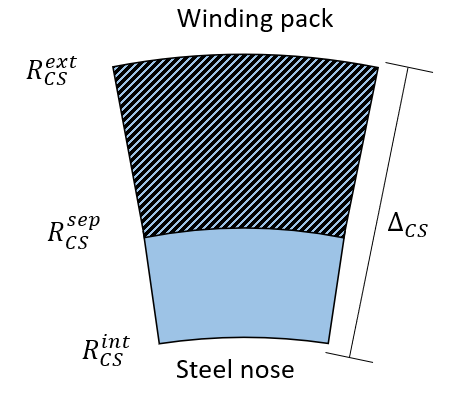}
		\caption{Schematic cross-section of the CS in the two layer approximation}
		\label{fig:2layerscs}
	\end{figure}
	
	Note that in the wedging configuration, the CS can be designed as several modules stacked together allowing one to improve plasma shaping capabilities and vertical stability control by varying the current in the different modules, as in ITER \cite{libeyre2009detailed}. In contrast, in the case of bucking, the CS contact surface with the TF coils must form a continuous interface along its entire length to ensure uniform force distribution, as implemented in JET \cite{rebut1981jet}. For simplicity, the CS is approximated as an infinite solenoid in this model and vertical stresses are neglected, as they are usually less critical than the hoop stress ~\cite{Fu2006,jong2009mechanical,nunio2019mechanical}. The Refined model (Section~\ref{D0FUS}) relaxes this assumption and accounts for the axial stress induced at the CS midplane by the radial fringe field at the solenoid ends.
	
	\subsubsection{Flux consumption criteria}
	\label{Flux}
	
	The magnetic flux requirement can be expressed as:
	
	\begin{equation}
		\Psi_{\text{Init}} + \Psi_{\text{Ramp-Up}} + \Psi_{\text{plateau}} =  \Psi_{\text{CS}} + \Psi_{\text{PF}}
		\label{eq:CS_flux}
	\end{equation}
	
	In Eq. \ref{eq:CS_flux}:
	
	• $\Psi_{\text{Init}}$ is the flux required to initiate the plasma. Integrating Faraday's law over the breakdown duration $t_{Init}$ gives:
	
	\begin{equation}
		\Psi_{\text{Init}} = 2\pi R_0 \, E_\varphi \, t_{BD} = 2\pi R_0 \, \mathcal{E}_{BD}
		\label{eq:Psi_PI}
	\end{equation}
	
	where $\mathcal{E}_{BD} = E_\varphi \, t_{BD}$ (in V.s/m) lumps all breakdown physics (fill pressure, error fields, impurity burn-through) into a single calibration parameter. This formulation captures the machine-size scaling: a larger torus requires more flux for the same breakdown conditions. For ITER ($R_0 = 6.2$ m, $\Psi_{\text{Init}} \approx 10$ Wb \cite{shimada2007chapter}), one obtains $\mathcal{E}_{BD} \approx 0.25$ V.s/m, consistent with the typical breakdown fields of 0.3 to 0.5 V/m during 0.3 to 0.5 s \ \cite{lloyd1991plasma}. In practice, $\Psi_{\text{Init}}$ represents less than 5\% of the total CS flux budget, so the result is only weakly sensitive to $\mathcal{E}_{BD}$.
	
	• The flux consumption to ramp up the plasma current is composed of an inductive and a resistive term:
	
	\begin{equation}
		\Psi_{\text{Ramp-Up}} = (1-f_h) \left[\underbrace{L_{\text{p}} \, I_p}_{\Psi_{\text{Ind}}} + \underbrace{C_{\text{Ejima}} \, \mu_0 R_0 \, I_p}_{\Psi_{\text{Res}}}\right]
		\label{eq:Psi_rampup}
	\end{equation}

	In D0FUS, a fraction $f_h \in [0,1]$ of the plasma current during the ramp-up phase can be attributed to non-inductive contributions from auxiliary heating and current drive systems. Both the inductive term $\Psi_{\rm Ind} = L_p I_p$ and the resistive term $\Psi_{\rm Res}$ are reduced proportionally, since both scale with the inductively-driven current fraction. The conservative default is $f_h = 0$, adopted in the EU-DEMO baseline (Table~\ref{tab:input_parameters}).
	
	The plasma self-inductance $L_p = L_{\text{ext}} + L_{\text{int}}$ is decomposed into an external part (field energy outside the plasma) and an internal part $L_{\text{int}} = \mu_0 R_0 \, l_i / 2$, where $l_i$ is the normalised internal inductance, evaluated from the Wesson fit~\cite{wesson2011tokamaks} $l_i \simeq \ln(1.65 + 0.89\,\nu)$ with $\nu$ the current peaking exponent. $L_{\text{ext}}$ is determined, at a given inverse aspect ratio $\varepsilon = \frac{a}{R_0}$ by the Hirshman and Neilson \cite{hirshman1986external} fit:
	\[
	L_{\text{ext}} = \mu_0 R_0 \, \frac{a_\varepsilon \, (1-\varepsilon)}{1 - \varepsilon + b_\varepsilon \, \kappa}
	\]
	with $\kappa$ the vertical elongation of the plasma at the last closed flux surface (LCFS) and :
	\[
	a_\varepsilon = (1 + 1.81\sqrt{\varepsilon} + 2.05\,\varepsilon)\ln\frac{8}{\varepsilon} - (2 + 9.25\sqrt{\varepsilon} - 1.21\,\varepsilon)
	\]
	\[
	b_\varepsilon = 0.73\sqrt{\varepsilon}\,(1 + 2\varepsilon^4 - 6\varepsilon^5 + 3.7\varepsilon^6)
	\]
	
	The resistive term $\Psi_{\text{Res}}$ follows the Ejima scaling \cite{duchateau2014conceptual,stambaugh2011fusion} with default value following the ITER design $C_{\text{Ejima}} = 0.45$.
	
	• The plateau magnetic flux \(\Psi_{\mathrm{plateau}}\) is estimated from the loop voltage \(V_{\mathrm{loop}}\) and the duration of the plateau \(t_\mathrm{plateau}\). 
	\begin{equation}
		\Psi_{\mathrm{plateau}} = V_{\mathrm{loop}} \, t_\mathrm{plateau}
	\end{equation}
	D0FUS decomposes the total plasma current as $I_p = I_{\mathrm{Ohm}} + I_b + I_{\mathrm{CD}}$, where $I_b$ is the bootstrap current (computed from the Sauter \emph{et al.}~\cite{sauter1999neoclassical} or Redl \emph{et al.}~\cite{redl2021new} neoclassical model), $I_{\mathrm{CD}}$ is the externally driven current, and $I_{\mathrm{Ohm}}$ is the remaining inductively driven (Ohmic) component. The loop voltage is then $V_{\mathrm{loop}} = R_{\mathrm{eff}} \, I_{\mathrm{Ohm}}$, where $R_{\mathrm{eff}}$ is the effective plasma resistance obtained by radially integrating the neoclassical conductivity over the plasma cross-section. In steady state configuration, the plasma current is entirely non-inductively driven and thus $\Psi_{\text{plateau}}$ is taken equal to 0.
	
	• The flux provided by an infinite solenoid (over a full swing) can be expressed as (details of the calculation in Appendix~\ref{appendixC}):
	\begin{equation}
		\Psi_{\text{CS}} = \frac{2 \pi B_{CS}}{3}\left((R_{\text{CS}}^{\text{ ext}})^2+R_{\text{CS}}^{\text{ ext}}R_{\text{CS}}^{\text{ sep}}+(R_{\text{CS}}^{\text{ sep}})^2\right)
		\label{Flux_CS_eq}
	\end{equation}

	where $B_{CS}$ is the magnetic field inside the CS at the beginning of the pulse. Eq.~\ref{Flux_CS_eq} expresses the flux that the CS hardware can deliver over a full swing, but only a fraction of this swing is actually available to drive the plasma current, the rest being reserved for plasma control during the discharge. To account for this, D0FUS introduces a parameter $f_{\rm swing}^{\rm usable} \in (0,1]$ such that the inductive flux available to the plasma is $f_{\rm swing}^{\rm usable} \, \Psi_{\rm CS}$. A default value of $f_{\rm swing}^{\rm usable} = 0.75$ is adopted, in line with EU-DEMO and ITER design practice.
	
	• The contribution of the PF coils is approximated by calculating the vertical magnetic field and the corresponding flux as in Ref.\cite{duchateau2014conceptual}:
	
	\[
	B_{\text{vert}} = \frac{\mu_0 I_{\text{p}}}{4\pi R_0} \left( \beta_p + \frac{l_i - 3}{2} + \log \frac{8R_0}{a\sqrt{\kappa}} \right)
	\]
	
	\begin{equation}
		\Psi_{\text{PF}} = B_{\text{vert}} \pi R_0^2
		\label{eq:Psi_PF_new}
	\end{equation}
	
	with the poloidal beta $\beta_p$ approximated from Ref.\cite{jean2011helios,duchateau2014conceptual}:
	
	\[
	\beta_p = \frac{4 \bar{n} k_B \bar{T} L^2}{\mu_0 I_p^2}
	\]
	
	where $\bar{T}$ and $\bar{n}$ are respectively the average electron temperature and density of the plasma, $k_B$ the Boltzmann constant and $L$ the length of the last closed flux surface.
	
	The integration of $B_{\rm vert}$ extends down to the magnetic axis because the PF coils enter the poloidal flux balance through two independent sizing criteria, each associated with a different phase of the scenario.
	
	The first criterion is set by the breakdown phase. The PF coils are configured to cancel the CS stray field over a broad region around $R_0$, producing the null-field zone required for plasma initiation~\cite{formisano2017analysis,devries2019breakdown}. This is precisely what makes the infinite-solenoid hypothesis (Eq.~\ref{Flux_CS_eq}) applicable to a finite CS. Without this active compensation, the flux delivered by the CS at $R_0$ would have to be multiplied by a coupling factor smaller than one to account for the radial leakage of the field outside the CS bore. The null-field criterion ensures that this leakage is exactly compensated up to $R_0$, so that the plasma effectively sees the on-axis flux of an infinite solenoid.
	
	The second criterion is set by the radial equilibrium of the plasma column. Once the plasma carries current, the PF coils must also generate the Shafranov vertical field $B_{\rm vert}$ that holds the column at $R_0$. This contribution is independent of the breakdown configuration and enters Eq.~\ref{eq:Psi_PF_new}. Because $B_{\rm vert}$ is approximately uniform across the plasma cross-section, its flux is integrated over the full disk up to the magnetic axis.
	
	Applying Ampère's theorem to an infinite solenoid of winding pack thickness $R_{\text{CS}}^{\text{ext}} - R_{\text{CS}}^{\text{sep}}$ carrying a uniform current density $J_{\rm CS}^{\rm wost}$ yields
	
	\begin{equation}
		B_{CS} = \mu_0 J_{\rm CS}^{\rm wost} (R_{\text{CS}}^{\text{ ext}} - R_{\text{CS}}^{\text{ sep}})
		\label{eq_BCS}
	\end{equation}
	
	The only remaining unknown being $R_{\text{CS}}^{\text{ sep}}$, we can now compute it by combining Eqs.~\ref{eq:CS_flux}, \ref{Flux_CS_eq} and~\ref{eq_BCS}:
	
	\begin{equation}
		R_{\text{CS}}^{\text{sep}} = \sqrt[3]{ R_{\text{CS}}^{\text{ext}^3} - \frac{3 \left| \Psi_{\text{Init}} + \Psi_{\text{Ramp-Up}} + \Psi_{\text{plateau}} - \Psi_{\text{PF}} \right|}{2 \pi \mu_0 J_{\rm CS}^{\rm wost}} }
	\end{equation}
	
	However, since $J_{\rm CS}^{\rm wost}$ depends on $B_{CS}$ (through the superconductor scaling laws detailed in Appendix~\ref{JBappendix}), the system must be solved iteratively. Initialising $B_{CS}$ with a thin-solenoid estimate $B_{CS} \approx \Psi_{CS} / (\pi (R_{CS}^{ext})^2)$ provides an accurate starting point, and convergence is reached within a few iterations.
	
	\subsubsection{Mechanical stress}
	
\begin{figure}
		\centering
		\begin{subfigure}[b]{1\linewidth}
			\centering
			\includegraphics[width=1\linewidth]{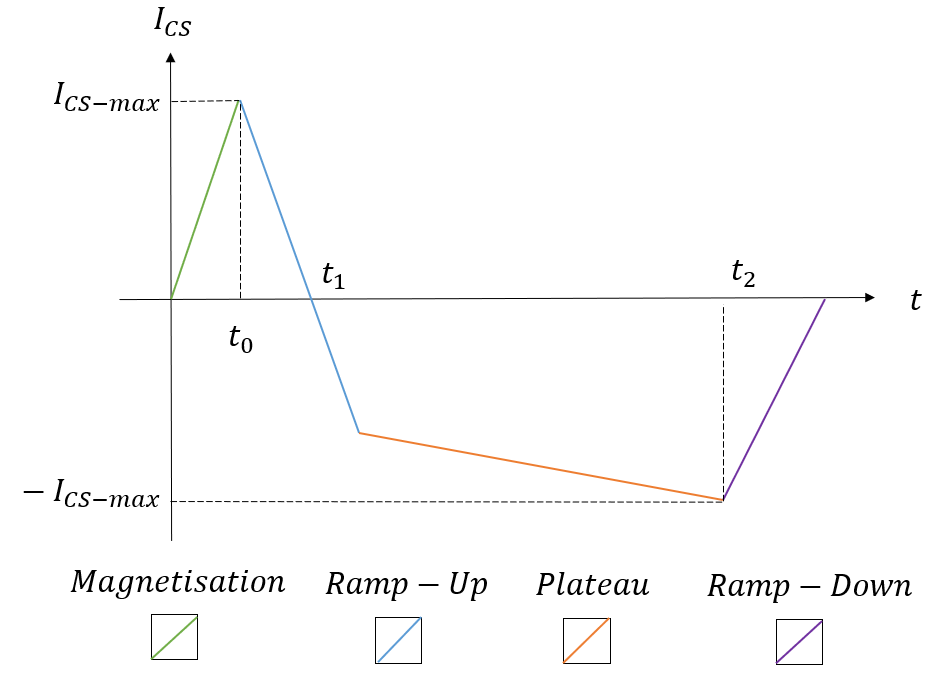}
			\caption{Schematic temporal evolution of the CS current during a typical pulsed discharge}
			\label{fig:icstemporal}
		\end{subfigure}
		
		\vspace{1em}
		
		\begin{subfigure}[b]{1\linewidth}
			\centering
			\includegraphics[width=1\linewidth]{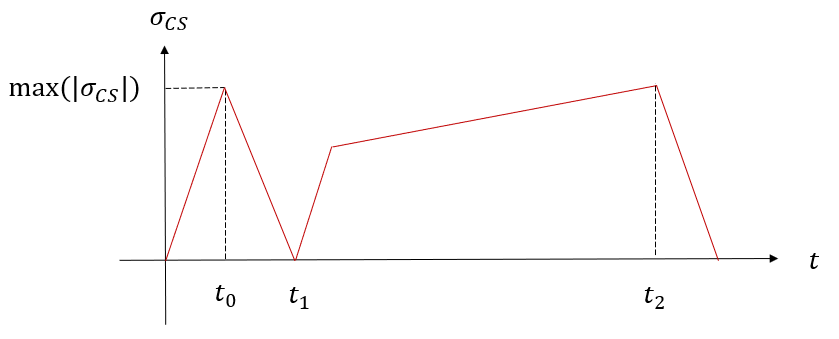}
			\caption{CS stress evolution during a typical discharge in wedging}
			\label{fig:sigmawedging}
		\end{subfigure}
		
		\vspace{1em}
		
		\begin{subfigure}[b]{1\linewidth}
			\centering
			\includegraphics[width=1\linewidth]{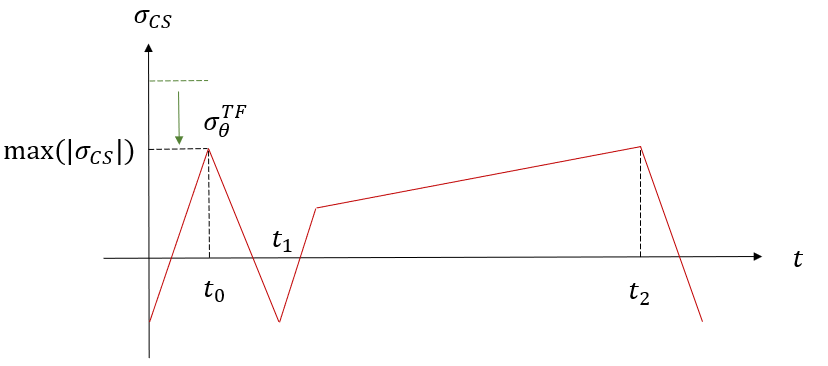}
			\caption{CS stress evolution during a typical discharge in "light" bucking}
			\label{fig:sigmabucking}
		\end{subfigure}
		
		\vspace{1em}
		
		\begin{subfigure}[b]{1\linewidth}
			\centering
			\includegraphics[width=1\linewidth]{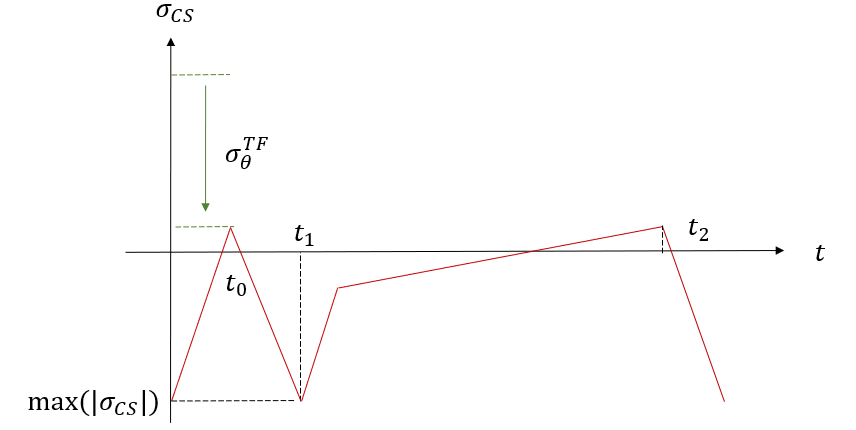}
			\caption{CS stress evolution during a typical discharge in "strong" bucking}
			\label{fig:sigmabuckingextreme}
		\end{subfigure}
		
		\caption{Temporal evolution of the CS current during a typical discharge (a), and of the resulting CS stresses for the wedging (b), light bucking (c), and strong bucking (d) configurations. Sign convention: tensile stresses are positive, compressive stresses are negative.}
	\end{figure}

	A temporal analysis of the forces acting on the CS is necessary to identify the limiting cases for sizing the CS. Fig. \ref{fig:icstemporal} illustrates the typical evolution of the current in the CS.
	
	The cycle begins with a pre-magnetization  phase where the CS current increases to \(I_{\text{CSmax}}\), followed by its discharge to initiate the plasma. During this phase, the current passes through zero at $t_1$ before reversing, enabling further ramp-up of the plasma current. During the plateau, \(I_p\) is then maintained at a constant value. The example plotted on Fig. \ref{fig:icstemporal} assumes that $I_p$ is at least partly driven by the CS during this phase. In a steady state scenario this is not the case, so that the slope of $I_{CS}$ is vanishing during this phase. However, the overall conclusion regarding mechanical stresses remain unchanged from Pulsed to Steady State operation. Finally, the current decreases back to \(I_{\text{CS}} = 0\) during the ramp-down phase.\\
	
	\textbf{a - Wedging}
	
	In the wedging configuration, the CS is free-standing: the TF coils are separated from the CS. Therefore, the CS needs to withstand its own $J \times B$ forces through hoop tension. As presented in Fig. \ref{fig:sigmawedging}, the critical moments occur at \(t_0\) and \(t_2\), when the CS reaches \(I_{\text{CS-max}}\) or \(- I_{\text{CS-max}}\).

	Within the thin cylinder approximation, the corresponding stress reads:
	
	\begin{equation}
		\sigma_{\theta} = \frac{P_{\text{CS}} R_{\text{CS}}^{\text{sep}}}{R_{\text{CS}}^{\text{sep}} - R_{\text{CS}}^{\text{int}}}
	\end{equation}
	with
	\begin{equation}
		P_{\text{CS}} = \frac{B_{\text{CS}}^2}{2 \mu_0}
		\label{P_CS_eq}
	\end{equation}
	
	Setting $\sigma_{\theta}$ = $\sigma_{\text{lim}}$, one can directly express $R_{\text{CS}}^{\text{int}}$:
	
	\begin{equation}
		R_{\text{CS}}^{\text{int}} = R_{\text{CS}}^{\text{sep}} \left(1-\frac{P_{CS}}{\sigma_{lim}}\right)
	\end{equation}
	
	\textbf{b - Bucking}
	
	In this configuration, the TF coils transfer their centering force to the CS, resulting in its compression.
	
	Two limiting instants could a priori dimension the CS. At $t_0$, the CS carries its maximum current: its own hoop stress $|\sigma_{\theta,\text{CS}}|$ (driven by $P_{\text{CS}}$) is partially relieved by the TF compression $|\sigma_{\theta,\text{TF}}|$ (driven by $P_{\text{TF}}$), giving a net stress $|\sigma_{\theta,\text{CS}}| - |\sigma_{\theta,\text{TF}}|$ (Fig.~\ref{fig:sigmabucking}). At $t_1$, the CS current vanishes and only the TF pre-compression remains, with a stress $|\sigma_{\theta,\text{TF}}|$ (Fig.~\ref{fig:sigmabuckingextreme}). The dimensioning case is the larger of the two. Comparing $P_{\text{CS}}$ (Eq.~\ref{P_CS_eq}) to $2 P_{\text{TF}}$ (Eq.~\ref{equation0}) shows that the CS own pressure dominates (referred to as 'light' bucking) only when $B_{\text{CS}} > \sqrt{2}\, B_{\text{max}}$, a condition that is unlikely to be met in tokamaks since $B_{\text{CS}}$ and $B_{\text{max}}$ are expected to be of comparable magnitudes. The TF pressure is therefore expected to dominate in essentially all relevant tokamak designs, a regime referred to as 'strong' bucking. Both branches are retained in D0FUS through a $\max(\cdot)$ comparison, but only the strong bucking case is detailed below.
	
	The hoop stress in the CS then reads:
	
	\begin{equation}
		|\sigma_{\theta}| = \frac{P_{\text{TF}}\, R_{\text{CS}}^{\text{sep}}}{R_{\text{CS}}^{\text{sep}} - R_{\text{CS}}^{\text{int}}}
	\end{equation}
	
	Setting $|\sigma_{\theta}| = \sigma_{\text{lim}}$, one obtains:
	
	\begin{equation}
		R_{\text{CS}}^{\text{int}} = R_{\text{CS}}^{\text{sep}} \left(1-\frac{P_{\text{TF}}}{\sigma_{\text{lim}}}\right)
		\label{equation_P_comparison}
	\end{equation}
	
	\textbf{c - Plug}
	\label{Pluging}
	
	The addition of a plug in the center of the CS as illustrated in Fig. \ref{fig:topviewtf} is an option of interest in the strong bucking case \cite{sorbom2015arc,wade2021cost}.
	
	The main advantage of this technique is that the plug experiences uniform radial pressure, producing a near-hydrostatic stress state ($\sigma_r \approx \sigma_\theta \approx -P$) in which the Tresca stress $|\sigma_r - \sigma_\theta| \approx 0$. The residual stresses are governed by local effects (material defects, geometric imperfections, etc.) and remain well below the allowable limits \cite{puthoff1969digital, yiannopoulos5stress, grigorenko2006solving}. The plug can therefore be assumed to withstand the applied loads without detailed dimensioning.
	
	The usefulness of such a plug logically depends on its stiffness, which determines the distribution between hoop  and radial stress in the CS. Indeed, a very soft plug will not significantly change the situation: the CS will still react the TF pressure through the vault effect, leading to large $\sigma_\theta$. Conversely, a very stiff plug will leave the CS to support only its $\sigma_r$, entirely transferring it to the plug.
	
\begin{figure*}
		\centering
		\includegraphics[width=1\linewidth]{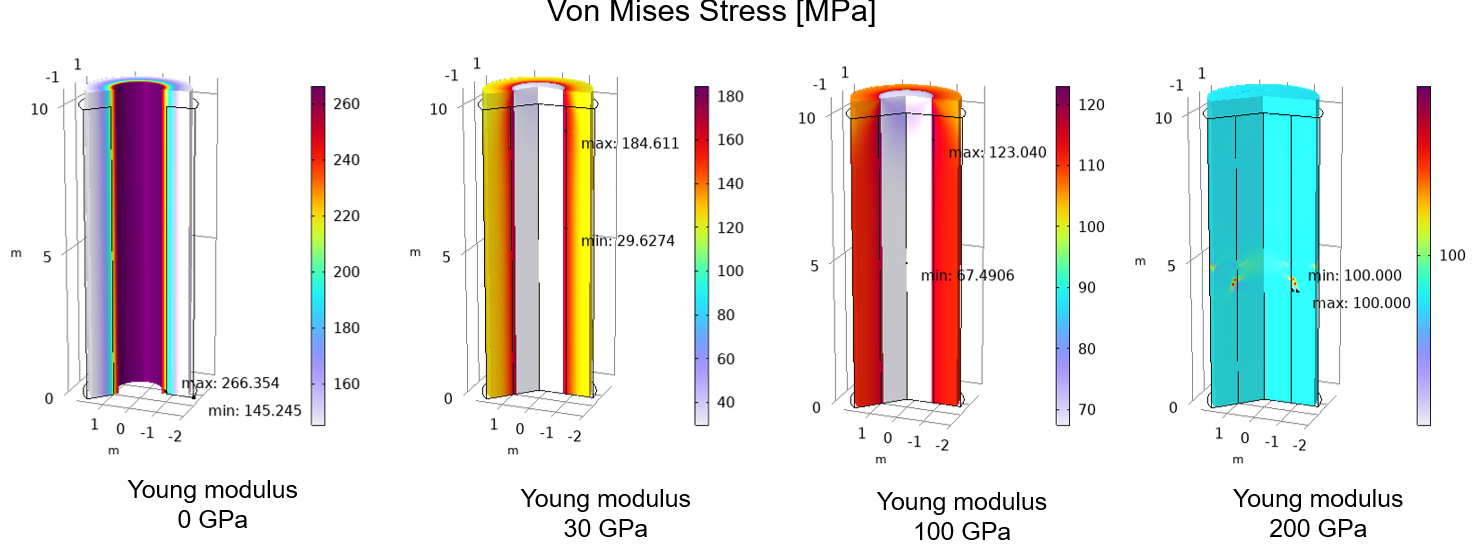}
		\caption{Von Mises stress from COMSOL simulations of a CS under 100~MPa external pressure, for varying plug Young's modulus.}
		\label{fig:comsolplug}
	\end{figure*}

	To illustrate this phenomenon and obtain relevant orders of magnitude, several simulations were carried out using COMSOL Multiphysics. An external pressure of 100 MPa (corresponding to the load exerted by a TF coil system at $B_{\max} \approx 16$~T) was applied to the outer radius of a typical CS (outer radius of 2 m and inner radius of 1 m) resembling that of ITER \cite{mitchell2011iter}. The CS material is assumed to be 316L steel, while the plug Young's modulus is varied as a free parameter. The influence of the Young's modulus on the maximum Von Mises stress (similar to Tresca stress) is then investigated (see Fig. \ref{fig:comsolplug}):
	
	One can see in Fig.~\ref{fig:comsolplug} that for a very soft plug (Young's modulus $\approx 0$~GPa), $\sigma_{\max} = max(\sigma_\theta)$, recovering the bucking case without a plug. Conversely, when the plug stiffness approaches that of steel (Young's modulus $= 200$~GPa), the stress becomes uniform with $\sigma_{\max} = \sigma_r$. Between these two extremes, a rapid gain is observed; $\sigma_{\max}$ is already reduced by 1/3 with a plug that offers low mechanical resistance (Young's modulus = 30 GPa). Above a Young's modulus of $\approx 100$~GPa, $\sigma_r$ is clearly the dominant component and the approximation $\sigma_{\max} \approx \sigma_r$ becomes acceptable.
	
	In practice, glass-fiber/epoxy composites (e.g., G10/G11) appear to be a good candidate for a central-solenoid plug because they offer very high electrical resistivity and low magnetic susceptibility (minimizing eddy current heating and magnetic interactions) while providing adequate mechanical strength and machinability\cite{kasen1980mechanical} \cite{li2019flashover} \cite{benzinger1980manufacturing}. The typical Young's modulus of these materials is 100 GPa \cite{kasen1980mechanical}.Since the CS winding pack is a composite of steel and non-structural materials (superconductor,copper, insulation, cooling), its effective Young's modulus is lower than that of pure 316L steel (200~GPa). A G10 plug ($\approx 100$~GPa) is therefore stiff enough relative to the CS to justify the approximation $\sigma_{\max} \approx \sigma_r$, as confirmed by the COMSOL simulations.
	
	Moreover, in the thin cylinder approximation, $\sigma_r$ is simply equal to $P_{TF}$, and thus does not depend at all on the steel thickness, suggesting that the latter could be very small. This illustrates that, to first order, the plug configuration relaxes the mechanical constraint to the simple condition $P_{\text{TF}} < \sigma_{\text{lim}}$, which is easily satisfied at all field levels considered here.
	
	\section{Refined model}
	\label{D0FUS}
	
	The Refined model provides more realistic thickness predictions at the cost of moderately increased complexity. It draws on classical analytical treatments of stress distributions in fusion magnets \cite{thome1982mhd, wilson1983superconducting, burkhard1975magnetic, gray1977electromechanical, johnson1977stress, swanson2022validation, boudes2025circe} and on their integration into system codes \cite{morris2015implications, morris2021preparing, reux2018demo, duchateau2014conceptual, giannini2023magnet}. Two key changes are introduced with respect to the Academic model: thin-cylinder stress theory is replaced by thick-cylinder (Lamé-Clapeyron) theory, and the idealised pure-conductor layer is replaced by a composite winding pack that includes the steel jacket of the conductor.
	
	We make use of the notation set in the academic model and presented in Fig. \ref{fig:2layerstf}.
	
	In this model, we consider the CS and TF coils as being composed of CICC. In reality, the TF coil can also be composed of radial plates in which the cables are embedded, but the sizing of the different thicknesses would not change.
	
	\subsection{TF winding pack}
	\label{TF_Complex}
	
	\subsubsection{Preliminary definitions}
	\label{Param_definition}
	
	Let us introduce some parameters that will be used throughout the model to characterize the CICC.
	
\begin{figure}
		\centering
		\includegraphics[width=1\linewidth]{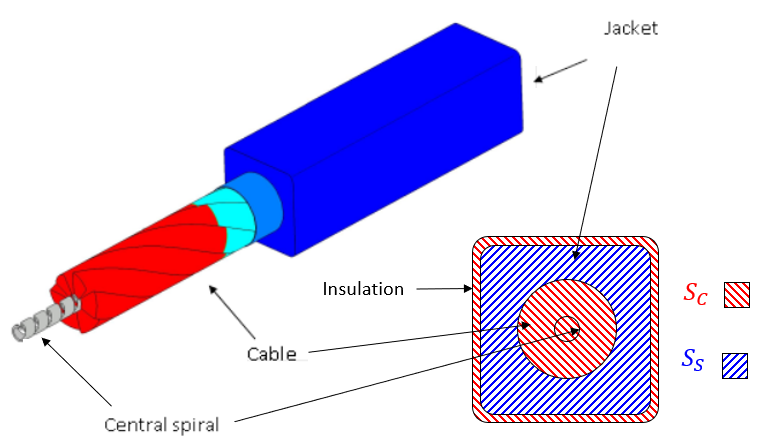}
		\caption{CICC illustration and surface definition: in red the non-steel cross-section on which $J^{\rm wost}$ is defined ($S_C$), and in blue the steel jacket sized for mechanical purposes ($S_S$)}
		\label{fig:wpmodelcicc_eclate}
	\end{figure}

	First, the cable fraction in the plane perpendicular to the $z$ axis is defined as:
	\[
	f_c = \frac{S_C}{S_{C} + S_{S}}
	\]
	with $S_S$ the steel jacket surface and $S_C$ the non-steel conductor surface (superconductor, copper stabiliser, helium cooling channels and insulation) in a given CICC, as presented in Fig.~\ref{fig:wpmodelcicc_eclate}. In the case where the whole cross-section of a given coil ($R,Z$ plane for the CS, $R,\theta$ plane for the TF coil winding pack) is filled in with identical CICC, then this fraction $f_c$ does not depend on the radial position and is sized to handle the worst loading case. Conversely, one might consider so-called grading techniques to concentrate the steel in areas where stresses are highest, resulting in a radial variation of $f_c$. This case is discussed in section \ref{grading}.
	
	The second parameter, $f_{z,\mathrm{WP}}$, is the fraction of the total tension force $F_z$ which is supported by the winding pack $F_{z,\mathrm{WP}}$:
	\[
	f_{z,\mathrm{WP}} = \frac{F_{z,\mathrm{WP}}}{F_z}
	\]
	This parameter, ranging from 1 (all the tension is held by the winding pack) to 0 (all the tension is held by the nose), affects the distribution of steel between the winding pack and the nose but has little impact on the total inboard leg thickness, as shown in Appendix~\ref{appendix_omega_sensitivity}. It would have been possible to consistently determine $f_{z,\mathrm{WP}}$ by comparing the steel surfaces of the two components and iterating until convergence, but the weak sensitivity of the final result on this parameter does not justify such an additional convergence loop. In wedging, the default $f_{z,\mathrm{WP}} = 1/2$ is therefore adopted, consistent with the ITER design~\cite{federici2026iter}. In bucking, it is logically set to $f_{z,\mathrm{WP}} = 1$ since there is no nose.
	
\begin{figure}
		\centering
		\begin{subfigure}[b]{0.65\linewidth}
			\centering
			\includegraphics[width=1\linewidth]{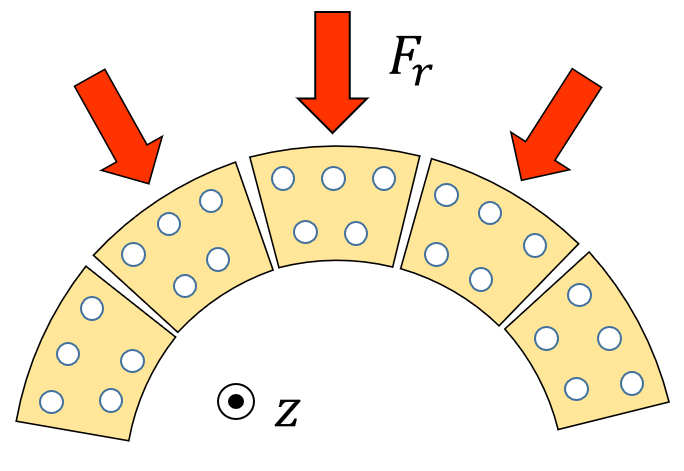}
			\caption{Vault created by the coils}
			\label{fig:surfacedilution_a}
		\end{subfigure}
		
		\begin{subfigure}[b]{0.65\linewidth}
			\centering
			\includegraphics[width=1\linewidth]{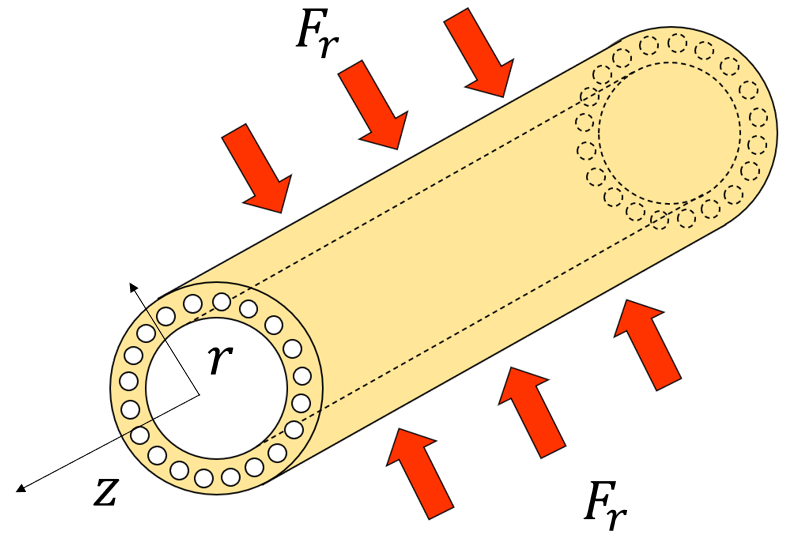}
			\caption{Cylindrical approximation of the coils}
			\label{fig:surfacedilution_b}
		\end{subfigure}
		
		\begin{subfigure}[b]{0.65\linewidth}
			\centering
			\includegraphics[width=1\linewidth]{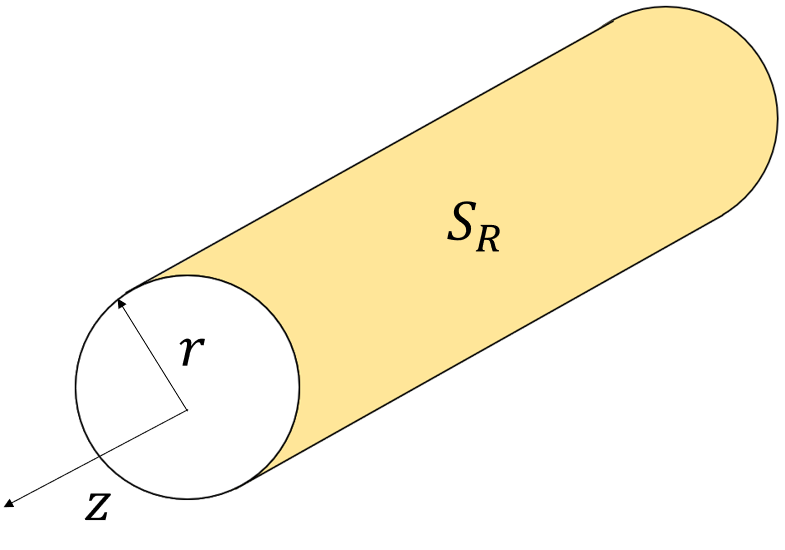}
			\caption{Complete cylindrical surface}
			\label{fig:surfacedilution_c}
		\end{subfigure}
		
		\begin{subfigure}[b]{0.65\linewidth}
			\centering
			\includegraphics[width=1\linewidth]{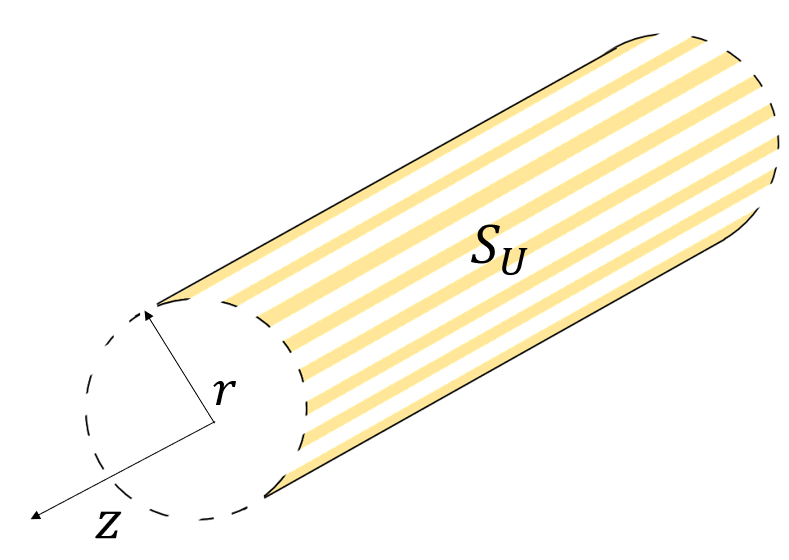}
			\caption{Useful cylindrical surface}
			\label{fig:surfacedilution_d}
		\end{subfigure}
		
		\caption{Cylinder with drilled hole approximation of the coil (a and b) and definition of the surfaces $S_R$ and $S_U$ used in the stress calculations (c and d).}
		\label{fig:surfacedilution}
	\end{figure}

	Finally, $f_u$ is defined by:
	\[
	f_u = \frac{S_U}{S_{R}}
	\]
	where $S_U$ is the useful steel surface that can be used to withstand the loads in the $r$ direction and $S_R$ the total cylindrical surface formed by the coils as presented in Fig. \ref{fig:surfacedilution}. Indeed, considering that the cable in the CICC cannot support any load (which is true for Low Temperature Superconductor (LTS, namely Nb3Sn and NbTi), but more debatable for High Temperature Superconductor (HTS, namely, REBCO) \cite{Godeke2006_Nb3Sn_review,Zhou2023_REBCO_mech,scanlan1980mechanical, barth2015electro}), they are mechanically equivalent to holes in the winding pack, leading to a radial stress concentration factor.
	
	A summary of these three fractions is given in Table~\ref{tab:fractions_recap} for reference throughout the rest of the paper.
	
	\begin{table}[ht]
		\centering
		\small
		\caption{Geometric and load-distribution fractions used throughout the Refined model.}
		\label{tab:fractions_recap}
		\begin{tabular}{cl}
			\toprule
			\textbf{Symbol} & \textbf{Definition} \\
			\midrule
			$f_c$               & Cable fraction: $S_C / (S_C + S_S)$ \\
			$f_u$               & Useful steel fraction for $\sigma_r$: $S_U / S_R$ \\
			$f_{z,\mathrm{WP}}$ & WP share of vertical tension: $F_{z,\mathrm{WP}} / F_z$ \\
			\bottomrule
		\end{tabular}
	\end{table}
	
	\begin{figure}
		\centering
		\includegraphics[width=1\linewidth]{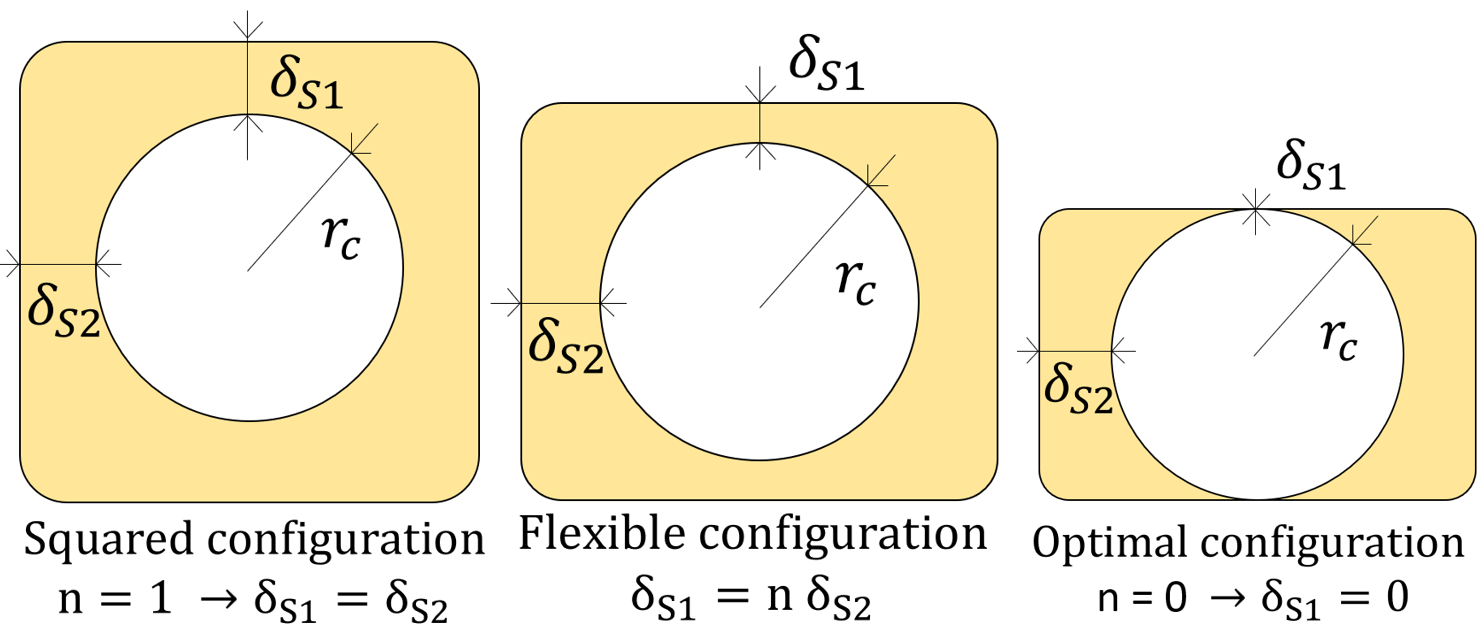}
		\caption{Conductor with parametrisation of its characteristic lengths in a generic, $n = 0$ and finally $n = 1$ cases.}
		\label{fig:surfacedilutionconductor}
	\end{figure}
	
	\subsubsection{Generation of \texorpdfstring{$B_{\text{max}}$}{Bmax}}
	
	Following the same procedure as in Eq.~\ref{equation_B_thin}, but without the thin-cylinder approximation (see Appendix~\ref{Annexe_B_thinlayer}), we obtain:
	
	\[
	B_{\text{max}} = \frac{\mu_0 f_c J_{\rm TF}^{\rm wost}}{2} \left( R_{\rm TF}^{\rm ext} - \frac{(R_{\rm TF}^{\rm sep})^2}{R_{\rm TF}^{\rm ext}} \right)
	\]
	
	leading to:
	
	\begin{equation}
		f_c = \frac{2 B_{\text{max}}}{\mu_0 J_{\rm TF}^{\rm wost}} \frac{R_{\rm TF}^{\rm ext}}{(R_{\rm TF}^{\rm ext})^2 - (R_{\rm TF}^{\rm sep})^2}
		\label{alpha_equation}
	\end{equation}
	
	Here, $J^{\rm wost}$ denotes the current density on the non-steel cross-section of the winding pack (cf.\ Section~2.1.1 and Appendix~\ref{JBappendix}).
	
	\subsubsection{Derivation of \texorpdfstring{$f_u$}{fu}}
	
\begin{figure}
		\centering
		\includegraphics[width=0.9\linewidth]{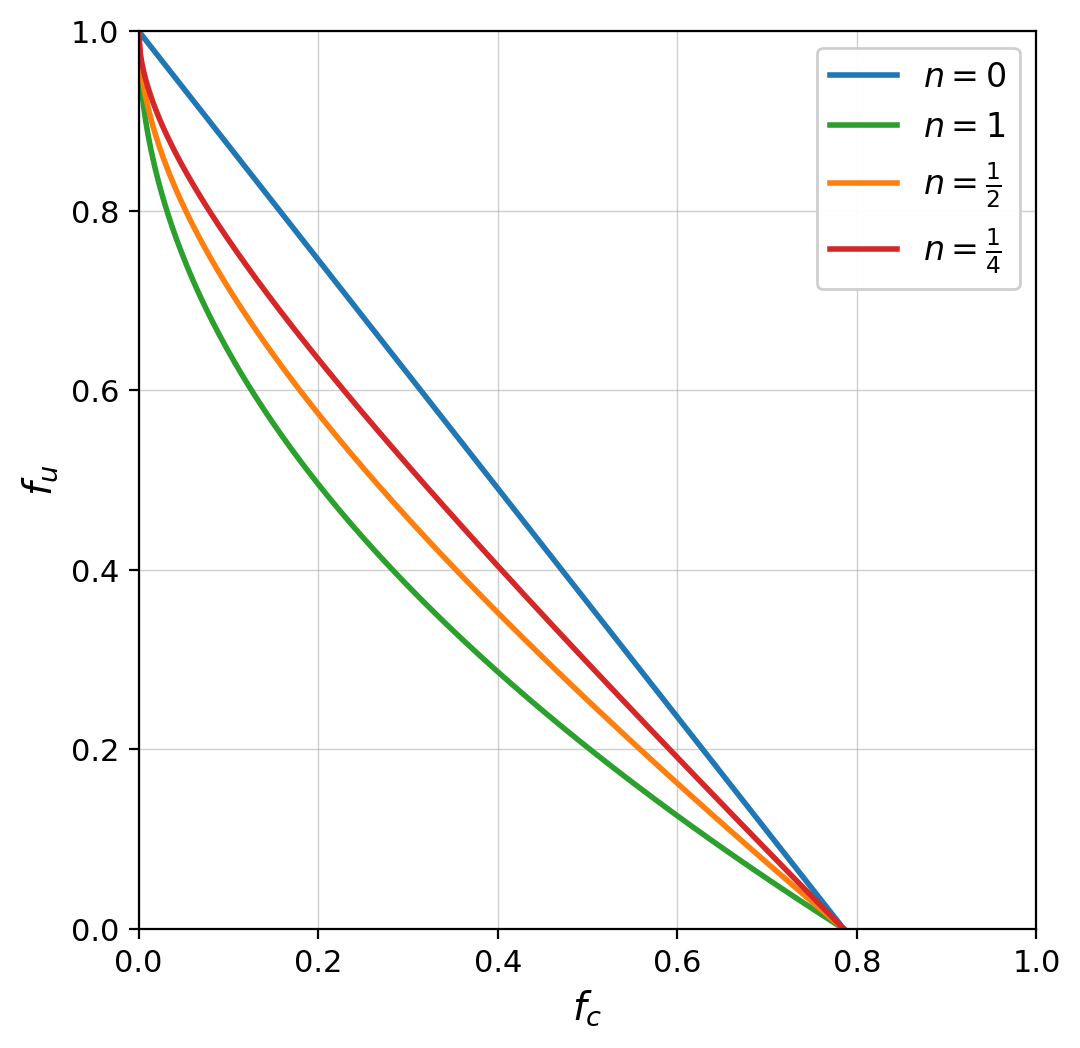}
		\caption{$f_u(f_c)$ considering different $n$ values using Eq. \ref{gamma}}
		\label{fig:figure-2025-06-11-105645}
	\end{figure}

	The variable $f_u$ can be expressed as a function of $f_c$ and of a conductor jacket asymmetry parameter $n$ as illustrated in Fig. \ref{fig:surfacedilutionconductor}. This parameter is defined as the ratio $\delta_{S1}/\delta_{S2}$, where $\delta_{S1}$ and $\delta_{S2}$ denote the steel thicknesses in the radial and toroidal directions, respectively. The usual value of $n$ ranges from 1 (corresponding to a square conductor) down to 0 (corresponding to a conductor with no steel in the radial direction, maximising the useful steel fraction $f_u$ for radial stress). The default value adopted in D0FUS is $n = 1$, corresponding to the standard round-in-square CICC geometry. This choice yields the lowest $f_u$ for a given $f_c$, and therefore the thickest coil predictions. Figure \ref{fig:figure-2025-06-11-105645} shows how different values of $n$ affect the resulting $f_u(f_c)$.
	
	\[
	f_u(f_c,n) =
	\]
	\begin{equation}
		\frac{\,2\pi + 4 f_c (n-1)\ -
			\sqrt{\bigl(2\pi + 4 f_c (n-1)\bigr)^2 - 4\pi\,(\pi - 4 f_c)}\,}
		{2\pi}
		\label{gamma}
	\end{equation}
	
	The derivation is given in Appendix~\ref{Annexe_gamma}.
	
	\subsubsection{Mechanical stress endurance}
	\label{gamma_usefull}
	
	Now considering a thick cylinder with homogeneous properties, the expressions of $\sigma_z$ and $\sigma_r$ are analogous to those derived in Section~\ref{Academic} except for the surface considerations.
	
	The surface used to estimate the most critical value of $\sigma_r$ corresponds to $S_U = f_u\, S_{\text{R}}$ (see Tab.~\ref{tab:fractions_recap}). Denoting $F_r$ the centering force applied to the inner legs of the TF coils, one then gets:
	
	\[
	\sigma_r = \frac{F_r}{S_U} = \frac{F_r / S_{\text{tot}}}{f_u} = \frac{P_{\text{TF}}}{f_u}
	\]
	
	with $P_{\text{TF}}$ the magnetic pressure as expressed in Eq.~\ref{equation0}.
	
	As far as the vertical stress is concerned, only the fraction $f_{z,\mathrm{WP}}$ of the tension $F_z$ is supported by the WP. Further accounting for the surface $(1 - f_c) S_{\text{tot}}$ ($S_{\text{tot}}$ being the total surface defined as $S_C + S_S$) occupied by steel in the plane perpendicular to the $z$ axis, and considering a uniform vertical stress, one gets:
	
	\begin{equation}
		\sigma_z = \frac{f_{z,\mathrm{WP}}\, F_z}{(1 - f_c) S_{\text{tot}}}
		\label{eq:sigma_z_uniform}
	\end{equation}
	\[
	= \frac{f_{z,\mathrm{WP}}}{(1-f_c)} \frac{B_{max}^2 (R_{\rm TF}^{\rm ext})^2}{((R_{\rm TF}^{\rm ext})^2 - (R_{\rm TF}^{\rm sep})^2) 2 \mu_0} \ln \left( \frac{R_0 + a + \Delta_{B}}{R_0 - a - \Delta_{B}} \right)
	\]
	For the $\sigma_\theta$ calculation (wedging case), departing from the thin cylinder approximation, the Lamé-Clapeyron theory yields a different expression detailed in Appendix~\ref{app:thin_wall_stress}: 
	
	\[
	\sigma_\theta = \frac{2 P_{TF} (R_{\rm TF}^{\rm ext})^2}{((R_{\rm TF}^{\rm ext})^2 - (R_{\rm TF}^{\rm sep})^2)}\frac{1 - f_u + n\, f_u}{n\, f_u}
	\]
	
	The $\frac{1 - f_u + n\, f_u}{n\, f_u}$ factor accounts for the useful steel surface in the $\theta$ direction (derivation in Appendix~\ref{Appendix_little_demo}). This expression is reported for completeness; as detailed below, $\sigma_\theta$ is not retained in the winding pack sizing.
	
	In both configurations (wedging and bucking), the hoop stress $\sigma_\theta$ in the winding pack is neglected: it is assumed to be predominantly recovered by the nose. In reality, $\sigma_\theta$ contributes to the winding pack stress (in ITER it is estimated that about 30$\%$ of the vault effect is taken by the WP) \cite{wilson1983superconducting}, but as discussed in Appendix~\ref{Wedgapproximation}, this mainly affects the steel distribution between the nose and the winding pack, not the total coil thickness. With $\sigma_\theta \approx 0$ in the winding pack (WP), and noting that $\sigma_r < 0$ (compression) while $\sigma_z > 0$ (tension), the Tresca criterion reduces to:
	
	\[
	\sigma_{\text{Tresca}} = \sigma_z - \sigma_r \leq \sigma_{\text{lim}}
	\]
	
	This expression holds for both mechanical architectures.\\
	
	Setting $\sigma_{\text{Tresca}} = \sigma_{\text{lim}}$ and solving for $R_{\rm TF}^{\rm sep}$ yields an implicit equation, since $f_u$ depends on $f_c$ (Eq.~\ref{gamma}), which itself depends on $R_{\rm TF}^{\rm sep}$ (Eq.~\ref{alpha_equation}). The solution is found numerically by iterating inward from $R_{\rm TF}^{\rm ext}$.
	
	\subsubsection{Cover}
	
\begin{figure}
		\centering
		\includegraphics[width=0.7\linewidth]{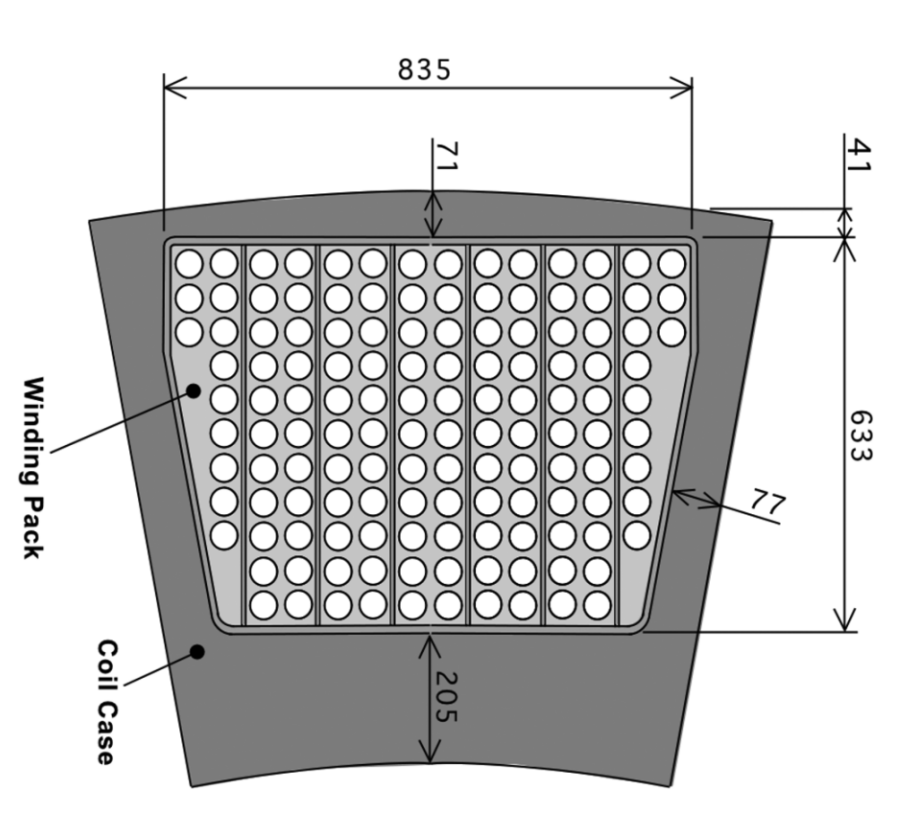}
		\caption{ITER TF cross-section taken from \cite{sborchia2008design}}
		\label{fig:ITER}
	\end{figure}

	Finally, a cover can be needed for reinforcement and high-voltage impregnation purposes (see Fig.~\ref{fig:ITER}). We choose a default thickness of 7~cm, following the ITER design (this thickness appears relatively consistent across all designs studied, ranging from 5 to 10~cm).
	
	\subsection{TF coil nose}
	\label{TF_nose}
	
	In the wedging configuration (the only one where a steel nose is considered), the centering force is conserved but applied on a smaller radius, so that the effective pressure at $R_{\text{TF}}^{\text{sep}}$ is:
	
	\[
	P'_{\text{TF}} = P_{\text{TF}} \frac{R_{\rm TF}^{\rm ext}}{R_{\text{TF}}^{\text{sep}}} = \frac{B_{\text{max}}^2}{2 \mu_0} \frac{R_0 - a - \Delta_{B}}{R_{\text{TF}}^{\text{sep}}}
	\]
	
	The nose is in azimuthal compression ($\sigma_\theta < 0$) and vertical tension ($\sigma_z > 0$). The Tresca criterion therefore reads:
	
	\[
	\sigma_{\text{Tresca}} = \sigma_z - \sigma_\theta \leqq \sigma_{\text{lim}}
	\]
	with:
	\[
	\sigma_\theta = -\frac{2 P'_{\text{TF}} \left(R_{\text{TF}}^{\text{sep}}\right)^2}{\left(R_{\text{TF}}^{\text{sep}}\right)^2 - (R_{\rm TF}^{\rm int})^2} 
	\]
	\[
	\sigma_z = (1-f_{z,\mathrm{WP}})\frac{B_{\text{max}}^2 \left(R_{\text{TF}}^{\text{sep}}\right)^2}{\left[\left(R_{\text{TF}}^{\text{sep}}\right)^2 - (R_{\rm TF}^{\rm int})^2\right] 2 \mu_0} \ln\left( \frac{R_0 + a + \Delta_{B}}{R_0 - a - \Delta_{B}} \right)
	\]
	
	Setting $\sigma_{\text{Tresca}} = \sigma_{\text{lim}}$ leads to:
	
	\[
	R_{\rm TF}^{\rm int} = 
	\sqrt{ \left(R_{\text{TF}}^{\text{sep}}\right)^2 - \frac{\left(R_{\text{TF}}^{\text{sep}}\right)^2}{\sigma_{\text{lim}}} \left[ 2 P'_{\text{TF}} + (1-f_{z,\mathrm{WP}}) \frac{B_{\text{max}}^2}{2 \mu_0} \ln\left( \frac{R_0 + a + \Delta_{B}}{R_0 - a - \Delta_{B}} \right) \right] }
	\]
	
	\subsection{CS winding pack}
	
	The CS winding pack is modelled analogously to the TF coil winding pack, using the same fractions $f_c$ and $f_u$ introduced in Section~\ref{Param_definition} (Tab.~\ref{tab:fractions_recap}). Since no CS nose is considered in this model, the parameter $f_{z,\mathrm{WP}}$ is not used.
	
	\subsubsection{Generation of \texorpdfstring{$\Psi$}{Psi}}
	
	Combining the flux balance (Eq.~\ref{eq:CS_flux}), the CS flux expression (Eq.~\ref{Flux_CS_eq}) and Ampère's theorem (Eq.~\ref{eq_BCS}) with the cable fraction $f_c$, one obtains:
	
	$$
	B_{CS} = \frac{3(\Psi_{\rm Init} + \Psi_{\rm plateau} + \Psi_{\rm Ramp\text{-}Up} - \Psi_{\rm PF})}{2 \pi ((R_{\rm CS}^{\rm ext})^2 + R_{\rm CS}^{\rm ext}R_{\rm CS}^{\rm int} + (R_{\rm CS}^{\rm int})^2)}
	$$
	
	$$
	f_c  =  \frac{B_{CS}}{\mu_0 J_{\rm CS}^{\rm wost} (R_{\rm CS}^{\rm ext} - R_{\rm CS}^{\rm int})}
	$$
	
	The only unknown is $R_{\rm CS}^{\rm int}$, since $J_{\rm CS}^{\rm wost}$ is determined from $B_{CS}$ via the superconductor scaling laws (Appendix~\ref{JBappendix}).
	
	\subsubsection{Axial stress at the CS midplane}
	\label{sec:CS_sigma_z}
	
	Unlike the Academic model, the Refined model accounts for the axial compressive stress induced by the radial component of the fringe field at the CS ends. In a finite-length solenoid, $\nabla \cdot B = 0$ implies a non-zero $B_r$ near the coil extremities. The resulting $J_\theta \times B_r$ force pushes the winding pack toward the midplane. Integrating from the free end ($z = h$, $\sigma_z = 0$) to the midplane ($z = 0$) yields (derivation in Appendix~\ref{appendix_sigma_z_CS}):
	
	\begin{equation}
		\sigma_{z}^{\rm smear} = -\frac{\mu_0 J_{\rm smear}^2\, h\, R_{\rm CS}^{\rm int}}{2} \left[\mathcal{L}(h) - \mathcal{L}(2h)\right]
		\label{eq:sigma_z_CS}
	\end{equation}
	
	with $J_{\rm smear} = f_c\, J_{\rm CS}^{\rm wost}$ the homogenised current density, $h = H_{\rm CS}/2$ the CS half-height, and:
	
	\[
	\mathcal{L}(\zeta) = \ln\frac{R_{\rm CS}^{\rm ext} + \sqrt{(R_{\rm CS}^{\rm ext})^2 + \zeta^2}}{R_{\rm CS}^{\rm int} + \sqrt{(R_{\rm CS}^{\rm int})^2 + \zeta^2}}
	\]
	
	The peak steel stress is $\sigma_z^{\rm steel} = \sigma_z^{\rm smear}/f_u$. This stress is compressive and typically one order of magnitude smaller than $\sigma_\theta$. It is computed using the CS current at the most critical instant identified in Section~\ref{Flux}: in wedging configuration and in light bucking, this corresponds to $I_{\rm CS} = I_{\rm CS,max}$. In strong bucking ($I_{\rm CS} = 0$), $\sigma_z$ vanishes.
	
	Note that a typical modular CS, where individual modules carry different and sometimes reversed currents, gives rise to a more complex stress distribution. Nonetheless, the monolithic solenoid approximation provides a reasonable first estimate of the relevant magnitudes.
	
	\subsubsection{Mechanical solution in wedging}
	
	The most critical hoop stress occurs at $R_{\text{CS}}^{\text{int}}$ (where $\sigma_r = 0$). The CICC is oriented in the $\theta$ direction, so the useful steel surface factor is $f_c$ (rather than $f_u$). Applying the Lamé-Clapeyron theory~\cite{LameClapeyron1833} for a thick-walled cylinder under internal magnetic pressure $P_{CS}$:
	
	\[
	\sigma_{\theta}^{\max}
	= \frac{1}{(1- f_c)} \frac{P_{CS}\bigl[(R_{\text{CS}}^{\text{ext}})^2+(R_{\text{CS}}^{\text{int}})^2\bigr]}
	{(R_{\text{CS}}^{\text{ext}})^2-(R_{\text{CS}}^{\text{int}})^2}
	\]
	
	Since $|\sigma_z^{\rm steel}| \ll |\sigma_\theta^{\max}|$, the Tresca criterion is dominated by the hoop stress. Setting $|\sigma_\theta^{\max} - \sigma_z^{\rm steel}| = \sigma_{\text{lim}}$ yields a polynomial equation of degree 7 in $R_{CS}^{\text{int}}$, solved numerically.
	
	\label{sec:CS_fatigue}
	In wedging and light bucking, the CS "breathes" at every plasma pulse, going from full hoop tension at maximum current to rest when discharged. This cyclic loading induces fatigue, whose actual impact depends on the steel microstructure and the number of cycles over the plant lifetime~\cite{jong2007iter, sarasola2020progress, sutcliffe2025magnet}. As a first approximation, $\sigma_{\text{lim}}$ is divided by a factor of 2 in the Tresca criterion, consistent with standard practice in preliminary design studies. In strong bucking and plug configurations, the CS remains in compression throughout the cycle, which tends to close rather than propagate cracks~\cite{elber1971significance, pippan2017fatigue, newman1981crack, shih1974study}; the fatigue knockdown is therefore not applied in these architectures.
	
	\subsubsection{Mechanical solution in bucking}
	
	As established in Section~\ref{sec:bucking}, the regime expected for tokamak designs is strong bucking, where the critical instant is $I_{\rm CS} = 0$ and only the TF pressure acts on the CS:
	
	\[
	\sigma_{\theta}^{\max} = \frac{1}{(1- f_c)} \frac{2 P''_{\text{TF}} (R_{\text{CS}}^{\text{ext}})^2}{(R_{\text{CS}}^{\text{ext}})^2 - (R_{\text{CS}}^{\text{int}})^2}
	\]
	
	where $P''_{\text{TF}}$ is the TF centering pressure transported to the CS outer surface, defined analogously to $P'_{\text{TF}}$ (Section~\ref{TF_nose}):
	\[
	P''_{\text{TF}} = \frac{B_{\text{max}}^2}{2 \mu_0} \frac{R_0 - a - \Delta_{B}}{R_{\text{CS}}^{\text{ext}}}
	\]
	
	The Tresca criterion reduces to $|\sigma_{\theta}^{\max} - \sigma_z^{\rm steel}| \leqq \sigma_{\text{lim}}$, solved numerically as in the wedging case. The light bucking branch is implemented in D0FUS for completeness through a $\max(\cdot)$ comparison but is not expected to be selected in tokamak designs.
	
	\subsubsection{Mechanical solution in plug}
	
	When the TF pressure dominates over the CS own pressure ($|P_{\rm CS} - P_{\rm TF}| \leq |P_{\rm TF}|$, the usual case as discussed in Section~\ref{sec:bucking}), the critical instant is $I_{\rm CS} = 0$ and the CS simply transmits the radial pressure to the plug. The dominant stress is $\sigma_r = P''_{\text{TF}}/f_u$, and the sizing equation is solved numerically as in previous configurations. In the rare opposite case, the bucking model of the previous section applies.
	
	\section{Benchmark}
	\label{Benchmark}
	
	The D0FUS Refined model is now benchmarked against reference codes and machine designs. We note that no systems or design code was found in the literature providing detailed magnet sizing in bucking or plug configurations. A cross-check is nonetheless possible against the published SPARC (bucking) and ARC (plug) designs, as reported in Sections~\ref{sec:TF_bench} and~\ref{sec:CS_bench}.
	
	\subsection{TF coil comparison with MADE}
	
	MADE (MAgnet Design Explorer) is a parametric optimisation tool developed for designing magnet systems for tokamaks \cite{giannini2023magnet}. It accounts for electromagnetic, structural, and superconducting constraints. MADE has been cross-checked against other pre-sizing tools such as the CEA magnet design codes (MADMACS \cite{torre2016tools,sutcliffe2025magnet}), and its outputs have been found consistent with detailed finite element analyses carried out downstream on specific design points, notably for the EU-DEMO TF and CS.
	
\begin{figure*}
		\centering
		\begin{subfigure}[b]{0.48\linewidth}
			\centering
			\includegraphics[width=\linewidth]{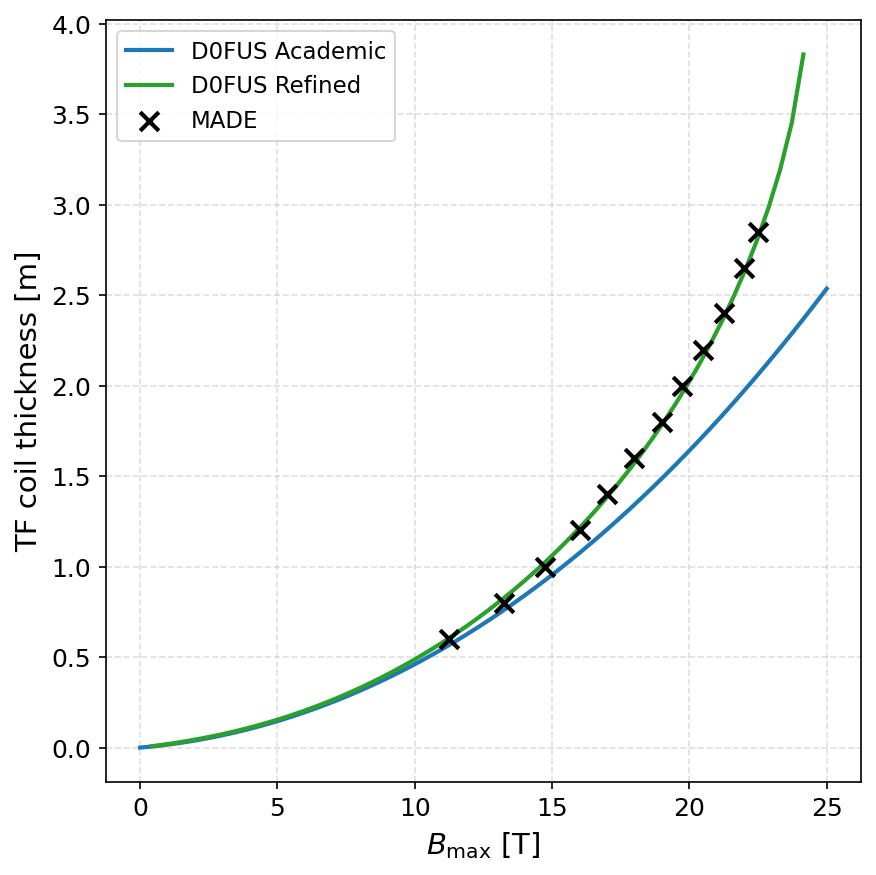}
			\caption{TF coil: scan in maximum field on the inner leg}
			\label{fig:Wedging}
		\end{subfigure}
		\hfill
		\begin{subfigure}[b]{0.48\linewidth}
			\centering
			\includegraphics[width=\linewidth]{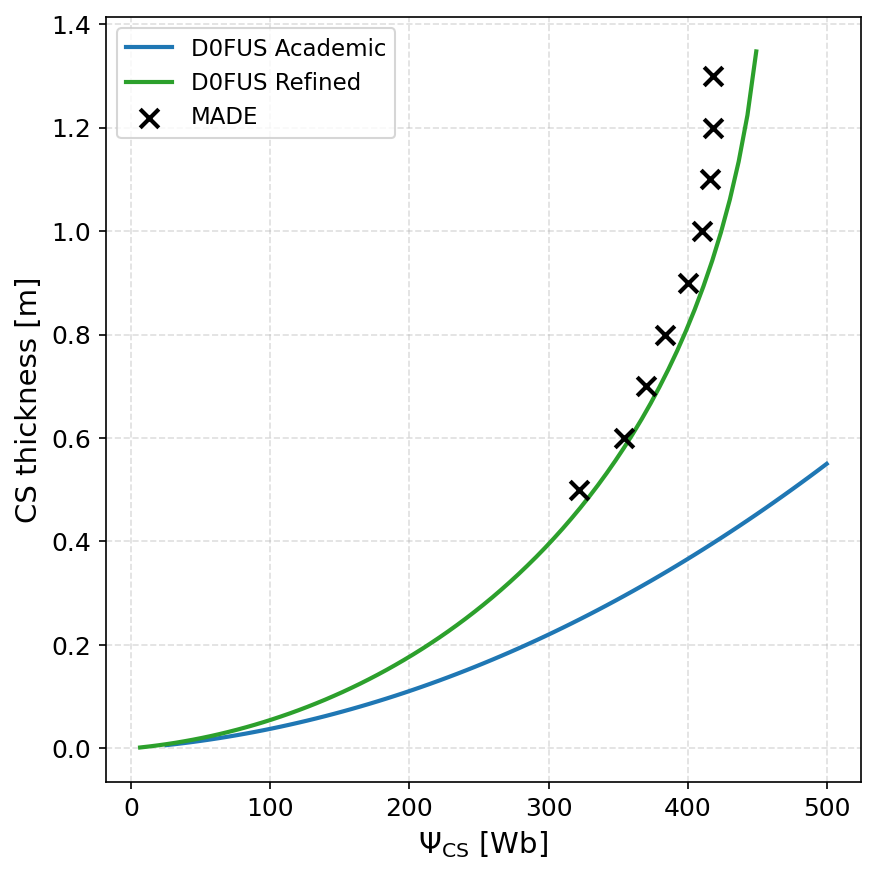}
			\caption{CS: scan in magnetic flux $\Psi_{\rm CS}$}
			\label{fig:WedgingCS}
		\end{subfigure}
		\caption{Comparison of the predicted thicknesses from the D0FUS Academic and Refined models with the MADE code in wedging configuration, using reference data from Ref.~\cite{giannini2023magnet}~fig.~19 for the TF coils and Ref.~\cite{sarasola2020progress}~fig.~2 for the CS.}
		\label{fig:MADE_benchmark}
	\end{figure*}

	To benchmark the D0FUS TF coil models against MADE, we base our analysis on Fig.~19 of Ref.~\cite{giannini2023magnet}. We perform the same magnetic field scan considering an outer radius of the TF coil inner leg of 4.3~m, a wedging architecture, a round-in-square CICC with HTS superconductor (T = 20~K), and an austenitic steel ($\sigma_{\text{lim}} = 867$~MPa). The resulting thicknesses are compared in Fig.~\ref{fig:Wedging}.
	
	One can observe a quasi-perfect agreement between the Refined model and MADE. on the other hand, the Academic model reproduces the qualitative trend well but is not quantitatively accurate at high magnetic field, as expected given its simplified assumptions, notably the lack of steel in the winding pack.
	
	\subsection{CS comparison with MADE}
	
	To benchmark the D0FUS CS models against MADE, we base our analysis on Fig.~2 of Ref.~\cite{sarasola2020progress} (at the time not yet named MADE). Note that the original figure is expressed in terms of the half-swing flux, whereas D0FUS uses the full-swing convention (Eq.~\ref{Flux_CS_eq}), the abscissa values of Fig.~\ref{fig:WedgingCS} are therefore twice those read from the original reference.
	
	We conduct the same magnetic flux scan, taking into account an outer radius of the CS of 2.7~m, a CS height $H_{\rm CS} = 17.92$~m (EU-DEMO baseline 2018 allocation \cite{sarasola2020progress}), a wedging architecture, an HTS superconductor, and austenitic steel with a fatigue-reduced allowable stress $\sigma_{\text{eff}} = 300$~MPa. The resulting thicknesses are presented in Fig.~\ref{fig:WedgingCS}.
	
	One can observe a good match between the Refined model and MADE, although slightly optimistic. On the other hand, the Academic model reproduces the qualitative trends well but is not quantitatively accurate, especially at high magnetic flux. The asymptotic behaviour visible at high flux in Fig.~\ref{fig:WedgingCS} reflects the nonlinear feedback loop discussed below.
	
	\subsection{TF coil comparison with reference designs}
	\label{sec:TF_bench}
	
	We now compare the Refined model predictions against constructed machines or highly refined designs of various sizes, magnetic fields and mechanical architectures. The results are presented in Table~\ref{tab:tf_coil_benchmark}. We used as D0FUS inputs the outer radius of the TF coil inner leg $(R_0 - a - \Delta_{B})$, the type of steel used, the cable current density $J_{\rm TF}^{\rm wost}$, and the target magnetic field, all taken from the literature.
	
	The most uncertain comparison concerns the two Commonwealth Fusion Systems (CFS) designs (ARC and SPARC), as no publication explicitly provides their winding pack configurations, thicknesses, or detailed coil characteristics. However, the comparison with these two machines is important, as they are the only available refined designs in bucking and plug configuration. Note that the CFS TF coils use a stack-in-plate design, which may allow the REBCO tapes to contribute to the mechanical strength of the winding pack, an effect ignored in D0FUS. For this comparison, we consider the PIT-VIPER conductor \cite{Sanabria2024PITVIPER} and expect to obtain correct orders of magnitude.
	
	Some values, which we could not directly find in the literature, have been estimated and are denoted by a tilde $\sim$.
	
	\begin{table*}[!htbp] 
		\centering
		\begin{tabular}{lcccccccc}
			\toprule
			& \textbf{ITER} & \textbf{EU-DEMO}  & \textbf{JT60-SA}  & \textbf{EAST}  & \textbf{ARC V1}  & \textbf{SPARC} \\
			&  \cite{sborchia2008design,mitchell2011iter,libeyre2009detailed} & \cite{federici2024relationship,eurodemo2017process} &  \cite{di2014overview,yoshida2010design,tsuchiya2008design} & \cite{chen20163d,wu2003east,chen2008design} &  \cite{sorbom2015arc,Sanabria2024PITVIPER,wang2024structure} &  \cite{creely2020overview,creely2024comment,hartwig2023sparc,Sanabria2024PITVIPER,wang2024structure,diazpacheco2025electromechanical}\\
			\midrule
			Inputs & & & & & & \\
			Configuration & Wedging & Wedging & Wedging & Wedging & Plug & Bucking \\
			$R_0$ (m) & 6.20 & 8.94 & 2.96 & 1.85 & 3.30 & 1.85 \\
			$a$ (m) & 2.00 & 2.88 & 1.18 & 0.45 & 1.10 & 0.57 \\
			$\Delta_{B}$ (m) & 1.10 & 1.82 & 0.36 & 0.15 & 0.89 & 0.18 \\
			Superconductor & Nb$_3$Sn & Nb$_3$Sn & NbTi & NbTi & REBCO & REBCO \\
			$J_{\rm TF}^{\rm wost}$ (MA/m$^2$) & 35 & 30 & $\sim$20 & 30 & $\sim$120 & $\sim$120 \\
			Steel & 316L & 316L & 316LN & 316L & CHSN01 & CHSN01 \\
			$\sigma_{\rm lim}$ (MPa) & 660 & 600 & 547 & 660 & 1000 & 1000 \\
			$B_{\text{max}}$ (T) & 11.8 & 10.6 & 5.65 & 5.8 & 23 & 20 \\
			\midrule
			TF coil thickness (m) & & & & & & \\
			D0FUS  & 0.91 & 0.98 & 0.45 & 0.25 & 0.79 & 0.38 \\
			Published & 0.90 & 0.96 & 0.41 & 0.25 & 0.64 & $\sim$0.35 \\
			$\Delta$ (m) & 0.01 & 0.02 & 0.04 & 0.00 & 0.15 & 0.03 \\
			$\Delta$ (\%) & 1 & 2 & 10 & 0 & 23 & 9 \\
			\bottomrule
		\end{tabular}
		\caption{TF coil benchmark.}
		\label{tab:tf_coil_benchmark}
	\end{table*}
	
	The benchmark demonstrates good agreement for all six machines, with predicted thicknesses close to the published values. The larger discrepancy for ARC ($\Delta = 0.15$~m) is expected given the substantial uncertainties on CFS coil characteristics discussed above.
	
	\subsection{CS comparison with reference designs}
	\label{sec:CS_bench}
	
	The same methodology has been employed to predict the CS thickness and magnetic field. The magnetic flux $\Psi_{\rm CS}$ values are derived from the literature or several BOBOZ runs (a CEA magnetostatic code based on the EFFI electromagnetic code \cite{sackett1978effi}). The fatigue knockdown defined in Section~\ref{sec:CS_fatigue} is applied in wedging and light bucking cases only.
	
	\begin{table*}[!htbp] 
		\centering
		\begin{tabular}{lcccccccc}
			& \textbf{ITER} & \textbf{EU-DEMO} & \textbf{JT60-SA} & \textbf{EAST}  & \textbf{ARC V1} & \textbf{SPARC} \\
			& \cite{sborchia2008design,libeyre2009detailed} & \cite{federici2024relationship,sarasola2023parametric,eurodemo2017process}&  \cite{di2014overview,yoshida2010design,tsuchiya2008design} & \cite{chen20163d,wu2003east,chen2008design} & \cite{sorbom2015arc,Sanabria2024PITVIPER,wang2024structure} & \cite{creely2020overview,creely2024comment,hartwig2023sparc,Sanabria2024PITVIPER,wang2024structure,diazpacheco2025electromechanical}\\
			\midrule
			Inputs & & & & & & \\
			Configuration & Wedging & Wedging & Wedging & Wedging & Plug & Bucking \\
			$R_0$ (m) & 6.20 & 8.94 & 2.96 & 1.85 & 3.30 & 1.85 \\
			$a$ (m) & 2.00 & 2.88 & 1.18 & 0.45 & 1.10 & 0.57 \\
			$\Delta_{B}$ (m) & 1.10 & 1.82 & 0.36 & 0.15 & 0.89 & 0.18 \\
			$\Delta_{TF}$ (m) & 0.90 & 0.96 & 0.41 & 0.25 & 0.64 & 0.35 \\
			Steel & JK2LB & 316L & 316LN & 316LN & CHSN01 & CHSN01 \\
			$\sigma_{\rm lim}$ (MPa) & 667 & 600 & 547 & 547 & 1000 & 1000 \\
			Superconductor & Nb$_3$Sn & Nb$_3$Sn & Nb$_3$Sn & NbTi & REBCO & REBCO \\
			$J_{\rm CS}^{\rm wost}$ (MA/m$^2$) & $\sim$45 & 60 & $\sim$45 & 45 & $\sim$120 & $\sim$120 \\
			$\Psi_{CS}$ (Wb) & 233 & 500 & 40 & 10 & 19.2 & 25.2 \\
			\midrule
			CS thickness & & & & & & \\
			D0FUS (m) & 0.75 & 0.76 & 0.37 & 0.10 & 0.24 & 0.29 \\
			Published (m) & 0.75 & 0.81 & 0.34 & 0.16 & 0.30 & $\sim$0.25\\
			$\Delta$ (m) & 0.00 & 0.05 & 0.03 & 0.06 & 0.06 & 0.04 \\
			$\Delta$ (\%) & 0 & 6 & 9 & 38 & 20 & 16 \\
			\midrule
			$B_{CS}$ results & & & & & & \\
			D0FUS (T) & 12.4 & 10.2 & 9.6 & 3.6 & 10.0 & 10.7 \\
			Published (T) & 13 & 11.4 & 8.9 & 4.5 & 13 & 25 \\
			$\Delta$ (T) & 0.6 & 1.2 & 0.7 & 0.9 & 3.0 & 14.3 \\
			$\Delta$ (\%) & 5 & 10 & 8 & 19 & 23 & 57 \\
		\end{tabular}
		\caption{CS benchmark.}
		\label{tab:cs_coil_benchmark}
	\end{table*}
	
	As shown in Table~\ref{tab:cs_coil_benchmark}, good agreement is obtained on both magnetic field and thickness across the six machines, with the correct orders of magnitude recovered, with the exception of ARC and SPARC where less public information is available on the CS design. For the compact-HTS machines (ARC and SPARC), the published total volt-second budget covers the combined CS+PF system, with no separate breakdown available. A few BOBOZ runs on representative layouts indicate a PF share of about 40\% for these architectures, larger than the typical 25\% of ITER-class designs~\cite{duchateau2014conceptual}. This correction is applied to derive $\Psi_{\rm CS}$ for ARC and SPARC. The apparent $\Delta$ on EAST corresponds to an absolute discrepancy of only 6~cm: percentage errors are not very meaningful at such small thicknesses.
	
	The CS sizing is inherently more sensitive to input assumptions than the TF sizing, owing to a strongly nonlinear feedback loop. Indeed, to gain even a few additional webers of flux, the peak field $B_{\rm CS}$ must be raised. A higher field increases the mechanical loads, requiring more structural material, and simultaneously degrades $J_{\rm CS}^{\rm wost}(B_{\rm CS})$, requiring more superconductor. Both effects thicken the winding pack, which encroaches on the bore and reduces the cross-sectional area available for flux generation, and so on. This self-reinforcing loop, illustrated by the asymptotic behaviour visible in Fig.~\ref{fig:WedgingCS}, makes the CS benchmark extremely sensitive to the input assumptions.
	
	\section{DEMO design space exploration}
	\label{Results}
	
	Having established and benchmarked the magnet models, we now explore the high-field design space for DEMO-class machines ($P_{\rm fus} \approx 2$~GW, $Q \approx 40$).
	
	\subsection{EU-DEMO baseline reproduction}
	\label{sec:methodology}
	
	The reference parameter set (Table~\ref{tab:input_parameters}, Appendix~\ref{appendixinput}) is taken from the EU-DEMO1~2017 baseline established by PROCESS~\cite{coleman2025definition, eurodemo2017process}. Using the same input parameters and physics assumptions (plasma geometry, safety factor definition, confinement scaling, bootstrap model, etc.), D0FUS reproduces the main plasma parameters to within a few percents, as summarised in Table~\ref{tab:benchmark_compact}. This design point serves as starting point for the subsequent scans.
	
	\begin{table}[htbp]
		\centering
		\caption{Key D0FUS vs PROCESS outputs for the EU-DEMO1 2017 baseline. Full inputs in Table~\ref{tab:input_parameters}.}
		\label{tab:benchmark_compact}
		\begin{tabular}{llcc}
			\toprule
			\textbf{Parameter} & \textbf{Symbol} & \textbf{PROC.} & \textbf{D0FUS}\\
			\midrule
			Plasma volume [m$^3$]                        & $V_p$            & 2466  & 2543\\
			Plasma current [MA]                          & $I_p$            & 19.08 & 19.23\\
			Safety factor                                & $q_{95}$         & 3.000 & 3.140\\
			Confinement time [s]                         & $\tau_E$         & 3.878 & 3.556\\
			Thermal energy [MJ]                          & $W_\mathrm{th}$  & 1251  & 1261\\
			Vol.\ avg.\ density [$10^{20}$\,m$^{-3}$]   & $\bar{n}_e$      & 0.791 & 0.704\\
			Energy gain                                  & $Q$              & 39.3  & 39.9\\
			Total radiated power [MW]                    & $P_\mathrm{rad}$ & 275   & 228\\
			Neutron wall load [MW/m$^2$]                 & $\Gamma_n$       & 1.036 & 1.066\\
			\bottomrule
		\end{tabular}
	\end{table} 
	
	Note that some discrepancies persist on current-profile-related quantities, notably the loop voltage and bootstrap fraction, which stem from identified differences in the underlying physics models (profile-integrated neoclassical conductivity~\cite{sauter1999neoclassical} in D0FUS vs the IPDG89 0D formula in PROCESS, and direct Sauter Eq.~5 bootstrap \cite{sauter1999neoclassical} vs a reformulation in PROCESS).
	
	\subsection{Reproduction of the reference \texorpdfstring{$R_0(B_0)$}{R0(B0)} curves}
	
\begin{figure}[ht]
		\centering
		\includegraphics[width=1\linewidth]{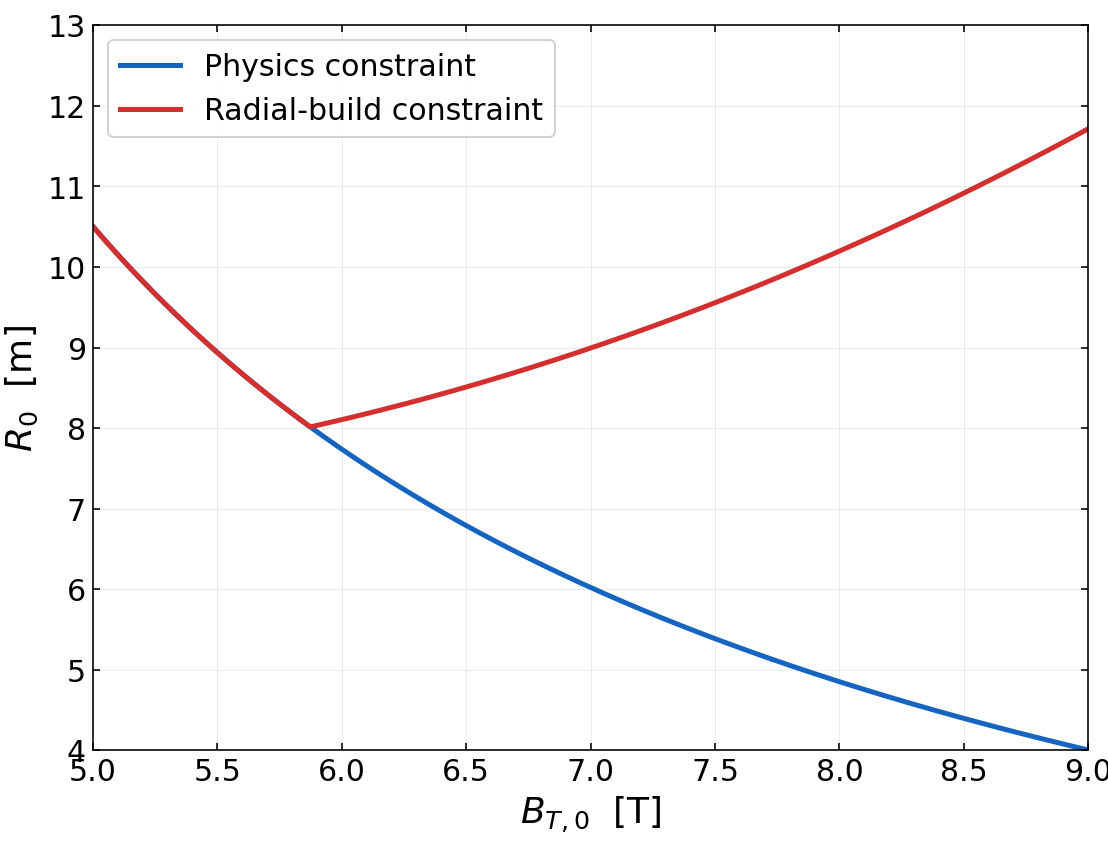}
		\caption{$R_0(B_0)$ curves from D0FUS for the reference configuration ($P_\mathrm{fus} \approx 2$~GW, $A = 3.1$, wedging, 316L, Nb$_3$Sn/REBCO). Blue: physics constraint ($q_{95}$, $f_G$ targets); red: radial-build constraint. To be compared with Fig.~4(i) of Ref.~\cite{federici2024relationship}.}
		\label{fig:federici_reference}
	\end{figure}

	To benchmark not only the anchor point but also the evolution of the design space as the toroidal field is increased, we adopt the methodology of Federici \emph{et al.}~\cite{federici2024relationship}. Starting from the EU-DEMO baseline, the peak TF $B_\mathrm{max}$ is varied from 8 to 25~T (covering $B_0$ from 5 to 9~T) at constant aspect ratio ($A = 3.1$), fusion power ($P_\mathrm{fus} \approx 2$~GW), plateau duration ($t_\mathrm{plateau} = 2$~h), safety factor limit ($q_{95} \geq 3$), and inboard blanket thickness ($\Delta_B = 1.4$~m). The superconductor is Nb$_3$Sn below 15~T and REBCO above. For each $B_\mathrm{max}$, the average temperature $\bar{T}$ is adapted to maintain the target Greenwald fraction $f_G = 1.1$. The resulting major radius $R_0$ is plotted as a function of $B_0$, producing two characteristic curves (Fig.~\ref{fig:federici_reference}):
	\begin{itemize}
		\item A \emph{physics} limited curve (blue), where $R_0$ is set only by the minimum machine size achieving the target $P_\mathrm{fus}$ for the given $q_{95}$ and $f_G$ constraints.
		\item A \emph{radial-build} limited curve (red), where $R_0$ is additionally constrained by the geometric closure of the central column (CS~+~TF inner leg must fit inside $R_0 - a - \Delta_B$).
	\end{itemize}
	The overall shape and trends are in good agreement with Fig.~4(i) of Ref.~\cite{federici2024relationship}. A systematic offset of approximately 0.4~m is observed on the radial-build curve, which is a direct consequence of the loop voltage discrepancy mentioned above.
	
	\subsection{Comparison with the four reference design points}
	\label{sec:federici_4points}
	
	Beyond the global $R_0(B_0)$ curve, Federici \emph{et al.}~\cite{federici2024relationship} report four specific TF designs sampled along their radial-build-limited curve at $B_0 = 5.5$, 7, 8 and 9~T. For each point, D0FUS is run at the same target $B_0$ on its own radial-build curve, and the predicted TF inner leg thickness $\Delta_\mathrm{TF}$ is compared to the value reported by Federici \emph{et al.} in Table~\ref{tab:federici_designs}.
	
	\begin{table}[ht]
		\centering
		\caption{D0FUS verification of the four TF design points reported by Federici \emph{et al.}~\cite{federici2024relationship} along their radial-build-limited curve. For each point, D0FUS is run at the same target $B_0$ on its own radial-build curve, and the resulting peak field $B_\mathrm{max}$ and TF inner leg thickness $\Delta_\mathrm{TF}$ are compared.}
		\label{tab:federici_designs}
		\begin{tabular}{c l c c c}
			\toprule
			Point & Source & $B_{\max}$ [T] & $B_0$ [T] & $\Delta_\mathrm{TF}$ [m] \\
			\midrule
			\multirow{2}{*}{(\#2)} & Federici \emph{et al.} & 12.0 & 5.50 & 0.98 \\
			& D0FUS    & 12.9 & 5.50 & 0.95 \\
			\midrule
			\multirow{2}{*}{(\#3)} & Federici \emph{et al.} & 14.7 & 7.00 & 1.78 \\
			& D0FUS    & 15.2 & 7.01 & 1.61 \\
			\midrule
			\multirow{2}{*}{(\#4)} & Federici \emph{et al.} & 16.4 & 8.00 & 2.19 \\
			& D0FUS    & 16.5 & 7.98 & 2.21 \\
			\midrule
			\multirow{2}{*}{(\#5)} & Federici \emph{et al.} & 17.6 & 9.00 & 3.22 \\
			& D0FUS    & 17.5 & 8.87 & 2.96 \\
			\bottomrule
		\end{tabular}
	\end{table}
	
	The TF inner leg thickness $\Delta_\mathrm{TF}$ is reproduced within 10\% across the four points: D0FUS predicts 0.95, 1.61, 2.21 and 2.96~m against 0.98, 1.78, 2.19 and 3.22~m for Federici \emph{et al.}, with no systematic bias.
	
	\subsection{D0FUS scans}
	
	To explore the design space, we now turn to two-dimensional $(R_0, a)$ maps. Restarting from the EU-DEMO baseline, we scan the major radius $R_0$ and minor radius $a$ on a $25 \times 25$ grid while imposing $P_\mathrm{fus} = 2$~GW and $Q = 40$. At each grid point, D0FUS converges a complete self-consistent design and its viability is assessed against both the plasma stability limits and the radial build constraint.
	
	The background colour indicates which stability limit is most restrictive (blue: Greenwald density limit, green: kink safety factor limit, red: Troyon $\beta_N$ limit), with darker shading indicating lower stability margin. The solid white contour marks the stability boundary. The solid black contour indicates the radial build limit, beyond which the CS and TF coil inner leg no longer fit within $R_0 - a - \Delta_B$. Dashed black iso-contours show the combined coil thickness $\Delta_\mathrm{TF} + \Delta_\mathrm{CS}$.
	
	The present system-code analysis focuses on the radial build as the most critical structural constraint and is not intended to define fully realistic power plant designs: other requirements (divertor heat flux, neutron wall load, maintenance access, cost, disruptions, etc.) may further restrict the accessible domain.
	
	\subsubsection{\texorpdfstring{$B_\mathrm{max}$}{Bmax} scan}
	\label{Bmax_impact}
	
\begin{figure}[!htbp] 
		\centering
		\begin{subfigure}[b]{0.9\linewidth}
			\includegraphics[width=\linewidth]{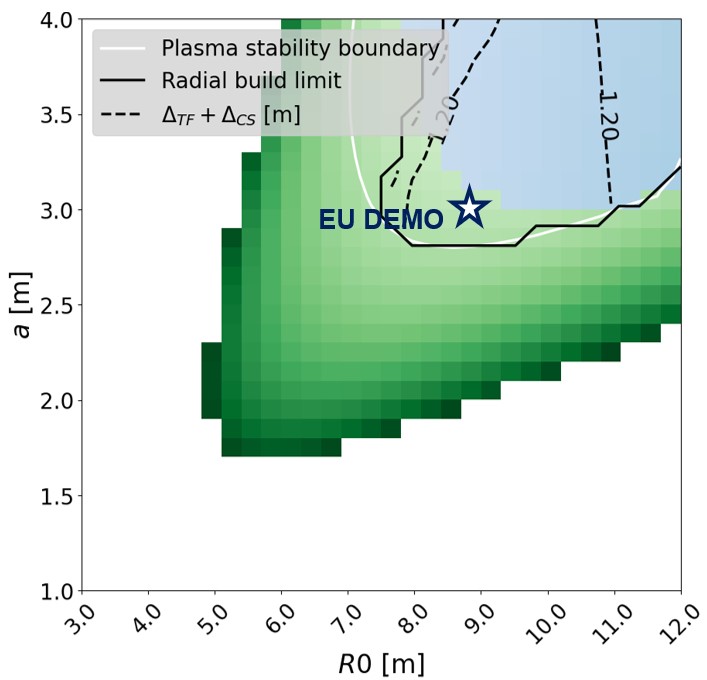}
			\caption{$B_\mathrm{max} = 10$~T, Nb$_3$Sn, wedging, 316L}
			\label{fig:scan-10-t}
		\end{subfigure}
		
		\begin{subfigure}[b]{0.9\linewidth}
			\includegraphics[width=\linewidth]{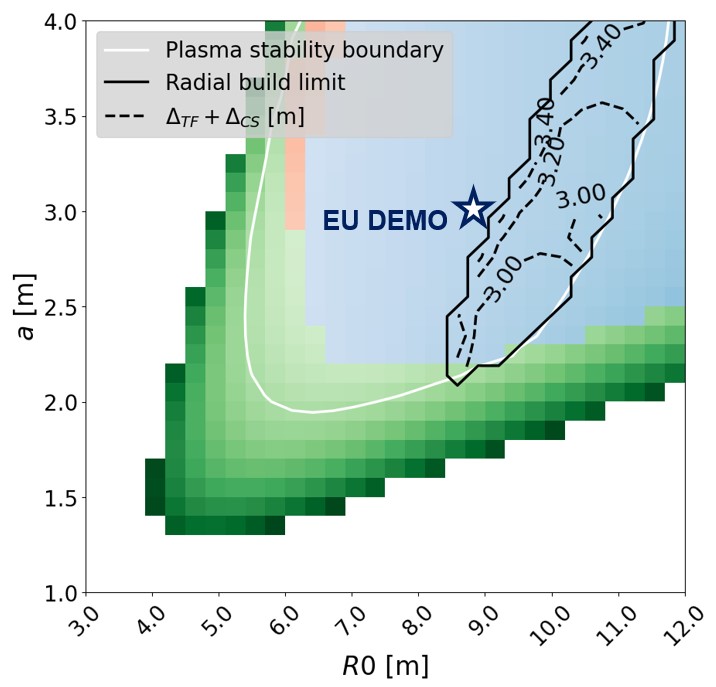}
			\caption{$B_\mathrm{max} = 15$~T, REBCO, wedging, 316L}
			\label{fig:scan-15-t}
		\end{subfigure}
		
		\begin{subfigure}[b]{0.9\linewidth}
			\includegraphics[width=\linewidth]{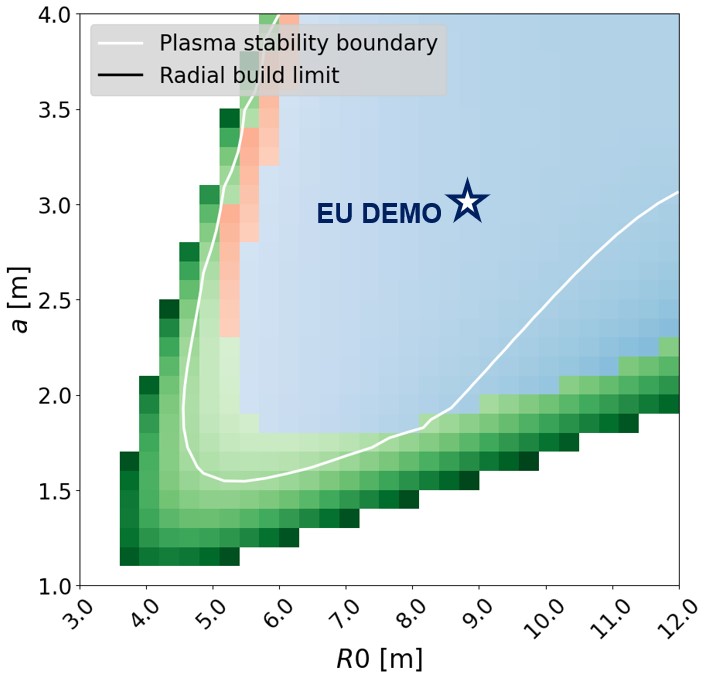}
			\caption{$B_\mathrm{max} = 20$~T, REBCO, wedging, 316L.}
			\label{fig:scan-20-t-wedging}
		\end{subfigure}
		
		\caption{D0FUS design maps based on the EU-DEMO 2017 reference parameters, scanning $B_\mathrm{max}$ over three values at constant $P_\mathrm{fus} = 2$~GW, all in the wedging/316L configuration.}
		\label{fig:scan_Bmax_progression}
	\end{figure}

	We explore the $(R_0, a)$ parameter space at three values of the peak toroidal field $B_\mathrm{max}$: 10~T (EU-DEMO class, Fig.~\ref{fig:scan-10-t}), 15~T (intermediate, Fig.~\ref{fig:scan-15-t}), and 20~T (high-field HTS, Fig.~\ref{fig:scan-20-t-wedging}), all in the reference wedging/316L configuration.
	
	At $B_\mathrm{max} = 10$~T, the radial-build constraint is barely visible: machines can be designed in nearly the entire plasma stability domain. As the field increases to 15~T, the plasma stability domain has expanded as already shown in \cite{auclair2025tokamak}, but the radial-build constraint now largely encroaches on this domain. At 20~T, the radial-build constraint is so restrictive that no viable design exists in the wedging configuration with 316L steel.
	
	\subsubsection{Advantage of large aspect ratio}
	\label{Aspect}
	
	A first observation from the 15~T scan (Fig.~\ref{fig:scan-15-t}) is the potential interest of machines with a high aspect ratio $A$. Increasing $B_\mathrm{max}$ from 10 to 15~T does not enable a smaller $R_0$ for the most compact feasible design, but it significantly reduces the minor radius $a$, and thus the volume and potential cost of the machine. Beyond this geometric gain, large aspect ratio designs ($A \approx 4$--$5$) bring two further benefits. On the plasma side, they offer a lower plasma current and reduced divertor heat flux (as detailed in Ref.~\cite{auclair2025tokamak}). On the mechanical side, at fixed $R_0$, a larger aspect ratio increases the radial space $R_0 - a - \Delta_B$ available for both the CS and the TF coils, easing their sizing. For the TF specifically, this is reinforced by a reduction of the logarithmic factor in $F_z$ (Eq.~\ref{equationFz}).
	
	At 20~T, however, even high aspect ratio designs are not viable in the reference configuration. Alternative strategies are needed and are discussed next.
	
	\subsubsection{Design lever hierarchy}
	
\begin{figure}[!htbp] 
		\centering
		\begin{subfigure}[b]{0.9\linewidth}
			\includegraphics[width=\linewidth]{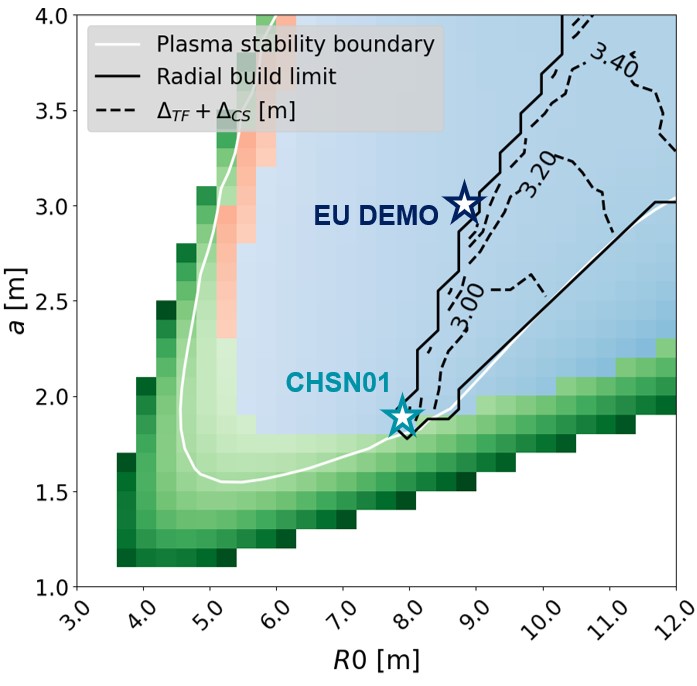}
			\caption{20~T REBCO reference and CHSN01.}
			\label{fig:scan-CHSN01-isolated}
		\end{subfigure}
		
		\begin{subfigure}[b]{0.9\linewidth}
			\includegraphics[width=\linewidth]{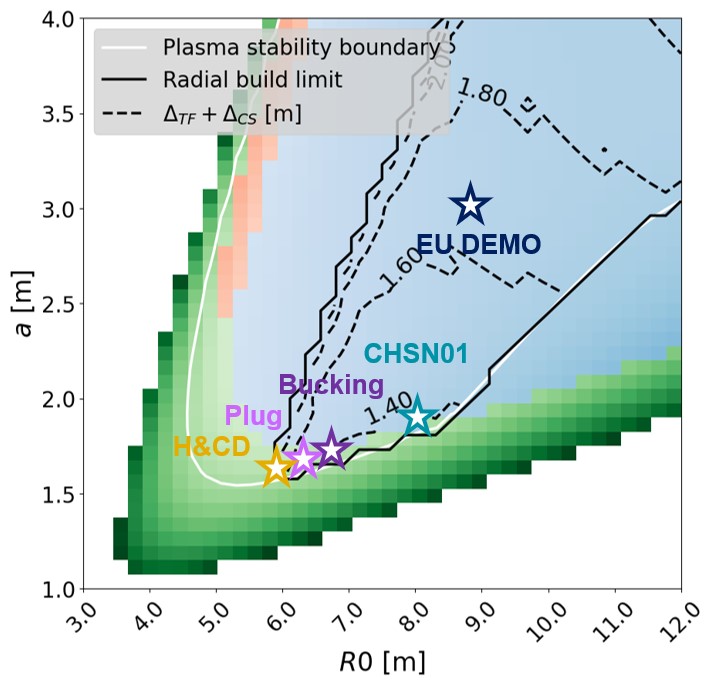}
			\caption{Cumulative addition of all first-order levers.}
			\label{fig:scan-1st-order-cumulative}
		\end{subfigure}
		
		\begin{subfigure}[b]{0.9\linewidth}
			\includegraphics[width=\linewidth]{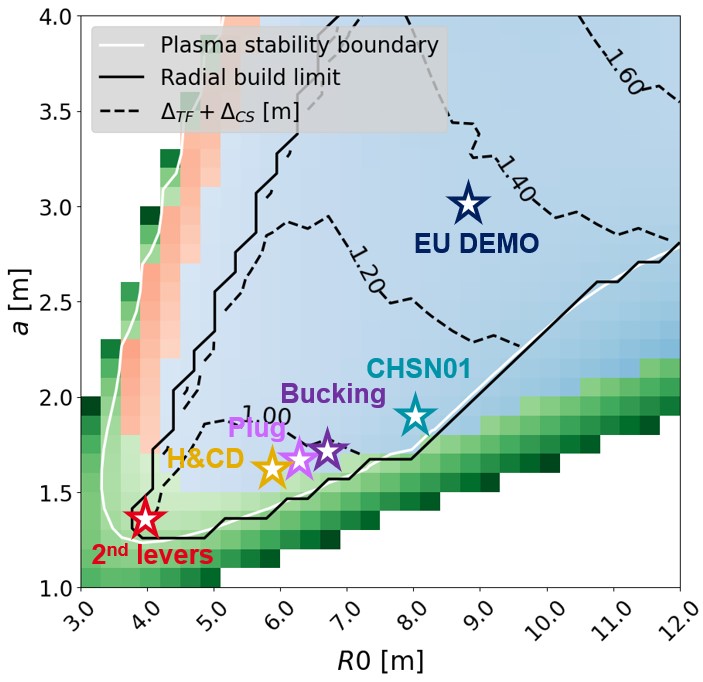}
			\caption{Cumulative addition of chosen second-order levers.}
			\label{fig:scan-ARC}
		\end{subfigure}
		
		\caption{Cumulative impact of design levers at $B_\mathrm{max} = 20$~T. Each star marks the most compact feasible $(R_0, a)$.}
		\label{fig:scan_levers}
	\end{figure}

	The design levers explored in the following sections fall into two categories according to their impact on the accessible design space. First-order levers have a dominant effect: each one, taken alone, broadens the accessible region significantly. They comprise high-strength steels (CHSN01), alternative mechanical configurations (bucking, plug), and reductions of the flux required from the CS (through heating and current-drive assistance during ramp-up, for example). At 15~T, any single first-order lever is sufficient to recover viable compact machines. At 20~T, a single lever reopens a feasible region within the plasma-stability domain, but reaching compact machines requires combining several of them as shown in the following sections. By comparison, second-order levers (conductor shape, radial grading, minimisation of the void fraction, etc.) provide smaller, but still non-negligible gains. Their cumulative impact is illustrated in Fig.~\ref{fig:scan_levers}, which serves as a visual roadmap for the rest of this section: each star marks the most compact feasible $(R_0, a)$ point reached after a given lever (or group of levers) has been added on top of the previous ones.
	
	To rank these levers quantitatively on a consistent basis, we complement the two-dimensional maps of Fig.~\ref{fig:scan_levers} with a dedicated one-dimensional sweep, hereafter referred to as the fixed-$a$ protocol. Starting from the 20~T reference configuration, we fix the minor radius at $a = 2.5$~m, disable the plasma-stability limits so that only the radial-build constraint drives feasibility, and sweep $R_0$ upward until the radial build becomes feasible. The smallest major radius passing this test is denoted $R_0^{\min}$. The sweep is repeated with each lever applied in isolation to the same reference, and the shift $\Delta R_0 = R_0^{\min,\,\mathrm{ref}} - R_0^{\min,\,\mathrm{lever}}$ is reported in Table~\ref{tab:levers_20T_fixed_a} at the end of Section~\ref{sec:ARC_like}.
	
	It should be emphasised that the reference case used here is deliberately pessimistic. At $B_\mathrm{max} = 20$~T in wedging/316L, no viable design exists inside the plasma-stability domain (Fig.~\ref{fig:scan-20-t-wedging}), and even without the stability constraint the radial build only opens up for $R_0 \gtrsim 12$~m at $a = 2.5$~m, a regime where both the machine and its coils are unreasonably large. In such a configuration, even modest structural improvements translate into large reductions of $R_0^{\min}$, so the magnitudes of the $\Delta R_0$ reported below (several metres for first-order levers) should be read as an indication of the relative strength of the levers.
	
	\subsubsection{Impact of steel properties}
	\label{Steel}
	
	A first option is to use high-performance steel. CHSN01 \cite{wu2025development,wang2026mechanical,wang2026mass} appears to be a very promising candidate offering a 50\% larger allowable Tresca stress ($\sigma_\mathrm{lim} \approx 1000$~MPa vs 660~MPa for 316L). As shown in Fig.~\ref{fig:scan-CHSN01-isolated}, this single change significantly relaxes the radial build limit compared to the reference wedging/316L case (Fig.~\ref{fig:scan-20-t-wedging}), re-opening the design space at 20~T even in wedging configuration.
	
	In the fixed-$a$ protocol (Table~\ref{tab:levers_20T_fixed_a}), this translates to $\Delta R_0 \approx 3.4$~m, the largest shift of any individual lever in this study. This must however be weighed against the maturity gap with 316L: CHSN01 is a recent steel grade, and further qualification on homogeneity, availability and weldability is needed.
	
	\subsubsection{Impact of the mechanical configuration}
	\label{Config}
	
	A second route to push the limit of the radial build and access more compact machines is to consider the diversity of mechanical configurations previously discussed (wedging, bucking, plug), illustrated alongside the other first-order levers in Fig.~\ref{fig:scan-1st-order-cumulative}.
	
	Firstly, transitioning from wedging to bucking significantly relaxes the radial build limit, with a downward shift of $\Delta R_0 \approx 2.45$~m in the fixed-$a$ protocol (Table~\ref{tab:levers_20T_fixed_a}). This compactness gain arises from an asymmetric redistribution of structural requirements between the TF coils and the CS. On the TF side, the inner leg becomes significantly thinner: in wedging, the centering force is reacted through the vault effect, which amplifies the stress and demands substantial structural thickness: in bucking, the TF coil simply transmits its radial centering force to the CS without stress amplification, resulting in a much more compact inner leg. On the CS side, the thickness increase is comparatively modest, because in the wedging configuration, the CS already contains structural material to withstand its own magnetic pressure. In bucking, this same material is repurposed to react the TF centering force. Since both loads are governed by magnetic pressures scaling respectively with $B_\mathrm{CS}^2$ and $B_\mathrm{max}^2$, and since these fields have similar orders of magnitude in compact configurations, the additional structural demand does not dramatically alter the CS dimensions. The net effect is a significant reduction in TF inner leg thickness with only a marginal increase in CS thickness, yielding a more compact overall design.
	
	Secondly, moving from bucking to plug also pushes back the radial build limit, with a further but marginal gain in $R_0$ (a total shift of $\Delta R_0 \approx 3.2$~m, i.e. an additional $\approx 0.75$~m beyond bucking alone). The mechanism is analogous but concerns the CS: it is no longer sized to withstand the TF pressure through a penalising vault effect, but merely to transmit it radially to the plug, which absorbs the load in a near-hydrostatic stress state.
	
	Thirdly, fatigue considerations bring an additional relaxation specific to the strong bucking and plug regimes. As established in Section~\ref{sec:CS_fatigue}, wedging and light bucking incur a factor 2 knockdown on $\sigma_{\text{lim}}$, since the CS cycles between full hoop tension at peak current and rest at flat-top. In strong bucking and plug, this penalty vanishes: the CS remains in compression throughout the discharge, which tends to close rather than propagate cracks. The effective allowable stress is therefore doubled, providing an additional lever towards a thinner CS and a more compact radial build.
	
	Bucking and plug thus appear promising for exploiting the expanded plasma stability domain. However, these configurations should be considered with caution, for reasons discussed in Section~\ref{sec:limitations}. The corresponding ``Bucking'' and ``Plug'' stars in Fig.~\ref{fig:scan-1st-order-cumulative} confirm this ordering.
	
	In the fixed-$a$ protocol (Table~\ref{tab:levers_20T_fixed_a}), bucking alone shifts $R_0^{\min}$ downward by $\Delta R_0 \approx 2.45$~m; adding the central plug brings this shift to $\Delta R_0 \approx 3.2$~m. The corresponding "Bucking" and "Plug" stars in Fig.~\ref{fig:scan-1st-order-cumulative} confirm this ordering.
	
	\subsubsection{Sensitivity to the CS flux demand}
	\label{sec:fh}
	
	A third lever, of the same order of magnitude, concerns the volt-second budget of the CS. In the EU-DEMO reference, the entire plasma current is conservatively assumed to be ramped up inductively ($f_h = 0$).
	
	Setting $f_h = 0.5$ attributes half of the plasma current ramp-up to current drive systems. This reduces $\Psi_{\rm Ramp\text{-}Up}$ proportionally (Eq.~\ref{eq:Psi_rampup}), relaxing the CS flux demand and permitting a more compact solenoid. In the fixed-$a$ protocol (Table~\ref{tab:levers_20T_fixed_a}), this single change yields $\Delta R_0 \approx 1.55$~m, of the same order as the gains offered by bucking or CHSN01 alone. Its cumulative effect, applied on top of CHSN01, bucking and plug, is visible as the "H\&CD" star in Fig.~\ref{fig:scan-1st-order-cumulative}.
	
	The current drive parameter $f_h$ is chosen here because it represents an explicit design choice with a transparent effect on the flux budget. The STEP power plant concept is a reference case for this approach: with severely restricted central solenoid space, STEP is designed to operate fully non-inductively throughout the majority of the ramp-up and flat-top phases, with the plasma current driven by electron cyclotron and electron Bernstein waves~\cite{freethy2024optimisation, meyer2024plasma}.
	
	Another potential source of gain is auxiliary heating during the current ramp-up, which raises the electron temperature and hence lowers the plasma resistivity, reducing the effective Ejima coefficient and the associated resistive flux consumption~\cite{imbeaux2011current, wakatsuki2019safety}.
	
	A third is the predicted bootstrap fraction, which can vary from one model to another or from one set of profile peaking assumptions to another. This spread, although not directly a design lever (even if profile shaping can in principle be exploited to raise it), translates directly into changes in $\Psi_{\rm Ind}$ and $\Psi_{\rm plateau}$, and therefore in the CS flux demand.
	
	\subsubsection{Impact of TF coil pre-compression}
	\label{Compression}
	
	Before turning to second-order levers, we briefly address pre-compression, often invoked as a structural lever but found to be insignificant at high field.
	
	This method consist in the use of pre-compression structures of TF coils such as Pre-Compression Rings (PCR) as in ITER, C-clamps~\cite{reccia2023iter}, or steel cables and tensioners. Good reviews of these concepts can be found in Refs.~\cite{bachmann2023influence, titus2013tf}. All these structures operate by pre-loading the TF coils to create an offset in tension: $F_z' = F_z - F_z^\mathrm{PC}$.
	
	Two criteria limit the pre-compression force $F_z^\mathrm{PC}$. The first is the maximum force that the pre-compression structure itself can deliver. Of order 30~MN per coil, i.e. about twice the ITER PCR performance~\cite{bachmann2023influence}, seems a good order of upper bound. 
	The second is the tolerable axial deformation of the TF coils before energisation, typically $\lesssim 2$~mm, beyond which alignment with the surrounding components can no longer be guaranteed~\cite{bachmann2023influence}.
	
	Even under these upper-bound/optimistic assumptions, the orders of magnitude remain insufficient to have a significant impact on the sizing at high fields. For $\sim$20 coils, a total pre-compression force of 600~MN represents only $\sim$5\% of the total tensile force at 20~T ($\sim$12~GN), and its effect on the accessible design space is accordingly negligible. Pre-compression nonetheless retains its role in keeping the winding pack in compression (to prevent conductor delamination) and in maintaining contact between adjacent TF coils (to preserve shear key efficiency)~\cite{bachmann2023influence,salpietro2007precompression}, but it does not enable more compact machines.
	
	\subsubsection{Impact of conductor shape}
	\label{shape}
	
	An alternative approach is to optimise the conductor shape to maximise the steel fraction in the most loaded regions of the jacket. D0FUS parametrises this through the asymmetry factor $n$, where $n = 1$ corresponds to a round-in-square conductor and $n \to 0$ to a round-in-rectangle (Fig.~\ref{fig:surfacedilutionconductor}). Reducing $n$ increases the effective steel area fraction $f_u(f_c, n)$ for a given cable fraction $f_c$ (see Tab.~\ref{tab:fractions_recap}), thereby lowering the stress concentration. For $n$ varying from 1 to 0.2, the resulting reduction in winding-pack thickness reaches $\sim$25\% at high fields in typical coil size. This is non-negligible, but below the gains offered by high-strength steels or alternative mechanical configurations. 
	
	Note that a conductor with $n \to 0$ is not realistic: a finite steel fraction must remain around each cable to carry residual hoop and shear stresses. The lower bound $n = 0.2$ adopted here is arbitrary; a dedicated mechanical study of the jacket would be needed to establish the actual minimum.
	
	In the fixed-$a$ protocol (Table~\ref{tab:levers_20T_fixed_a}), $n_\mathrm{TF} = 0.2$ yields $\Delta R_0 \approx 1.0$~m.
	
	\subsubsection{Impact of TF radial structural grading}
	\label{grading}
	
	Further potential exists through radial structural grading, i.e.\ varying the cable fraction $f_c$ (introduced in Subsection~\ref{Param_definition}) radially to saturate the Tresca criterion at every radius (see Appendix~\ref{appendix_grading}) (as opposed to \emph{superconductor} grading, in which the conductor type itself is varied along the radius to track the field profile, e.g.\ the CFETR TF coil~\cite{hao2022conductor}; only structural grading is considered here). In the non-graded model, $f_c$ is sized to withstand the peak $\sigma_r$ which occurs at $R = R_{\rm TF}^{\rm int}$, and all other radii operate below the stress limit, over-provisioning steel. The graded model redistributes this margin, assigning more superconductor material and less steel to the outer winding pack and reducing the total thickness. The maximum thickness reduction that can be expected is of order $\sim 25\%$ of the winding pack, in typical TF coil geometries, as discussed in Appendix~\ref{appendix_grading}. Note that these estimates are upper bounds: manufacturing constraints may limit grading applicability.
	
	This grading strategy is applicable to the TF coils but not straightforwardly to the CS. As shown in Ref. \cite{mitchell2002stress}, decreasing the steel fraction in the outer region of a graded solenoid induces undesirable radial tension in the coil insulation. This effect is intrinsic to solenoid geometry (with the likely exception of the strong bucking case), where the hoop stress changes sign between inner and outer turns. The TF coil inner leg does not suffer from this issue because the radial stress is monotonically compressive from $R_\mathrm{TF}^\mathrm{int}$ to $R_\mathrm{TF}^\mathrm{ext}$ in both wedging and bucking configurations.
	
	In the fixed-$a$ protocol (Table~\ref{tab:levers_20T_fixed_a}), enabling TF coil grading yields $\Delta R_0 \approx 0.65$~m, the smallest shift among the levers considered here.
	
	\subsubsection{Further secondary levers}
	
	Beyond the conductor shape and radial grading discussed above, several additional second-order levers can contribute to relieving the radial build constraint:
	
	\begin{itemize}
		\item \textbf{Reducing the plateau duration $t_\mathrm{plateau}$} directly lowers the resistive flux $\Psi_\mathrm{plateau} = V_\mathrm{loop}\,t_\mathrm{plateau}$ that the CS must supply, and therefore relaxes the solenoid sizing. This lever might only be admissible when the CS is kept in compression throughout the cycle, that is, in bucking or plug configurations, so that fatigue cycling of the CS does not become too important~\cite{sarasola2020progress}. Compatibility with continuous electricity production also sets a lower bound on $t_\mathrm{plateau}$, unless a thermal energy storage stage is integrated into the balance of plant to bridge the dwell phase~\cite{manta2024negative}.
		
		\item \textbf{Reducing the high-field-side blanket thickness $\Delta_B$} moves the TF winding pack closer to the plasma and raises $B_0$ at fixed $B_\mathrm{max}$, at the potential cost of margins on the tritium breeding ratio~\cite{hong2019optimal}. The ARC pilot plant illustrates how far this lever can be pushed: thanks to its FLiBe molten salt immersion blanket, ARC reaches $\Delta_B \approx 0.8$~m~\cite{sorbom2015arc}.
		
		\item \textbf{Reducing the copper stabiliser cross-section} may be feasible for HTS magnets through no-insulation (NI) or partial-insulation (PI) winding architectures~\cite{hahn2010hts, mitchell2021superconductors}, where current bypass through turn-to-turn contact provides intrinsic quench protection without requiring a large stabiliser. This route is pursued at fusion scale by the SPARC tokamak~\cite{hartwig2023sparc} (NI) and by the STEP project~\cite{nasr2024magnetic} (PI).
		
		\item \textbf{Sharing part of the mechanical load between the steel jacket and the conductor bundle} (superconductor and copper) would also relax the case sizing. This approach is arguably more accessible with REBCO than with low-temperature superconductors, thanks to its tape (ribbon) form factor and to the substantial mechanical contribution of the Hastelloy/stainless-steel substrate on which the REBCO film is deposited~\cite{Zhou2023_REBCO_mech, barth2015electro, zhao2022structural}.
	\end{itemize}
	
	\subsubsection{Combined impact}
	\label{sec:ARC_like}
	
	The cumulative path through the first-order levers, on top of the wedging/316L reference, is shown in Fig.~\ref{fig:scan-1st-order-cumulative}. Each star shows the smallest possible design as CHSN01, bucking, plug, and H\&CD ($f_h = 0.5$) are added in turn. By the time all four levers are active, viable designs are recovered well into the plasma stability domain: in the fixed-$a$ protocol (Table~\ref{tab:levers_20T_fixed_a}), $R_0^{\min}$ drops from $12.10$~m down to $6.85$~m, a reduction of $\sim 5$~m relative to the reference.
	
	Adding chosen (and easy to implement in D0FUS) second-order levers on top of this configuration completes the picture: optimised conductor shape ($n = 0.2$), radial grading, reduced inboard blanket ($\Delta_B = 0.8$~m, with a slight shift of the plasma stability boundary through the resulting change of $B_0$ at fixed $B_\mathrm{max}$), and a shortened pulse (15~min). Several of these choices, in particular the thin inboard blanket and the shortened pulse, are also adopted in the ARC V2 pilot plant~\cite{sorbom2015arc, creely2024aps, creely2025eps}, from which the reference values used here are taken. Fig.~\ref{fig:scan-ARC} shows the resulting map at $B_\mathrm{max} = 20$~T, where the "2$^\mathrm{nd}$ levers" star marks the most compact feasible point of this combined configuration. The design space is dramatically expanded, and nearly the entire plasma stability domain is now accessible, including compact designs with $R_0 < 4$~m.
	
\begin{figure}[ht]
		\centering
		\includegraphics[width=1.0\linewidth]{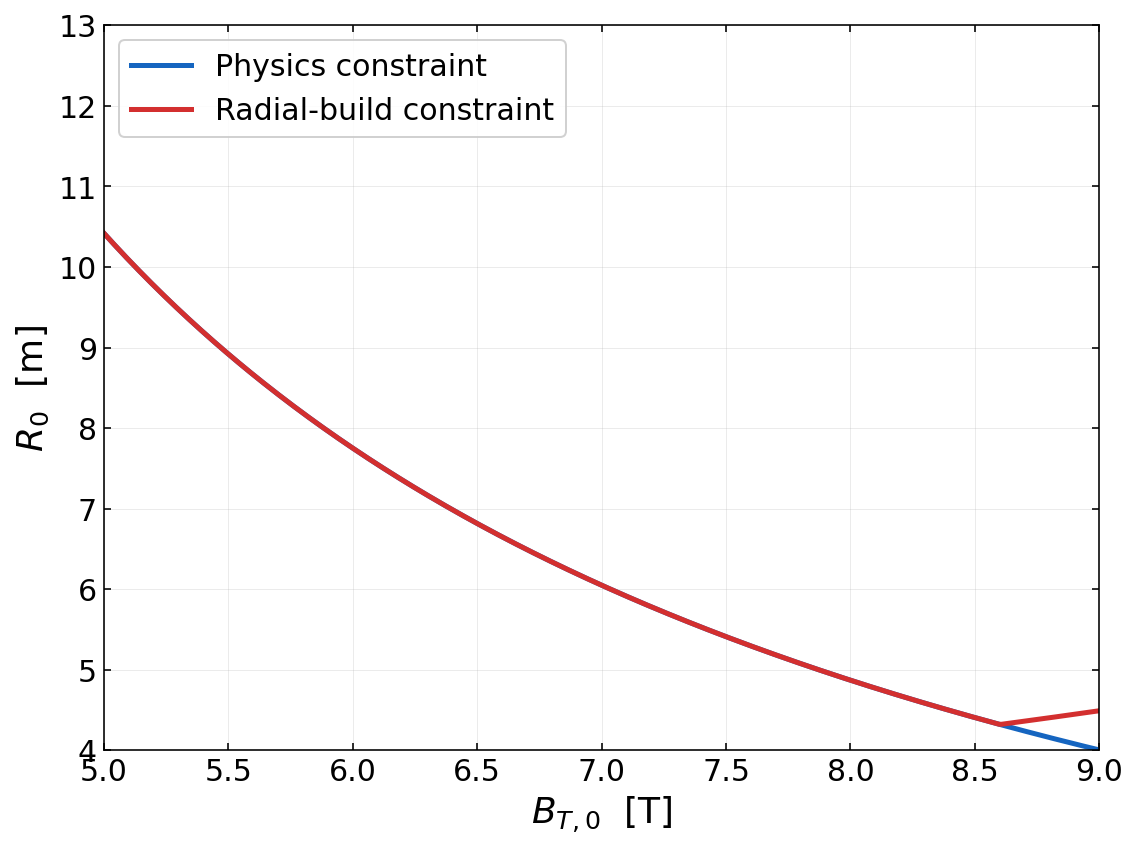}
		\caption{$R_0(B_0)$ curves for the ARC-like combined configuration. The radial-build curve (red) closely follows the physics curve (blue) down to $R_0 \approx 4$~m; the inversion occurs beyond $B_0 \approx 8.5$~T. To be compared with Fig.~\ref{fig:federici_reference}.}
		\label{fig:federici_ARC}
	\end{figure}

	The corresponding $R_0(B_0)$ curve (Fig.~\ref{fig:federici_ARC}) confirms the qualitative change: the radial-build curve now closely follows the physics curve over nearly the entire plotted range ($B_0$ up to 9~T, $R_0$ down to $\sim$4~m). The radial build starts becoming the limiting factor only at very high fields ($B_0 > 8.7$~T).
	
	\begin{table*}[ht]
		\centering
		\caption{Individual and cumulative impact of design levers on the radial-build limit at $B_\mathrm{max} = 20$~T. Each lever is applied to the 20~T REBCO / wedging / 316L reference configuration with the minor radius held fixed at $a = 2.5$~m. Plasma-stability limits are removed so that only the radial-build boundary drives feasibility. $R_0^{\min}$ is the smallest major radius for which the TF inner leg and the central solenoid both fit inside $R_0 - a - \Delta_B$. The "Isolated" columns report the effect of each lever applied alone to the reference; the "Combined" columns report the effect of stacking all levers from the first row down to the current one. The first four levers are first-order, the last four are second-order.}
		\label{tab:levers_20T_fixed_a}
		\begin{tabular}{lrrrr}
			\toprule
			\multirow{2}{*}{Lever} & \multicolumn{2}{c}{Isolated} & \multicolumn{2}{c}{Combined} \\
			\cmidrule(lr){2-3} \cmidrule(lr){4-5}
			& $R_0^{\min}$ [m] & $\Delta R_0$ [m] & $R_0^{\min}$ [m] & $\Delta R_0$ [m] \\
			\midrule
			Reference$^\dagger$ & 12.10 & --- & 12.10 & --- \\
			\midrule
			\makecell[l]{High-strength steel \\ \textit{CHSN01}}                      & \phantom{0}8.70 & $3.40$ & \phantom{0}8.70 & $3.40$ \\
			\makecell[l]{Mechanical configuration \\ \textit{Bucking}}                & \phantom{0}9.65 & $2.45$ & \phantom{0}7.40 & $4.70$ \\
			\makecell[l]{Mechanical configuration \\ \textit{Bucking + Plug}}                   & \phantom{0}8.90 & $3.20$ & \phantom{0}7.30 & $4.80$ \\
			\makecell[l]{Current drive assistance \\ \textit{$f_h = 0.5$}}            & 10.55           & $1.55$ & \phantom{0}6.85 & $5.25$ \\
			\midrule
			\makecell[l]{Shortened pulse \\ \textit{15~min}}                          & 10.85           & $1.25$ & \phantom{0}6.25 & $5.85$ \\
			\makecell[l]{Reduced inboard blanket \\ \textit{$\Delta_B = 0.8$~m}}      & 10.80           & $1.30$ & \phantom{0}5.50 & $6.60$ \\
			\makecell[l]{Optimised conductor shape \\ \textit{$n_\mathrm{TF} = 0.2$}} & 11.10           & $1.00$ & \phantom{0}5.20 & $6.90$ \\
			\makecell[l]{TF radial grading \\ \textit{enabled}}                       & 11.45           & $0.65$ & \phantom{0}5.15 & $6.95$ \\
			\bottomrule
			\multicolumn{5}{l}{\footnotesize $^\dagger$ The reference $R_0^{\min} = 12.10$~m lies outside the plasma stability domain.} \\
		\end{tabular}
	\end{table*}
	
	\subsubsection{Saturation of cumulative gains}
	\label{sec:saturation}
	
	One can observe in the combined column of Table~\ref{tab:levers_20T_fixed_a} that the gains saturate. The first lever, here CHSN01, alone accounts for about half of the total reduction in $R_0^{\min}$, and the cumulative gains tend to saturate as further levers are stacked.
	
\begin{figure}[ht]
		\centering
		\includegraphics[width=\linewidth]{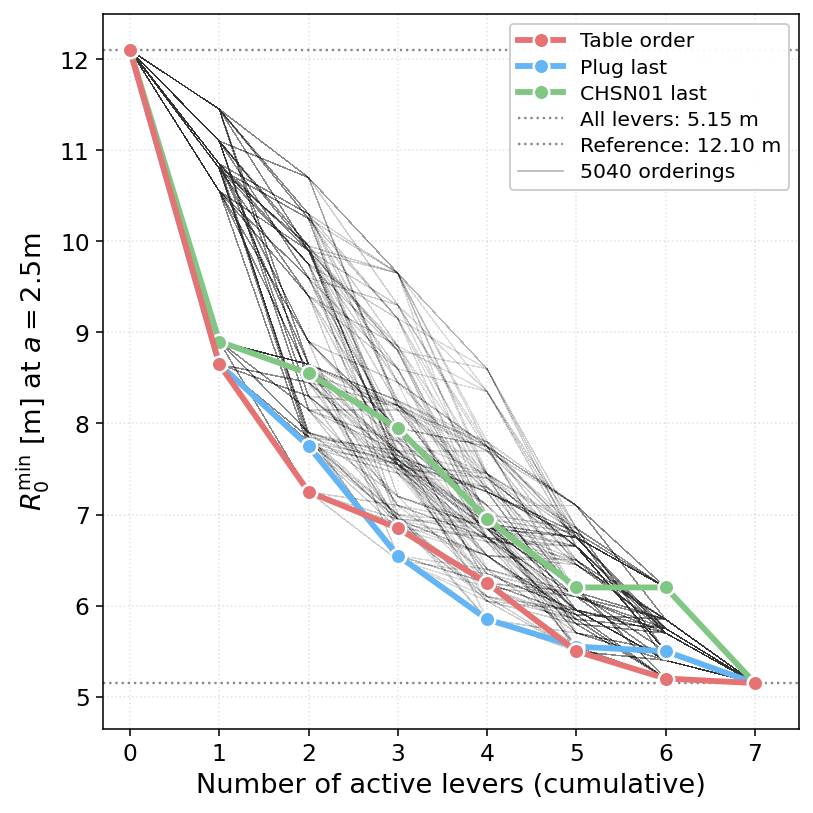}
		\caption{Cumulative $R_0^{\min}$ at $a = 2.5$~m, $B_\mathrm{max} = 20$~T, as seven design levers are activated one by one. The red curve follows the order of Table~\ref{tab:levers_20T_fixed_a}. The blue and green curves use the same order but with the plug and CHSN01, respectively, deferred to the last activation. The light grey envelope spans all $7! = 5040$ activation orders. Plasma-stability limits are disabled so that only the radial-build boundary drives feasibility.}
		\label{fig:lever_saturation}
	\end{figure}

	This is related to the fact that the successive activation of levers gradually reduces the steel fraction until the current-carrying capacity becomes the governing constraint. In this regime, the mandatory cross-section for the conductor limits any further minimization, as the design reaches the incompressible limit of the cable components.
	Figure~\ref{fig:lever_saturation} illustrates this saturation by tracing $R_0^{\min}$ as a function of the number of active levers. For clarity, only seven individual levers are retained, with bucking and plug grouped under a single mechanical-upgrade lever corresponding to the plug configuration (their isolated impacts on $R_0^{\min}$ being of similar magnitude, see Table~\ref{tab:levers_20T_fixed_a}). The red curve follows the order of the table; the blue curve uses the same order but with the plug activated last; the green curve activates CHSN01 last. The light grey envelope behind them spans all $7! = 5040$ activation orders. All paths start at $R_0^{\min} = 12.10$~m and converge to the same asymptote at $R_0^{\min} \approx 5.15$~m once the seven levers are active.
	
	A practical consequence is that, in the saturated regime, individual levers become highly substitutable, but not all to the same extent. The blue curve drops the plug to the last activation: its penultimate point ($R_0^{\min} = 5.50$~m) corresponds to a fully optimised wedging design, only 35~cm above the all-levers asymptote. The green curve drops CHSN01 instead: its penultimate point sits at $R_0^{\min} = 6.20$~m, about 1.05~m above the asymptote, three times the plug penalty. CHSN01 is therefore the least substitutable lever, consistent with its first-order impact in Table~\ref{tab:levers_20T_fixed_a}; the plug, conversely, can often be deferred at limited cost when other levers are available. A high-field power-plant design can accordingly afford to drop the plug entirely, provided the remaining levers compensate.
	
	\subsection{Radial build constraints favour compact designs}
	\label{Compact}
	
	A trend stands out from the D0FUS maps: at fixed $B_\mathrm{max}$, the cumulative inboard thickness $\Delta_\mathrm{TF} + \Delta_\mathrm{CS}$ (dashed iso-contours) decreases as one moves towards more compact machines (smaller $a$ and $R_0$). High fields and large major radius may not simultaneously be accessible: at large $R_0$, the coils may become prohibitively thick (as already pointed out by Federici \emph{et al.}~\cite{federici2024relationship} and Bachmann \emph{et al.}~\cite{bachmann2023influence}). The pathway to exploit high fields therefore likely passes through compact designs. The next subsections show that this trend holds across all three mechanical configurations, although for different reasons.
	
	\subsubsection{Wedging}
	
	In wedging configuration, the TF coil thickness $\Delta_\mathrm{TF}$ is thinner at smaller $R_0$. This is due to the vault effect: by transferring the centering radial pressure $P$ to an azimuthal stress $\sigma_\theta \approx P R_{TF} / \Delta_\mathrm{TF}$ (using the thin cylinder approximation for convenience, but the result is more general), it amplifies the former by the ratio $R_0/\Delta_\mathrm{TF}$. Consequently, small values of $R_0$ permit small values of $\Delta_\mathrm{TF}$.
	
	Conversely, the CS tends to thicken in compact machines. To generate the same $\Psi_\mathrm{CS}$ with a smaller bore, $B_\mathrm{CS}$ must increase (Eq.~\ref{Flux_CS_eq}), which in turn increases $\Delta_\mathrm{CS}$ (Eq.~\ref{eq_BCS}) and the associated $J \times B$ forces. The competition between these two opposite trends is, however, dominated by the TF coil thickness reduction: the sum $\Delta_\mathrm{TF} + \Delta_\mathrm{CS}$ decreases in compact machines, as visible on the Figs.~\ref{fig:scan-CHSN01-isolated} and ~\ref{fig:scan-10-t}.
	
	\subsubsection{Bucking}
	
	In bucking configuration, the TF coil also thins at small $R_0$, but more modestly than in wedging. Since $\sigma_\theta = 0$, the Tresca criterion reduces to $\sigma_z + |\sigma_r|$. The radial stress $\sigma_r$ is driven by $P_\mathrm{TF} = B_\mathrm{max}^2/(2\mu_0)$, which is independent of $R_0$, so that the Tresca criterion can be seen, at fixed $B_{max}$, as a criterion on $\sigma_z$. The analysis in Appendix~\ref{Dependence_sigma_z} shows that this $\sigma_z$ criterion can be fulfilled with a smaller $\Delta_\mathrm{TF}$ when $R_0$ is smaller.
	
	The CS, almost always in strong bucking (Section~\ref{sec:bucking}), is sized against the TF coil compression through the vault effect ($\sigma_\theta \approx P_\mathrm{TF}\,R_\mathrm{CS}/\Delta_\mathrm{CS}$) that also favours compact configurations. Both thicknesses thus decrease simultaneously at small $R_0$.
	
	\subsubsection{Plug}
	
	When adding a central plug, the TF coil behaves as in bucking. The CS now transfers its radial load onto the central plug, so it requires much less structural steel and its thickness is essentially set by the superconductor cross section required to generate $\Psi_\mathrm{CS}$. The radial build is dominated by the TF reduction with $R_0$, as illustrated in Fig.~\ref{fig:scan-ARC}.
	
	In addition, a purely geometric effect of $a$ becomes visible on $\Delta_\mathrm{CS}$: at fixed $R_0$, a smaller $a$ enlarges the CS bore and lowers the $B_\mathrm{CS}$ required to generate $\Psi_\mathrm{CS}$, yielding a thinner CS.
	
	\subsection{Limitations}
	\label{sec:limitations}
	
	The analyses presented in this paper are performed at the system-code level and carry inherent limitations.
	
	Firstly, the models are purely in-plane and cannot address effects requiring 3D structural analysis. The bucking configuration in particular raises challenging out-of-plane force management issues. The toroidal component of the Lorentz force ($I_\mathrm{TF} \times B_\mathrm{pol}$) creates an overturning moment on the TF inner leg \cite{titus2002provisions}. In wedging, this force is partly recovered by toroidal friction between adjacent coils. In bucking, the coils are no longer in toroidal contact and the out-of-plane loads must be managed by dedicated structures. The present work assumes that these forces \emph{can} be managed, and aims to help decide whether the resulting radial build gains are worth pursuing.
	
	Secondly, beyond the management of out-of-plane loads, both bucking and plug configurations raise practical engineering challenges. A modular CS, with independent current control in each module, can be desirable for plasma shaping purposes~\cite{huguet2001iter}, but is difficult to implement in these architectures. The central issue being the routing of helium cooling lines and electrical feeds, which is already constrained in bucking by the direct TF-CS mechanical contact, and potentially impractical in the plug configuration where it must pass through a solid central cylinder. Moreover, this TF-CS interface, idealised as direct contact in our model, must accommodate a dedicated structural component (such as the JET sliding cylinder~\cite{rebut1976jet}) with specific surface, friction, and load-transfer properties, which also govern the long-term resistance to scuffing and progressive CS erosion under the cyclic TF torsion.
	
	Thirdly, this study has only addressed the mechanical aspects of the radial build. Other constraints, most notably divertor heat flux, neutron wall load and maintenance access, can become critical in compact machines. To illustrate the divertor side, Table~\ref{tab:Psep_R0} reports the proxy $P_\mathrm{sep}/R_0$ for three designs at $P_\mathrm{fus} \approx 2$~GW: the conventional EU-DEMO 2017 baseline and the two compact 20~T configurations of Table~\ref{tab:levers_20T_fixed_a}. This ratio enters the simplest scaling of the parallel heat flux at the divertor entrance, $q_\parallel \propto P_\mathrm{sep}/(R_0\,\lambda_q)$. Absolute values depend on the conventions adopted for $P_\mathrm{sep}$ and do not translate directly into heat fluxes, but their ratios consistently rank design difficulty: compact high-field designs aggravate divertor difficulty by a factor of 1.5 to 2 compared to the EU-DEMO baseline. A more refined analysis~\cite{siccinio2019figure} confirms that such compact configurations lie near the lower edge of the viable divertor design space, and discusses possible mitigation strategies.
	
	\begin{table}[ht]
		\centering
		\caption{Divertor heat-flux proxy $P_\mathrm{sep}/R_0$ for the EU-DEMO 2017 baseline and the two compact 20~T designs of Table~\ref{tab:levers_20T_fixed_a} (at $a = 2.5$~m).}
		\label{tab:Psep_R0}
		\begin{tabular}{lrrr}
			\toprule
			Design & $P_\mathrm{sep}/R_0$ [MW/m] \\
			\midrule
			EU-DEMO 2017 baseline       & 29.5 \\
			First-order levers combined & 34.1 \\
			All levers combined         & 51.2 \\
			\bottomrule
		\end{tabular}
	\end{table}
	
	Lastly, as previously demonstrated, exploiting high fields in a compact machine requires a combination of innovations (REBCO, high-strength steels, alternative mechanical configurations, conductor optimisation), each carrying its own development risk, and at very different levels of maturity. Bucking, plug, and high-strength steels have been rarely studied at the power plant scale and even more rarely tested, making them inherently higher-risk options. In fact, wedging with 316L/JK2LB steel and no grading is the baseline for most existing superconducting tokamaks (ITER, JT-60SA, EAST, KSTAR, etc.), with the notable exceptions of JET~\cite{rebut1976jet} and early ITER designs~\cite{no1999final, mitchell1999iter, titus2002provisions, titus1998analysis, titus1995structural}, both in bucking configuration. Note that this situation should evolve in the near future: SPARC~\cite{creely2020overview, creely2024comment, diazpacheco2025electromechanical} will certainly provide the first modern bucking design, while BEST~\cite{zhu2025electromagnetic, wang2023study, wang2024structure, wang2026mechanical, wang2026mass} plans to use CHSN01 steel in both its TF and CS magnets, offering the first validation of this steel grade in a fusion magnet system.
	
	\section{Summary and conclusion}
	
	Two mechanical models for predicting the radial build of tokamak TF coils and CS have been implemented in the D0FUS system code.
	
	The Academic model (Section~\ref{Academic}), based on the thin-cylinder approximation, provides closed-form expressions that capture the essential physics governing coil thicknesses. While not quantitatively accurate at high fields, it is valuable for building physical intuition. The Refined model (Section~\ref{D0FUS}) introduces thick-cylinder stress theory, a composite CICC winding pack, and self-consistent CS axial stress. Benchmarked against six reference machines and the MADE magnet design code (Section~\ref{Benchmark}), it demonstrates good predictive capability across a wide range of configurations.
	
	Applied to the exploration of the high-field design space, the analysis yields the following findings.
	
	At low fields ($B_\mathrm{max} \leq 12$~T achievable with LTS), the radial build is not a limiting constraint. As the field increases, the radial build boundary progressively encroaches on the stability domain, first favouring high aspect ratio designs and then shrinking the accessible region further.
	
	A central finding is that, for a 2~GW fusion power machine in the reference wedging/316L configuration, this encroachment becomes complete at high field: no viable design remains for $B_\mathrm{max} > 20$~T (at least for $R_0 < 12$~m). The main levers to overcome this limitation are high-strength steels (CHSN01), alternative mechanical architectures (bucking, plug), and reductions of the effective CS flux demand (through current drive assistance during ramp-up, for example). Each of these first-order levers, taken alone, substantially broadens the accessible region, whereas secondary levers (conductor shape, radial grading, etc.) provide more modest improvements.
	
	When all favourable design levers are combined (Section~\ref{sec:ARC_like}), compact machines ($R_0 \approx 4$~m) become structurally accessible at $B_\mathrm{max} = 20$~T. Note that, at high field, a power plant may have no choice but to be relatively compact, since too large an $R_0$ may lead to prohibitively thick coils, as detailed in Section~\ref{Compact}.
	
	Achieving a compact design, however, does not require the simultaneous combination of all favourable levers. The analysis of the combinations of these levers shows that their cumulative impact saturates (Section~\ref{sec:saturation}): each additional lever brings diminishing returns. A design programme can therefore afford to drop one or several levers without significantly compromising the achievable compactness, provided the remaining ones compensate. The selection between them is then governed by a risk-versus-impact balance, which we now address.
	
	The identified levers do not all carry equal risk. High-strength steels such as CHSN01 offer arguably the most favourable balance of impact and risk: selectively upgrading the case material in the most stressed regions yields a large and predictable reduction of the radial build, with no fundamental physics uncertainty, the residual risk being on qualification and availability~\cite{zhu2025electromagnetic, wang2023study, wang2024structure, wang2026mechanical, wang2026mass}. Reducing the effective CS flux demand constitutes a second lever whose potential impact is non-negligible yet remains to be quantified through dedicated scenario studies. Bucking was used in JET~\cite{rebut1981jet}, considered in preliminary ITER designs~\cite{no1999final}, and is the published baseline for SPARC~\cite{creely2020overview, creely2024comment, diazpacheco2025electromechanical}: its structural benefit is substantial, and may even become a necessity in pulsed power plants, where fatigue can otherwise become a limiting factor over the plant lifetime~\cite{sutcliffe2025magnet}. The out-of-plane load question, however, has not yet been fully addressed at the high field power plant scale and requires further analysis. The plug concept introduces additional complexity (helium and electrical feedthroughs through the central column, assembly, maintenance) for a more modest gain, and is therefore not recommended by the authors. The second-order levers (conductor shaping and radial grading at least) carry low risk and minimal additional cost: there is no reason to ignore them in any detailed design effort.
	
	Ultimately, the choice between these design approaches reduces to a risk allocation problem, with technological choices driving the economic ones. Conventional designs (Nb$_3$Sn, wedging, 316L) rely on mature technologies but concentrate the risk on the size and scale-up of large structures and materials volumes, with direct consequences on capital cost and economic competitiveness. Compact high-field designs (REBCO, CHSN01, bucking architectures, etc.) bet on the combined development of several coupled technologies and/or architectural choices, each currently at low maturity, in exchange for a smaller and potentially more economically attractive machine (Section~\ref{sec:limitations}).
	
	\section*{Code availability}
	\label{sec:code_availability}
	
	D0FUS is open-source under the CeCILL-C license and hosted at \url{https://github.com/IRFM/D0FUS}. All numerical results presented in this paper were obtained with the version tagged \texttt{paper-mechanical-NF2026-Auclair} on this repository, which freezes the code state at the time of submission.
	
	\section*{Acknowledgements}
	
	The authors warmly thank Jean-François Artaud, Clarisse Bourdelle, Lorenzo Giannini, Francesco Maviglia and Andrea Quartararo for the many insightful discussions that have nourished this work.\\
	
	This work is financially supported by the French government's "France 2030" initiative through the "ANR" (National Research Agency) in the framework of the SupraFusion PEPR Program and its SF-Plant research project ANR-24-EXSF-0005.
	
	% Appendices
	\appendix
	\renewcommand{\thesubsection}{\Alph{subsection}} % Les sous-sections deviennent A, B, C...
	\section*{Appendix}
	\addcontentsline{toc}{section}{Appendix}
	
	\subsection{\texorpdfstring{$R_{\text{int}}^{TF}$}{R\_int\^TF} thin cylinder expression}
	\label{Annexe_A_thinlayer}
	
	\subsubsection{Wedging 1st order}
	\label{1rst_order_wedging}
	
	Starting from the Tresca expression,
	\[
	\sigma_{\text{Tresca}} = |\sigma_\theta - \sigma_z|,
	\]
	and inserting the $\sigma$ expressions, one obtains
	\[
	\sigma_{\text{Tresca}} = \frac{B_{\max}^2}{2 \mu_0 } \frac{R_{\text{TF}}^{\text{sep}}}{R_{\text{TF}}^{\text{sep}}-R_{\text{TF}}^{\text{int}}} + \frac{B_0^2 R_0^2 \ln \left( \frac{R_0 + a + \Delta_B}{R_0 - a - \Delta_B} \right)}{2 \mu_0 \left((R_{\text{TF}}^{\text{sep}})^2 - (R_{\text{TF}}^{\text{int}})^2\right)}
	\]
	Using $B_0 R_0 \approx B_{\max} R_{\text{TF}}^{\text{sep}}$ (thin cylinder approximation) and simplifying leads to
	\[
	\sigma_{\text{Tresca}} =
	\]
	\[
	\frac{B_{\max}^2}{2\mu_0}
	\left[
	\frac{R_{\text{TF}}^{\text{sep}}}{R_{\text{TF}}^{\text{sep}}-R_{\text{TF}}^{\text{int}}}
	+ \frac{\left(R_{\text{TF}}^{\text{sep}}\right)^2}{\left(R_{\text{TF}}^{\text{sep}}\right)^2-\left(R_{\text{TF}}^{\text{int}}\right)^2}
	\ln \left( \frac{R_0 + a + \Delta_B}{R_0 - a - \Delta_B} \right)
	\right]
	\label{eq:sigma}
	\]
	Defining $\delta = R_{\text{TF}}^{\text{sep}} - R_{\text{TF}}^{\text{int}}$ and assuming $\delta \ll R_{\text{TF}}^{\text{sep}}$, we have
	\[
	\left(R_{\text{TF}}^{\text{sep}}\right)^2 - \left(R_{\text{TF}}^{\text{int}}\right)^2 
	= \left(R_{\text{TF}}^{\text{sep}}-R_{\text{TF}}^{\text{int}}\right)\left(R_{\text{TF}}^{\text{sep}}+R_{\text{TF}}^{\text{int}}\right)
	\simeq 2\,\delta\, R_{\text{TF}}^{\text{sep}},
	\]
	so that the Tresca stress reduces to
	\[
	\sigma_{\text{Tresca}} 
	\simeq \frac{B_{\max}^2 R_{\text{TF}}^{\text{sep}}}{2\mu_0 \delta}
	\left(1+\frac{1}{2}\ln \left( \frac{R_0 + a + \Delta_B}{R_0 - a - \Delta_B} \right)\right)
	\]
	and therefore
	\[\boxed{
		R_{\text{TF}}^{\text{int}} 
		\simeq R_{\text{TF}}^{\text{sep}}
		- \frac{B_{\max}^2 R_{\text{TF}}^{\text{sep}}}{2\mu_0 \sigma_{\text{Tresca}}}
		\left(1+\frac{1}{2}\ln \left( \frac{R_0 + a + \Delta_B}{R_0 - a - \Delta_B}\right)\right)
	}
	\]
	We remind that expression is valid as a first-order approximation in the thin-wall limit $\delta \ll R_{\text{TF}}^{\text{sep}}$.
	
	\subsubsection{Bucking 1st order}
	\label{1rst_order_bucking}
	
	Starting from
	\[
	\sigma_{\text{Tresca}}
	= \frac{B_{\max}^2}{2\mu_0}
	+ \frac{B_{\max}^2 (R_{\text{TF}}^{\text{sep}})^2}{2\mu_0\bigl((R_{\text{TF}}^{\text{sep}})^2-(R_{\text{TF}}^{\text{int}})^2\bigr)}
	\ln\!\left(\frac{R_0+a+\Delta_B}{R_0-a-\Delta_B}\right),
	\]
	and applying the same thin-wall approximation as in Appendix~\ref{1rst_order_wedging} (i.e.\ $(R_{\text{TF}}^{\text{sep}})^2 - (R_{\text{TF}}^{\text{int}})^2 \simeq 2\,\delta\,R_{\text{TF}}^{\text{sep}}$ with $\delta = R_{\text{TF}}^{\text{sep}} - R_{\text{TF}}^{\text{int}}$), the Tresca stress reduces to
	\[
	\sigma_{\text{Tresca}} \simeq \frac{B_{\max}^2}{2\mu_0}\left[1 + \frac{R_{\text{TF}}^{\text{sep}}}{2\,\delta}\ln\!\left(\frac{R_0+a+\Delta_B}{R_0-a-\Delta_B}\right)\right]
	\]
	The internal radius is therefore
	\[
	\boxed{\,R_{\text{TF}}^{\text{int}}
		= R_{\text{TF}}^{\text{sep}}\left[1 - \frac{B_{\max}^2}{4\mu_0\left(\sigma_{\text{Tresca}}-\frac{B_{\max}^2}{2\mu_0}\right)}\ln\!\left(\frac{R_0+a+\Delta_B}{R_0-a-\Delta_B}\right)\right]}
	\]
	
	\subsection{Determination of \texorpdfstring{$B_{TF}$}{B\_TF}}
	\label{Annexe_B_thinlayer}
	
	We denote by $R_{\rm TF}^{\rm sep}$ the inner boundary of the current-carrying annulus, i.e.\ the interface between the winding pack and the steel nose.
	
	Applying Ampère's theorem with a uniform current density $J_{\rm TF}^{\rm wost}$,
	\[
	\oint \mathbf{B}\cdot d\mathbf{l} = \mu_0 \int_0^{2\pi} d\varphi \int_{R_{\rm TF}^{\rm sep}}^{R} r\, dr\, J_{\rm TF}^{\rm wost},
	\]
	and solving for $B(R)$ yields
	\[
	B(R) = \frac{\mu_0 J_{\rm TF}^{\rm wost}}{2}\left(R - \frac{(R_{\rm TF}^{\rm sep})^2}{R}\right)
	\]
	Evaluating at $R = R_{\rm TF}^{\rm ext}$ gives the thick-cylinder TF coil peak field,
	\[
	\boxed{%
		B_{\max} = B(R_{\rm TF}^{\rm ext}) = \frac{\mu_0 J_{\rm TF}^{\rm wost}}{2}\left(R_{\rm TF}^{\rm ext} - \frac{(R_{\rm TF}^{\rm sep})^2}{R_{\rm TF}^{\rm ext}}\right)
	}
	\]
	In the thin-shell limit $\Delta R = R_{\rm TF}^{\rm ext} - R_{\rm TF}^{\rm sep} \ll R_{\rm TF}^{\rm ext}$, the right-hand side can be factorised as
	\[
	B_{\max} = \frac{\mu_0 J_{\rm TF}^{\rm wost}}{2 R_{\rm TF}^{\rm ext}}\left(R_{\rm TF}^{\rm ext} - R_{\rm TF}^{\rm sep}\right)\left(R_{\rm TF}^{\rm ext} + R_{\rm TF}^{\rm sep}\right),
	\]
	and using $R_{\rm TF}^{\rm ext} + R_{\rm TF}^{\rm sep} \simeq 2\,R_{\rm TF}^{\rm ext}$, one obtains
	\[
	\boxed{%
		B_{\max} \approx \mu_0 J_{\rm TF}^{\rm wost}\,(R_{\rm TF}^{\rm ext} - R_{\rm TF}^{\rm sep})
	}
	\]
	
	\subsection{Determination of \texorpdfstring{$F_z$}{Fz}}
	\label{appendixB}
	
	The TF coils are themselves the source of the toroidal field $\vec B_\varphi = B_\varphi(R)\,\hat e_\varphi$ with $B_\varphi(R) = B_0 R_0 / R$. In the limit of a large number of coils, each inner leg can be idealised as a thin azimuthal current sheet across which $B_\varphi$ is discontinuous: $B_\varphi(R)$ on the plasma side and zero just beyond. The force per unit length on such a current sheet is given by the cross product of the current with the average of the fields on either side \cite{jackson1998classical}, so that
	\[
	d\vec F = I\,d\vec\ell \times \frac{\vec B_{\text{in}} + \vec B_{\text{out}}}{2} = \frac{1}{2}\,I\,d\vec\ell \times \vec B_\varphi,
	\]
	where $\vec B_\varphi$ is evaluated on the plasma-facing side and $I$ is the current per coil.
	
	Consider a planar coil of arbitrary shape lying in the $(R,z)$ plane, with inner leg at $R_{\text{in}} = R_0 - a - \Delta_B$ and outer leg at $R_{\text{out}} = R_0 + a + \Delta_B$. An element of its contour has components $d\vec\ell = dR\,\hat e_R + dz\,\hat e_z$, and
	\[
	d\vec\ell \times \vec B_\varphi = B_\varphi(R)\,(dR\,\hat e_z - dz\,\hat e_R)
	\]
	Integrating the vertical component along the upper half of the contour (from $R_{\text{in}}$ at $z = 0$ up to the top of the coil and back down to $R_{\text{out}}$ at $z = 0$) yields
	\[
	F_z = \frac{I}{2} \int_{R_{\text{in}}}^{R_{\text{out}}} \frac{B_0 R_0}{R}\,dR = \frac{I\,B_0 R_0}{2}\,\ln\!\left(\frac{R_{\text{out}}}{R_{\text{in}}}\right)
	\]
	The result depends only on the extreme radii $R_{\text{in}}$ and $R_{\text{out}}$, not on the specific shape of the contour in between, confirming the shape-independence noted in \cite{freidberg2015designing}.
	
	Applying Ampère's theorem to the toroidal solenoid formed by the $N_{\text{coil}}$ coils gives $B_0 = \mu_0 N_{\text{coil}} I / (2\pi R_0)$, so that $I = 2\pi R_0 B_0 / (\mu_0 N_{\text{coil}})$. Substituting yields the vertical force per coil,
	\[
	\boxed{%
		F_z = \frac{\pi}{\mu_0\,N_{\text{coil}}}\,B_0^2\,R_0^2\,\ln\!\left(\frac{R_0 + a + \Delta_B}{R_0 - a - \Delta_B}\right)
	}
	\]
	
	An equivalent derivation, reaching the same logarithmic dependence on the inner and outer radii, is given in Ref. \cite[Page 109, Eq.3.2]{thome1982mhd}.
	
	\subsection{Determination of \texorpdfstring{$\sigma_\theta$}{sigma\_theta} in the TF coil}
	\label{app:thin_wall_stress}
	
	The notation follows that introduced in Section~2.1.
	
	\subsubsection{Thick-wall solution}
	
	The Lamé-Clapeyron solution for stresses in a thick-walled cylinder under axisymmetric loading \cite{Clapeyron1829,timoshenko1970elasticity} writes
	\[
	\begin{cases}
		\sigma_r(r) = A - \dfrac{B}{r^2}, \\
		\sigma_\theta(r) = A + \dfrac{B}{r^2},
	\end{cases}
	\]
	where $A$ and $B$ are integration constants fixed by the boundary conditions. For a TF coil inner leg in wedging, these are zero internal pressure $\sigma_r(R_{\text{TF}}^{\text{int}}) = 0$ and external pressure $\sigma_r(R_{\text{TF}}^{\text{ext}}) = -P_{\text{TF}}$. Solving the linear system yields
	\[
	A = -P_{\text{TF}}\,\frac{(R_{\text{TF}}^{\text{ext}})^2}{(R_{\text{TF}}^{\text{ext}})^2 - (R_{\text{TF}}^{\text{int}})^2}, \qquad
	B = -P_{\text{TF}}\,\frac{(R_{\text{TF}}^{\text{ext}})^2\,(R_{\text{TF}}^{\text{int}})^2}{(R_{\text{TF}}^{\text{ext}})^2 - (R_{\text{TF}}^{\text{int}})^2}
	\]
	The hoop stress reaches its maximum at the inner face ($r = R_{\text{TF}}^{\text{int}}$), giving
	\begin{equation}
		\boxed{%
			|\max(\sigma_\theta^{\text{thick}})| 
			= \left|A + \frac{B}{(R_{\text{TF}}^{\text{int}})^2}\right|
			= P_{\text{TF}}\,\frac{2\,(R_{\text{TF}}^{\text{ext}})^2}{(R_{\text{TF}}^{\text{ext}})^2 - (R_{\text{TF}}^{\text{int}})^2}
		}
		\label{eq:hoop_stress_thick}
	\end{equation}
	
	\subsubsection{Thin-wall approximation}
	
	In the thin-wall limit $\Delta_R \equiv R_{\text{TF}}^{\text{ext}} - R_{\text{TF}}^{\text{int}} \ll R_{\text{TF}}^{\text{int}}$, factorising gives $(R_{\text{TF}}^{\text{ext}})^2 - (R_{\text{TF}}^{\text{int}})^2 \simeq 2\,\Delta_R\,R_{\text{TF}}^{\text{ext}}$, so that Eq.~\ref{eq:hoop_stress_thick} reduces to
	\begin{equation}
		\boxed{%
			\sigma_\theta^{\text{thin}} \approx \frac{P_{\text{TF}}\,R_{\text{TF}}^{\text{ext}}}{R_{\text{TF}}^{\text{ext}} - R_{\text{TF}}^{\text{int}}}
		}
		\label{eq:hoop_stress_thin}
	\end{equation}
	
	\subsection{Determination of \texorpdfstring{$\Psi_{\text{CS}}$}{Psi\_CS}}
	\label{appendixC}
	
	We compute the flux through the CS cross-section assuming an infinitely long solenoid and a uniformly distributed current density, so that $B_z(r) = B_{CS}$ for $r \le R_{\text{CS}}^{\text{int}}$ and decreases linearly from $B_{CS}$ to zero across the winding pack, i.e.\ $B_z(r) = B_{CS}\,(R_{\text{CS}}^{\text{ext}} - r) / (R_{\text{CS}}^{\text{ext}} - R_{\text{CS}}^{\text{int}})$ for $R_{\text{CS}}^{\text{int}} \le r \le R_{\text{CS}}^{\text{ext}}$. Integrating,
	\begin{equation}
		\begin{aligned}
			\Psi_{CS} &= 2\pi \int_0^{R_{\text{CS}}^{\text{ext}}} B_z(r)\,r\,dr \\
			&= 2\pi B_{CS} \left[ \frac{(R_{\text{CS}}^{\text{int}})^2}{2} + \frac{1}{R_{\text{CS}}^{\text{ext}} - R_{\text{CS}}^{\text{int}}} \int_{R_{\text{CS}}^{\text{int}}}^{R_{\text{CS}}^{\text{ext}}} (R_{\text{CS}}^{\text{ext}} - r)\,r\,dr \right] \\
			&= \frac{\pi B_{CS}}{3} \left[ (R_{\text{CS}}^{\text{ext}})^2 + R_{\text{CS}}^{\text{ext}} R_{\text{CS}}^{\text{int}} + (R_{\text{CS}}^{\text{int}})^2 \right]
		\end{aligned}
	\end{equation}
	
	A full swing of the CS provides a flux
	\[
	\boxed{%
		\Psi'_{CS} = \frac{2\pi B_{CS}}{3} \left[ (R_{\text{CS}}^{\text{ext}})^2 + R_{\text{CS}}^{\text{ext}} R_{\text{CS}}^{\text{int}} + (R_{\text{CS}}^{\text{int}})^2 \right]
	}
	\]
	
	\subsection{Geometrical dependence of \texorpdfstring{$\Delta_{\rm TF}$}{Delta\_TF} in bucking}
	\label{Dependence_sigma_z}
	
	This appendix uses the Academic model assumptions. In bucking, the radial stress $\sigma_r = B_{\max}^2/(2\mu_0)$ is set by the magnetic pressure alone and does not depend on $R_0$ or $a$. Saturating the Tresca criterion $\sigma_z + |\sigma_r| = \sigma_{\rm lim}$ therefore fixes $\sigma_z$ independently of the geometry,
	\begin{equation}
		\sigma_z \;=\; \sigma_{\rm lim} - \frac{B_{\max}^2}{2\mu_0}
		\label{eq:sigmaz_allowable}
	\end{equation}
	
	Using the notation $R_{\rm in} = R_0 - a - \Delta_B$ and $R_{\rm out} = R_0 + a + \Delta_B$ introduced in Appendix~\ref{appendixB}, the thin-cylinder approximation $B_0 R_0 \approx B_{\max}\,R_{\rm in}$ recasts Eq.~\ref{equationFz} as
	\begin{equation}
		F_z \;=\; \frac{\pi B_{\max}^2}{\mu_0\,N_{\rm coil}}\,R_{\rm in}^2\,\ln(R_{\rm out}/R_{\rm in})
		\label{eq:Fz_compact}
	\end{equation}
	
	In the thin-wall limit $\Delta_{\rm TF} \ll R_{\rm in}$, the steel cross-section of the inboard leg that carries the tension is
	\begin{equation}
		S \;=\; \pi\bigl(R_{\rm in}^2 - (R_{\rm in}-\Delta_{\rm TF})^2\bigr) \;\approx\; 2\pi\,R_{\rm in}\,\Delta_{\rm TF}
		\label{eq:area_thin}
	\end{equation}
	
	Substituting Eqs.~\ref{eq:Fz_compact} and \ref{eq:area_thin} in Eq.~\ref{equation1} yields
	\begin{equation}
		\sigma_z \;=\; \frac{N_{\rm coil}\,F_z}{2\,S} \;\approx\; \frac{B_{\max}^2}{4\mu_0}\,\frac{R_{\rm in}\,\ln(R_{\rm out}/R_{\rm in})}{\Delta_{\rm TF}}
		\label{eq:sigmaz_thin}
	\end{equation}
	
	Combining Eqs.~\ref{eq:sigmaz_allowable} and \ref{eq:sigmaz_thin} and solving for $\Delta_{\rm TF}$,
	\begin{equation}
		\boxed{\,\Delta_{\rm TF} \;=\; \frac{B_{\max}^2}{4\mu_0\,\sigma_z}\,R_{\rm in}\,\ln(R_{\rm out}/R_{\rm in})}
		\label{eq:DTF_final}
	\end{equation}
	
	The geometrical factor $R_{\rm in}\,\ln(R_{\rm out}/R_{\rm in})$ is strictly increasing with $R_0$ at fixed $a$.
	
	\subsection{Axial stress at the CS midplane}
	\label{appendix_sigma_z_CS}
	
	Consider a solenoid of inner radius $R_i$, outer radius $R_e$, half-height $h$, carrying a uniform current density $J$. The on-axis field at position $z$ along the solenoid axis is \cite{montgomery1969solenoid}
	\begin{equation}
		B_z(0,z) = \frac{\mu_0 J}{2}\left[(z+h)\,\mathcal{L}(z+h) - (z-h)\,\mathcal{L}(z-h)\right]
		\label{eq:Bz_axis}
	\end{equation}
	where $\mathcal{L}(\zeta) = \ln\!\left(\dfrac{R_e + \sqrt{R_e^2 + \zeta^2}}{R_i + \sqrt{R_i^2 + \zeta^2}}\right)$ is a decreasing geometric function. At the midplane and at the top end,
	\[
	B_z(0,0) = \mu_0 J h\,\mathcal{L}(h), \qquad B_z(0,h) = \mu_0 J h\,\mathcal{L}(2h)
	\]
	
	Inside the bore ($r < R_i$), the paraxial approximation $B_z(r,z) \approx B_z(0,z)$~\cite{humphries1990charged} combined with $\nabla\cdot B = 0$ yields the radial field at the bore surface
	\begin{equation}
		B_r(R_i,z) \approx -\frac{R_i}{2}\frac{dB_z(0,z)}{dz}
		\label{eq:Br_bore}
	\end{equation}
	We extend $B_r \simeq B_r(R_i,z)$ across the winding pack, accurate for $h \gtrsim 2\,R_i$ (which covers all fusion CS) and conservative for thicker windings.
	
	The azimuthal current density $J$ crossing this radial field produces an axial body force $f_z = J B_r$. Integrating from the free end ($z = h$, where $\sigma_z = 0$) to the midplane gives
	\[
	\sigma_z(z) = \frac{JR_i}{2}\left[B_z(0,h) - B_z(0,z)\right]
	\]
	At the midplane ($z = 0$), $\mathcal{L}(2h) < \mathcal{L}(h)$, so the stress is compressive,
	\begin{equation}
		\boxed{\sigma_z^{\rm smear} = -\frac{\mu_0 J^2 h R_i}{2}\left[\mathcal{L}(h) - \mathcal{L}(2h)\right]}
		\label{eq:sigma_z_CS_full}
	\end{equation}
	which is a smeared stress over the full CS cross-section; the peak stress in the steel is $\sigma_z^{\rm steel} = \sigma_z^{\rm smear}/f_u$.
	
	This formula agrees within $\sim$20\% with the exact treatment of~\cite[Sec.~3.5.2]{iwasa2009casestudies} for typical fusion CS.
	
	\subsection{Determination of the current density \texorpdfstring{$J^{\rm wost}$}{Jwost}}
	\label{JBappendix}
	
	The current density of a superconducting magnet can be defined at several scales, depending on which cross-section is considered: the superconducting material alone ($J_{\rm SC}$, the intrinsic critical current density), the non-copper cross-section ($J_{\text{non-Cu}}$, which also includes the substrate for REBCO or the matrix for Nb$_3$Sn), the strand or tape including its additional stabilizer material ($J_{\rm strand}$ or $J_{\rm tape}$), the cable with its additional stabilizer and cooling dedicated space, the full conductor with jacket and insulation, or the entire coil including additional structural material (i.e. casing).
	
	In D0FUS, the relevant quantity is $J^{\rm wost}$ ("without steel"), defined on the cross-section that includes everything except the structural steel. The conductor is sized under the assumption that it operates at its critical current, $J_{\rm op}(B,T_{\rm calc}) = J_{\text{non-Cu}}(B,T_{\rm calc})$, where $T_{\rm calc} = T_{\rm op} + \Delta T_{\rm margin}$ is a design temperature embedding a stability margin above the actual operating temperature $T_{\rm op}$ (see end of this appendix). The computation of $J^{\rm wost}$ proceeds in two stages: first, the non-copper critical current density $J_{\text{non-Cu}}$ is evaluated from material-specific scaling laws (Sections~\ref{sec:NbTi} to~\ref{sec:REBCO}); then, $J_{\text{non-Cu}}$ is diluted through successive material layers (copper stabilizer, cooling channels, insulation) to obtain $J^{\rm wost}$, as described in Section~\ref{sec:Jwost_calc}. An important aspect of this dilution is that the copper fraction is not an arbitrary input: it is determined by quench protection requirements through the Maddock adiabatic hot-spot criterion.
	
	\begin{figure}
		\centering
		\begin{subfigure}[b]{0.4\textwidth}
			\centering
			\includegraphics[width=\linewidth]{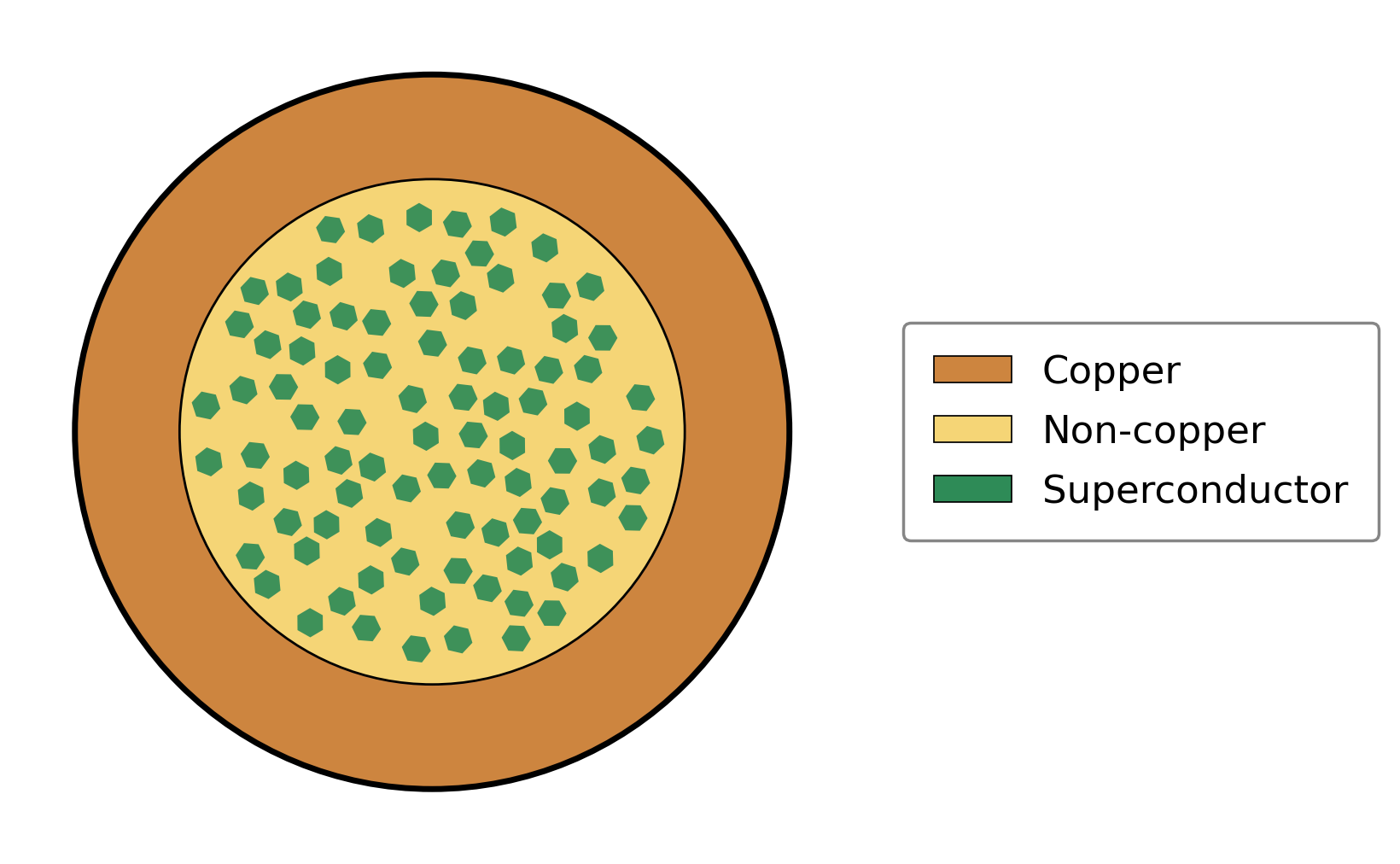}
			\caption{SC strand: superconducting filaments (green) embedded in a copper matrix (orange). $J_{\rm SC}$ is defined on the filament cross-section only.}
			\label{fig:strand}
		\end{subfigure}
		
		\vspace{0.3cm}
		
		\begin{subfigure}[b]{0.4\textwidth}
			\centering
			\includegraphics[width=\linewidth]{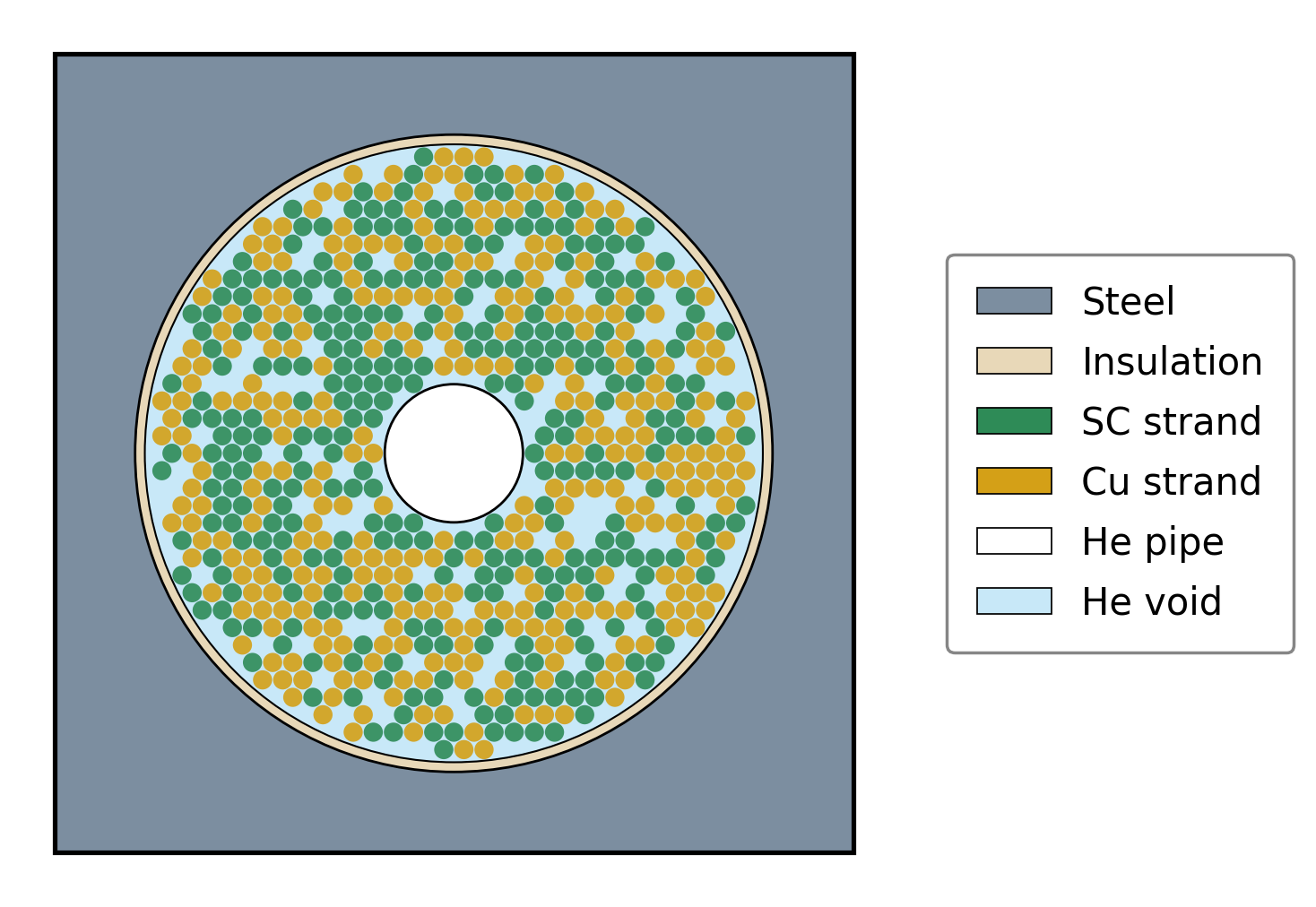}
			\caption{CICC: SC and Cu strands cabled together around a central He pipe, jacketed in steel. $J^{\rm wost}$ is defined on everything inside the steel jacket.}
			\label{fig:cicc}
		\end{subfigure}
		
		\vspace{0.3cm}
		
		\begin{subfigure}[b]{0.5\textwidth}
			\centering
			\includegraphics[width=\linewidth]{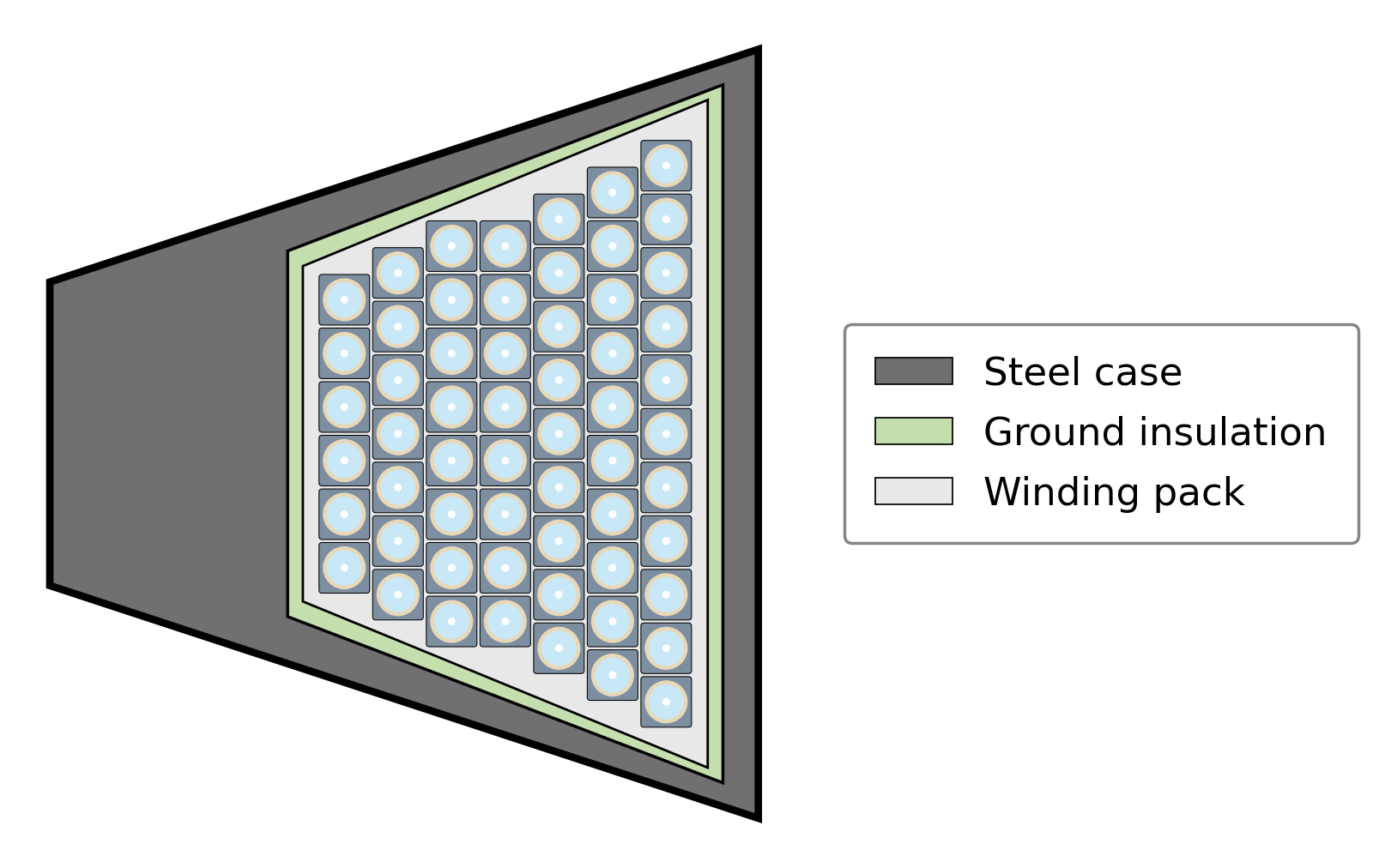}
			\caption{TF coil inboard leg: a winding pack of CICC turns held by a thick steel case. $J_{\rm coil}$ is defined on the full cross-section, steel case included.}
			\label{fig:coil}
		\end{subfigure}
		\caption{Hierarchy of current density definitions in a low-temperature superconducting CICC, from the strand-level superconducting filaments (a) up to the full TF coil inboard leg (c). At each level, a fraction of the cross-section is dedicated to non-current-carrying material (copper stabilizer, helium, insulation, structural steel), so that the current density progressively decreases as the reference surface grows. The quantity used in D0FUS is $J^{\rm wost}$, defined on the CICC non-steel cross-section (everything inside the steel jacket of (b)).}
		\label{fig:schemadensitecourant}
	\end{figure}
	
	The three panels of Fig.~\ref{fig:schemadensitecourant} represent successive nesting levels: each strand of (a) is one of the green or orange disks of (b), and each square cell of (c) is one CICC of (b). At each level, the current density seen from the outside drops because the reference cross-section grows while the conducting material (the green filaments of (a)) stays the same. Schematically:
	\begin{equation}
		J_{\rm SC} \;\longrightarrow\; J_{\text{non-Cu}} \;\longrightarrow\; J_{\rm strand} \;\longrightarrow\; J^{\rm wost} \;\longrightarrow\; J_{\rm coil}
		\label{eq:Jcascade}
	\end{equation}
	with each arrow corresponding to a "dilution" by the local fraction of useful conductor surface. D0FUS stops the cascade at $J^{\rm wost}$ because the steel jacket of the CICC is sized independently for mechanical purposes (Section~\ref{D0FUS}), and is not a free parameter of the conductor design.
	
	\begin{figure}
		\centering
		\includegraphics[width=1\linewidth]{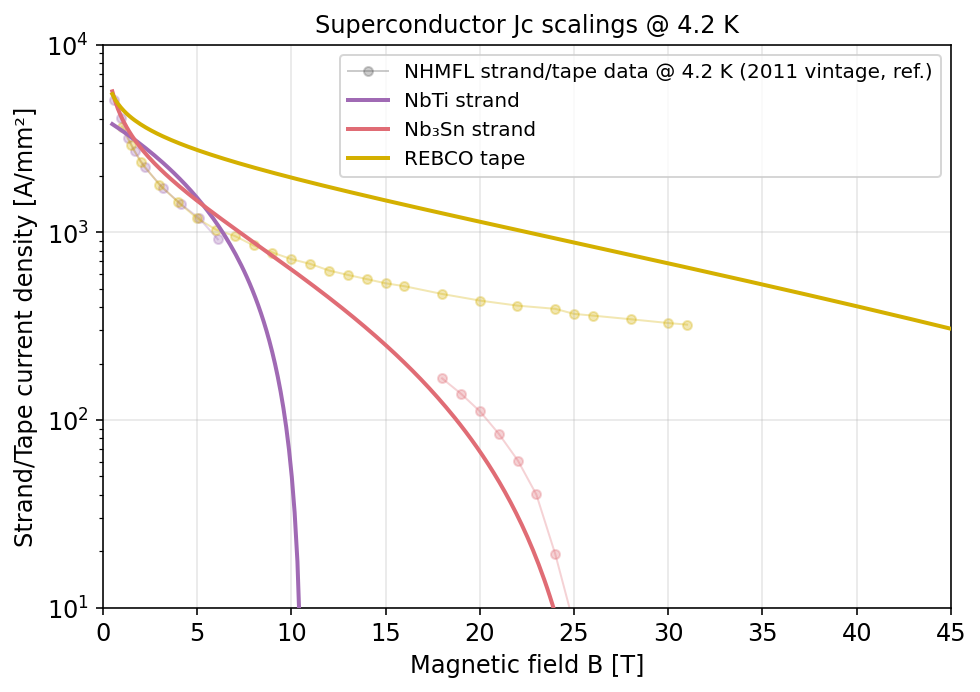}
		\caption{Strand and tape current density of the three superconductor scalings implemented in D0FUS, evaluated at 4.2~K. NHMFL maglab data~\cite{maglab2024} (transparent, 2011 vintage) is overlaid as an experimental reference.}
		\label{fig:jc}
	\end{figure}
	
	The scaling laws used for $J_{\text{non-Cu}}$ differ between conductor technologies. D0FUS implements the ITER/EU-DEMO NbTi parametrization~\cite{corato2016common}, the EU-DEMO WST Nb$_3$Sn scaling~\cite{bottura2009jc, corato2016common}, and for REBCO the Senatore \emph{et al.}\ (2024) pinning-force scaling~\cite{senatore2024rebco} calibrated on modern tapes (with the older Fleiter/CERN (2014) parametrization~\cite{fleiter2014rebco, bajas2022ship} also available).
	
	\subsubsection{NbTi model}
	\label{sec:NbTi}
	
	The NbTi scaling follows the ITER/EU-DEMO parametrization~\cite{corato2016common}:
	\begin{equation}
		J_{\text{non-Cu}}(B, T) = \frac{C_0}{B} \left(1 - t^{1.7} \right)^\gamma \cdot b^\alpha \cdot (1 - b)^\beta \quad \left[\text{A/mm}^2\right]
	\end{equation}
	where $t = T/T_{c0}$ is the reduced temperature, $b = B/B_{c2}(T)$ is the reduced magnetic field with $B_{c2}(T) = B_{c2,0}(1 - t^{1.7})$ the upper critical field beyond which superconductivity is lost. The exponents $\alpha$ and $\beta$ control the field dependence, while $\gamma$ governs the temperature roll-off near $T_{c0}$. The corresponding parameters are given in Table~\ref{tab:NbTi_params}.
	
	\begin{table}[ht!]
		\centering
		\begin{tabular}{lcl}
			\toprule
			\textbf{Parameter} & \textbf{Value} & \textbf{Unit} \\
			\midrule
			$T_{c0}$   & 9.03     & K \\
			$B_{c2,0}$ & 14.61    & T \\
			$C_0$      & 168512   & A$\cdot$T/mm$^2$ \\
			$\alpha$   & 1.0      & - \\
			$\beta$    & 1.54     & - \\
			$\gamma$   & 2.1      & - \\
			\bottomrule
		\end{tabular}
		\caption{NbTi scaling parameters (ITER/EU-DEMO)~\cite{corato2016common}}
		\label{tab:NbTi_params}
	\end{table}
	
	\subsubsection{\texorpdfstring{Nb$_3$Sn}{Nb3Sn} model}
	\label{sec:Nb3Sn}
	
	The Nb$_3$Sn scaling follows the EU-DEMO WST (Western Superconducting Technologies) strand parametrization~\cite{bottura2009jc, corato2016common} and accounts for the sensitivity to mechanical strain $\varepsilon$, which is particularly important for this brittle material. The strain function reads:
	\begin{equation}
		s(\varepsilon) = 1 + \frac{C_{a1}}{1 - C_{a1}\,\varepsilon_{0a}} \left(\sqrt{\varepsilon_{0a}^2} - \sqrt{\varepsilon^2 + \varepsilon_{0a}^2}\right)
	\end{equation}
	This strain function modifies both the critical temperature and the upper critical field:
	\begin{equation}
		T_{c0}^* = T_{cm}\,s(\varepsilon)^{1/3}, \quad B_{c2}^* = B_{c2m}\,s(\varepsilon)\,(1 - t^{1.52})
	\end{equation}
	with the reduced temperature $t = T/T_{c0}^*$ and reduced field $b = B/B_{c2}^*$. The critical current density is then expressed as:
	\begin{equation}
		J_{\text{non-Cu}}(B,T,\varepsilon) = \frac{C}{B}\,s(\varepsilon)\,(1 - t^{1.52})\,(1 - t^2)\,b^p\,(1 - b)^q \quad \left[\text{A/mm}^2\right]
	\end{equation}
	The exponents $p$ and $q$ govern the field dependence, while the temperature dependence is captured by the $(1 - t^{1.52})(1 - t^2)$ factor. Parameters for the EU-DEMO WST strand are given in Table~\ref{tab:Nb3Sn_params}.
	
	\begin{table}[ht!]
		\centering
		\begin{tabular}{lcc}
			\toprule
			\textbf{Parameter} & \textbf{Value} & \textbf{Unit} \\
			\midrule
			$T_{cm}$           & 16.34                & K \\
			$B_{c2m}$          & 33.24                & T \\
			$C$                & 83075                & A$\cdot$T/mm$^2$ \\
			$C_{a1}$           & 50.06                & - \\
			$\varepsilon_{0a}$ & $3.12\times10^{-3}$  & - \\
			$p$                & 0.593                & - \\
			$q$                & 2.156                & - \\
			\bottomrule
		\end{tabular}
		\caption{Nb$_3$Sn scaling parameters (EU-DEMO WST strand)~\cite{corato2016common}}
		\label{tab:Nb3Sn_params}
	\end{table}
	
	\subsubsection{REBCO model}
	\label{sec:REBCO}
	
	The default D0FUS model follows the Dew-Hughes pinning-force scaling of Senatore \emph{et al.}~\cite{senatore2024rebco}, calibrated on modern tapes (Fujikura FESC 2019 and SuperOx 2019). At a reference operating point $(B_{\rm ref}, T_{\rm ref})$ where $J_{\text{non-Cu}}$ is known from transport measurements, the scaling reads:
	
	\begin{equation}
		J_{\text{non-Cu}}(B,T) = J_{\text{non-Cu},\rm ref}\,\exp\!\left(-\frac{T - T_{\rm ref}}{T^*}\right) \frac{f_p(B,T)}{f_p(B_{\rm ref},T_{\rm ref})}
	\end{equation}
	
	where the pinning-force shape function (proportional to $F_p/B$) is:
	
	\[
	f_p(B,T) = b^{\,p-1}\,(1-b)^q, \qquad b = B / B_{\rm irr}(T)
	\]
	
	and the irreversibility field decreases with temperature as $B_{\rm irr}(T) = B_{\rm irr,0}\,(1-(T/T_c)^{n_1})^{n_2}$. Parameters for Fujikura FESC 2019 tapes (used by default in D0FUS) are given in Table~\ref{tab:REBCO_params_senatore}. The worst-case orientation, with the $B$ field normal to the tape ($\theta = 0$), is assumed.
	
	\begin{table}[ht!]
		\centering
		\begin{tabular}{lcc}
			\toprule
			\textbf{Parameter} & \textbf{Value} & \textbf{Unit} \\
			\midrule
			$T_{c}$                       & 93.0           & K \\
			$B_{\rm irr,0}$               & 187            & T \\
			$n_1$, $n_2$                  & 0.40, 1.0      & - \\
			$p$, $q$                      & 0.77, 4.5      & - \\
			$T^*$                         & 22             & K \\
			$J_{\text{non-Cu},\rm ref}$               & 2000           & A/mm$^2$ \\
			$(B_{\rm ref},\,T_{\rm ref})$ & (19~T,\,4.2~K) & - \\
			\bottomrule
		\end{tabular}
		\caption{REBCO scaling parameters (Senatore 2024, Fujikura FESC 2019 tape)~\cite{senatore2024rebco}}
		\label{tab:REBCO_params_senatore}
	\end{table}
	
	\subsubsection{Benchmark}
	
	At 4.2~K, the three scalings cover complementary field ranges (Fig.~\ref{fig:jc}): NbTi up to about 9~T, Nb$_3$Sn up to about 14~T, and REBCO beyond 20~T. The NHMFL maglab data~\cite{maglab2024} overlaid in Fig.~\ref{fig:jc} is in quantitative agreement with the NbTi and Nb$_3$Sn scalings. For REBCO, the D0FUS curve sits about a factor of two above the NHMFL SuperPower SP26 reference. This vertical offset is accounted for by the substantial improvement in REBCO tape performance over the past 15 years, the D0FUS scaling being anchored on modern Fujikura FESC 2019 and SuperOx 2019 transport measurements~\cite{senatore2024rebco}. The field dependence trend itself remains in good agreement with the maglab data.
	
	Table~\ref{tab:Jc_validation} summarizes a benchmark against ITER conductor specifications and commercial tape data.
	
	\begin{table*}[htbp]
		\centering
		\caption{Benchmark of $J_{\text{non-Cu}}$ scaling laws against reference specifications}
		\label{tab:Jc_validation}
		\begin{tabular}{lcccc}
			\toprule
			\textbf{Superconductor} & \textbf{Conditions} & \textbf{Calculated} & \textbf{Reference} & \textbf{Source} \\
			\midrule
			NbTi (ITER PF)   & 5~T, 4.2~K        & 3057 A/mm$^2$ & $\sim$2900 A/mm$^2$ & \cite{devred2012nbti}\\
			Nb$_3$Sn (ITER TF) & 11.8~T, 4.2~K   & 879 A/mm$^2$  & $\sim$900 A/mm$^2$ & \cite{devred2012nb3sn} \\
			REBCO (Fujikura 2019) & 19~T, 4.2~K, $B\perp$ & 2000 A/mm$^2$ & $\sim$2000 A/mm$^2$ & \cite{senatore2024rebco}\\
			\bottomrule
		\end{tabular}
	\end{table*}
	
	\subsubsection{\texorpdfstring{$J^{\rm wost}$}{Jwost} calculation}
	\label{sec:Jwost_calc}
	
	The non-steel cross-section of a Cable-In-Conduit Conductor is decomposed hierarchically. A fraction $f_{\rm In}$ is occupied by insulation and $f_{\rm CC}$ by a dedicated helium cooling channel, the remainder is the strand bundle.
	
	Within the strand bundle, the strands are packed with an interstitial helium void fraction $f_{\rm void}$, typically $\approx 0.33$ for LTS, and $\approx 0$ for HTS following the PIT-VIPER design~\cite{Sanabria2024PITVIPER}. The copper stabilizer is present both inside the strands or tape (within the matrix) and as separate pure copper components (strands or tapes) cabled in. All of this stabilizer cross-section is represented by the copper-to-non-copper ratio $r_{\rm Cu/non-Cu} = S_{\rm Cu}/S_{\text{non-Cu}}$.
	
	The ratio $r_{\rm Cu/non-Cu}$ is not a free parameter: it is set by quench protection through the Maddock adiabatic hot-spot criterion~\cite{maddock1969,wilson1983superconducting}, which expresses an energy balance between Joule heating in the copper and the enthalpy rise of the conductor materials up to a hot-spot temperature limit $T_{\rm hs}$. In the case of a current plateau (detection and hold time) followed by a pure exponential decay, this reads:
	\begin{equation}
		\int_{T_{\rm He}}^{T_{\rm hs}} \frac{\rho_{\rm mat}(T)\,c_{v,\rm mat}(T)}{\rho_{\rm Cu}(T,B,\rm RRR)}\,\mathrm{d}T
		= J_{\rm Cu,0}^{\,2}\!\left(\tau_h + \frac{\tau_d}{2}\right)
		\label{eq:maddock}
	\end{equation}
	where $\rho_{\rm mat}\,c_{v,\rm mat}$ is the volumetric heat capacity of the composite, $\rho_{\rm Cu}$ the copper resistivity, $\tau_h$ the detection and hold time, and $\tau_d = 2\,W_{\rm mag}/(I_0\,V_{\rm max})$ the exponential decay time of the discharge circuit, with $V_{\rm max}$ the maximum allowable terminal voltage during fast discharge. Default D0FUS values are $T_{\rm hs} = 250$~K (usually considered equivalent to a 150~K real hot-spot when taking into account the enthalpy of the structural material), $\mathrm{RRR} = 100$, $V_{\rm max} = 10$~kV (consistent with the ITER fast-discharge design~\cite{sborchia2008design}), and $\tau_h = 3$~s (LTS) or 10~s (HTS), with 2 TF coils per dump resistor. The corresponding module in D0FUS is adapted from the CEA magnet design code MADMACS~\cite{zani2019parametric}.
	
	The engineering current density is then obtained by combining the successive dilution fractions:
	\begin{equation}
		J^{\rm wost} = J_{\text{non-Cu}} \times f_{\text{non-Cu}} \times (1-f_{\rm void}) \times (1-f_{\rm In}-f_{\rm CC})
		\label{eq:Jwost}
	\end{equation}
	where $f_{\text{non-Cu}} = 1/(1 + r_{\rm Cu/non-Cu})$ is the non-copper fraction of a strand.
	
	\subsubsection{Operating margins}
	
	Three distinct temperatures are used in D0FUS:
	\begin{itemize}
		\item $T_{\rm He}$ is the nominal helium temperature at saturation (1 bar), $T_{\rm He} = 4.2$~K.
		\item $T_{\rm op}$ is the actual operating temperature of the conductor. The CICC channels are fed with supercritical helium at high pressure ($\sim 5-10$~bar), shifting the operating point to $T_{\rm op}$ a few tenths of K above $T_{\rm He}$ (typically 0.3 to 0.6 K depending on coolant pressure and circuit losses).
		\item $T_{\rm calc} = T_{\rm op} + \Delta T_{\rm margin}$ is the design temperature at which $J_{\text{non-Cu}}$ is evaluated for conductor sizing. Following EU-DEMO design rules~\cite{corato2016common}, $\Delta T_{\rm margin}$ embeds a temperature margin keeping the current-sharing temperature $T_{\rm cs}$ well above $T_{\rm op}$, with $\Delta T_{\rm margin} = 1.7$~K for NbTi, 1.5~K for Nb$_3$Sn, and 5.0~K for REBCO.
	\end{itemize}
	
	In addition, two material-specific operating conditions are assumed (see Sections~\ref{sec:Nb3Sn} and~\ref{sec:REBCO}): an effective strain of $\varepsilon = -0.6\%$ for Nb$_3$Sn, and the worst-case perpendicular field orientation ($\theta = 0$) for REBCO.
	
	\subsection{Determination of \texorpdfstring{$f_u(f_c;n)$}{fu(fc,n)}}
	\label{Annexe_gamma}
	
	We consider a cable of circular cross-section of radius $r_c$ inside a conductor of rectangular cross-section (Fig.~\ref{fig:surfacedilutionconductor}), of horizontal width $2(r_c + \delta_{S_2})$ and vertical height $2(r_c + \delta_{S_1})$, with $\delta_{S_1} = n\,\delta_{S_2}$ and $0 \leq n \leq 1$.
	
	The circular and total areas are:
	\[
	S_{\mathrm{circle}} = \pi\,r_c^2,
	\qquad
	S_{\mathrm{total}} = 4(r_c + \delta_{S_2})(r_c + \delta_{S_1})
	\]
	
	From the definition of $f_u$, considering a force acting in the vertical direction and the associated load-bearing section in the horizontal direction:
	\[
	f_u = \frac{\delta_{S_2}}{r_c + \delta_{S_2}} \;\Longrightarrow\; \delta_{S_2} = \frac{f_u\,r_c}{1 - f_u}
	\]
	Substituting into $\delta_{S_1} = n\,\delta_{S_2}$ gives $\delta_{S_1} = n\, f_u\, r_c/(1-f_u)$. Inserting both into $S_{\mathrm{total}}$ leads to:
	\[
	S_{\mathrm{total}} = \frac{4r_c^2}{(1 - f_u)^2}\bigl(1 + f_u (n - 1)\bigr)
	\]
	
	Therefore,
	\[
	\boxed{
		f_c(f_u, n) = \frac{\pi(1 - f_u)^2}{4(1 + f_u (n - 1))}
	}
	\]
	
	This expression can be inverted to obtain $f_u$ as a function of $f_c$:
	\[
	\boxed{
		\begin{gathered}
			f_u(f_c,n) = \\[6pt]
			\frac{2\pi + 4 f_c (n-1) \pm \sqrt{\bigl(2\pi + 4 f_c (n-1)\bigr)^2 - 4\pi(\pi - 4 f_c)}}{2\pi}
		\end{gathered}
	}
	\]
	with the physically meaningful root corresponding to the "$-$" branch.
	
	\subsection{Determination of \texorpdfstring{$f_u'(f_u ; n)$}{fu'(fu,n)}}
	\label{Appendix_little_demo}
	
	The stress concentration factor in the horizontal direction, denoted $f_u'$, is given by:
	\[
	f_u' = \frac{\delta_{S_1}}{r_c + \delta_{S_1}}
	\]
	Combining this with the vertical-direction expression $f_u = \delta_{S_2}/(r_c + \delta_{S_2})$ and the relation $\delta_{S_1} = n\,\delta_{S_2}$, one obtains:
	
	\[
	\boxed{
		f_u' = \frac{n\, f_u}{1 - f_u + n\, f_u}
	}
	\]
	
	\subsection{D0FUS reference inputs}
	\label{appendixinput}
	
	Table~\ref{tab:input_parameters} summarises the reference input parameters
	used throughout this study, calibrated on the EU-DEMO1 2017 baseline
	(PROCESS~v1.0.10~\cite{kovari2016process},
	run~EU~2NDSKT~v1.0~\cite{eurodemo2017process}) after aligning all input conventions between
	the two codes. This single parameter set defines the starting point from
	which all subsequent scans are derived. Parameters marked with $\dagger$
	vary across configurations in Section~\ref{Results}; all others remain fixed.
	
	\begin{table*}[!htbp] 
		\caption{Reference input parameters calibrated on the EU-DEMO1 2017
			baseline~\cite{eurodemo2017process}. $\dagger$: varied
			across configurations in Section~\ref{Results}.}
		\label{tab:input_parameters}
		\small
		\setlength{\tabcolsep}{4pt}
		\begin{minipage}[t]{0.47\textwidth}
			\begin{tabular}{llcl}
				\toprule
				\textbf{Parameter} & \textbf{Sym.} & \textbf{Value} & \textbf{Unit} \\
				\midrule
				\multicolumn{4}{l}{\textit{Geometry and power}} \\
				Fusion power         & $P_\mathrm{fus}$  & 1998    & MW  \\
				Major radius         & $R_0$             & 8.9   & m   \\
				Minor radius         & $a$               & 2.8   & m   \\
				Inboard radial build & $\Delta_{B}$      & 1.40   & m   \\
				Elongation model     & ---               & Wenninger \emph{et al.}~\cite{wenninger2015} & --- \\
				Plasma geometry      & ---               & Elliptical & --- \\
				\midrule
				\multicolumn{4}{l}{\textit{Confinement and transport}} \\
				Energy conf.\ scaling & ---             & IPB98(y,2)~\cite{ipb1999}     & ---  \\
				H-factor              & $H$             & 1.1        & ---  \\
				$q_{95}$ formula      & ---             & ITER\_1989~\cite{uckan1990} & ---  \\
				Bootstrap model       & ---             & Sauter \emph{et al.}~\cite{sauter1999neoclassical}    & ---  \\
				Trapped fraction      & ---             & ASTRA~\cite{fable_astra}     & ---  \\
				\midrule
				\multicolumn{4}{l}{\textit{Plasma profiles}} \\
				Average temperature        & $\bar{T}$                & 12.82 & keV \\
				Density peaking            & $\nu_n$                  & 1.0   & --- \\
				Temperature peaking        & $\nu_T$                  & 1.45  & --- \\
				Pedestal radius            & $\rho_\mathrm{ped}$      & 0.94  & --- \\
				Ped.\ density fraction     & $n_\mathrm{ped}/\bar{n}$ & 0.78 & --- \\
				Ped.\ temperature fraction & $T_\mathrm{ped}/\bar{T}$ & 0.43 & --- \\
				\midrule
				\multicolumn{4}{l}{\textit{Plasma composition and radiation}} \\
				Effective charge       & $Z_\mathrm{eff}$                   & 2.18                          & ---  \\
				Impurity species       & ---                                 & Xe, W                         & ---  \\
				Impurity fractions     & $f_\mathrm{Xe},f_\mathrm{W}$       & $3.5{\times}10^{-4}$,         & ---  \\
				&                                     & $5{\times}10^{-5}$            &      \\
				Core rad.\ boundary    & $\rho_\mathrm{rad,core}$           & 0.75                          & ---  \\
				Core rad.\ fraction    & ---                                 & 0.6                           & ---  \\
				Synchrotron reflection & $r_\mathrm{syn}$                   & 0.6                           & ---  \\
				$\alpha$ conf.\ factor & $C_\alpha$                         & 10                            & ---  \\
				\bottomrule
			\end{tabular}
		\end{minipage}
		\hfill
		\begin{minipage}[t]{0.50\textwidth}
			\begin{tabular}{llcl}
				\toprule
				\textbf{Parameter} & \textbf{Sym.} & \textbf{Value} & \textbf{Unit} \\
				\midrule
				\multicolumn{4}{l}{\textit{Magnets$^\dagger$}} \\
				Peak TF field      & $B_\mathrm{max}$              & 10.5                & T   \\
				Superconductor     & ---                           & Nb$_3$Sn            & --- \\
				Helium temperature & $T_\mathrm{He}$               & 4.75                & K   \\
				Temperature margin & $\Delta T_\mathrm{Nb_3Sn}$   & 1.5                 & K   \\
				Conductor current  & $I_\mathrm{cond}$             & 90                  & kA  \\
				Strain on SC       & $\varepsilon$                 & $-6.6{\times}10^{-3}$ & --- \\
				\midrule
				\multicolumn{4}{l}{\textit{CICC helium fractions}} \\
				Cooling channel fraction & $f_\mathrm{CC}$ & 0.10 & --- \\
				Void fraction         & $f_\mathrm{void}$    & 0.30 & --- \\
				Insulation fraction   & $f_\mathrm{In}$      & 0.15 & --- \\
				\midrule
				\multicolumn{4}{l}{\textit{Mechanical configuration$^\dagger$}} \\
				Radial build model   & ---                  & Refined  & --- \\
				Type of architecture & ---                  & Wedging  & --- \\
				Steel grade          & ---                  & 316L     & --- \\
				CS fatigue knockdown & ---                  & 2.0      & --- \\
				\midrule
				\multicolumn{4}{l}{\textit{Operation and current drive}} \\
				Operation mode      & ---                     & Pulsed   & ---                        \\
				Plateau duration    & $t_\mathrm{plateau}$    & 7200     & s                          \\
				Auxiliary heating   & $P_\mathrm{aux}$        & 50       & MW                         \\
				CD efficiency       & $\gamma_\mathrm{CD}$    & 0.30     & $10^{20}$\,A\,W$^{-1}$m$^{-2}$ \\
				Wall-plug efficiency & $\eta_\mathrm{WP}$    & 0.40     & ---                        \\
				Ramp-up CD fraction & $f_h$                & 0        & ---                        \\
				CS swing usable fraction & $f_\mathrm{swing}^\mathrm{usable}$ & 0.75 & --- \\
				\midrule
				\multicolumn{4}{l}{\textit{Stability limits}} \\
				$\beta_N$ limit    & ---                       & 2.9 & --- \\
				$q_{95}$ limit     & ---                       & 3.0 & --- \\
				Greenwald limit    & $f_\mathrm{GW,lim}$       & 1.2 & --- \\
				\bottomrule
			\end{tabular}
		\end{minipage}
	\end{table*}
	
	\subsection{Non-wedged winding-pack hypothesis}
	\label{Wedgapproximation}
	
	The D0FUS Refined model assumes that the toroidal centring load is reacted by the steel nose alone, the WP transmitting only radial stress ($\sigma_\theta^\mathrm{WP} \approx 0$). This appendix tests that simplification on an ITER-like reference case by sweeping the share of the load reacted by the WP.
	
	We define a wedging fraction $\alpha_\mathrm{WP} \in [0, 1]$ as the share of the tangential reaction moment absorbed by the WP, the remainder being reacted by the nose:
	\begin{equation}
		\alpha_\mathrm{WP} \;=\; \frac{\langle \sigma_\theta \rangle_\mathrm{WP} \, \Delta_\mathrm{WP}}{\langle \sigma_\theta \rangle_\mathrm{WP} \, \Delta_\mathrm{WP} + \langle \sigma_\theta \rangle_\mathrm{nose} \, \Delta_\mathrm{nose}},
		\label{eq:alpha_WP_def}
	\end{equation}
	with $\langle \cdot \rangle$ a layer-averaged value. The limit $\alpha_\mathrm{WP} = 0$ corresponds to the D0FUS hypothesis (the nose reacts the entire load) and $\alpha_\mathrm{WP} = 1$ to the opposite limit (the WP reacts the entire load).
	
	The radial elastic problem is solved with the analytical multilayer thick-cylinder solver CIRCE~\cite{boudes2025circe}, which provides closed-form expressions for the displacement and stress fields in concentric layers under combined boundary loading and a Lorentz body force $f_r(r) = J_\theta(r)\, B_z(r)$.
	
	The validation strategy is to sweep $\alpha_\mathrm{WP}$ across the full $[0, 1]$ interval and, at each value, use CIRCE to compute the stress field that is consistent with the prescribed level. The geometry $(R_\mathrm{sep}, \Delta_\mathrm{nose})$ is then optimised to minimise the total radial extent $\Delta_\mathrm{nose} + \Delta_\mathrm{WP}$ subject to the Tresca criterion saturated in both the nose steel and the WP steel. The procedure therefore produces, at every $\alpha_\mathrm{WP}$, a fully self-consistent design where mechanical equilibrium, the Tresca limit, and the ampere-turn constraint are simultaneously satisfied.
	
	\begin{figure}[ht]
		\centering
		\includegraphics[width=1.0\linewidth]{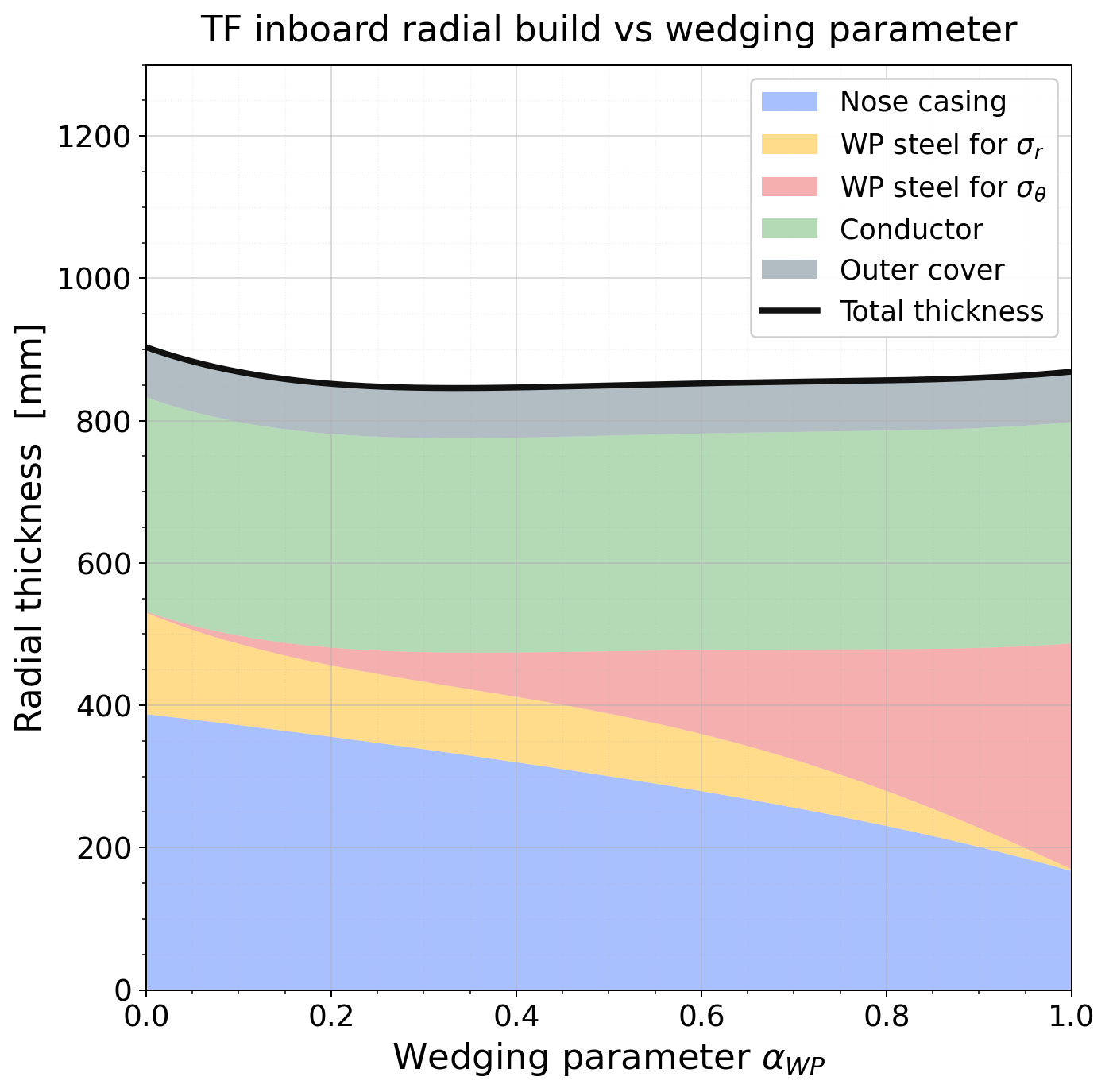}
		\caption{Optimal radial build of the inboard TF leg as a function of $\alpha_\mathrm{WP}$, computed with the CIRCE multilayer solver on ITER Q=10 reference parameters. The WP steel is split visually into a transmissive part ($\propto 1-\alpha_\mathrm{WP}$) and a vault part ($\propto \alpha_\mathrm{WP}$); this decomposition is conceptual since a single piece of steel reacts $\sigma_r$, $\sigma_\theta$ and $\sigma_z$ simultaneously.}
		\label{fig:wedging_alpha}
	\end{figure}
	
	Figure~\ref{fig:wedging_alpha} shows the optimised radial build for the ITER Q=10 parameter set of Table~\ref{tab:tf_coil_benchmark}. The total thickness $\Delta_\mathrm{tot}$ is essentially flat, varying by less than 5\% over the full $\alpha_\mathrm{WP}$ interval.
	
	Treating the WP as non-wedged ($\sigma_\theta^\mathrm{WP} \approx 0$) is therefore a reasonable approximation for sizing the total radial extent of the inboard TF leg. This justifies the modelling choice of Section~\ref{D0FUS} and removes the need to introduce $\alpha_\mathrm{WP}$ as an additional design degree of freedom in D0FUS.
	
	\subsection{Sensitivity to the WP/nose tension partition}
	\label{appendix_omega_sensitivity}
	
	The Refined model parameter $f_{z,\mathrm{WP}} \in [0,1]$ defined in Section~\ref{Param_definition} sets the fraction of the total vertical tension $F_z$ borne by the winding pack, the complementary fraction $1 - f_{z,\mathrm{WP}}$ being reacted by the steel nose. This appendix quantifies the impact of $f_{z,\mathrm{WP}}$ on the predicted radial build.
	
\begin{figure}[ht]
		\centering
		\includegraphics[width=1.0\linewidth]{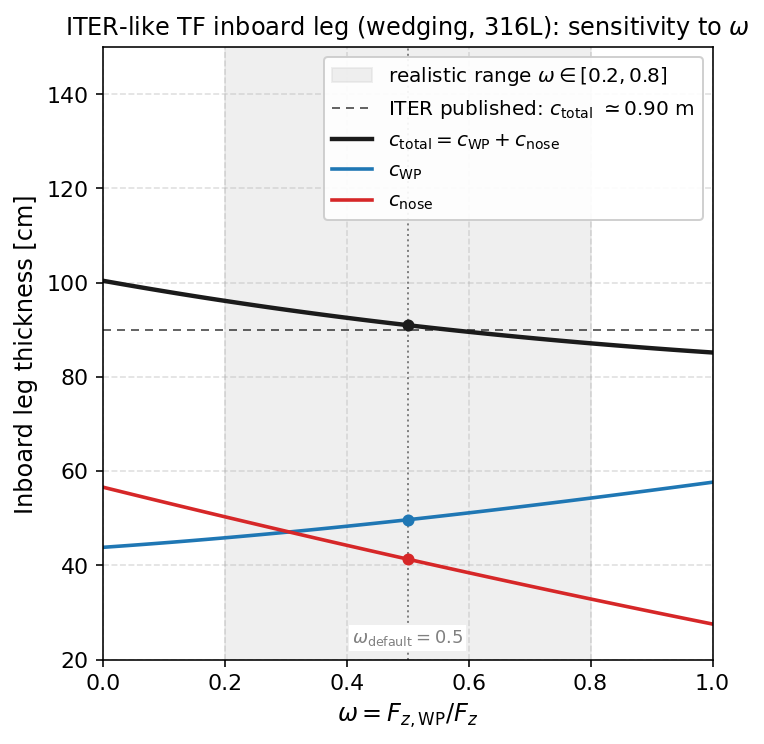}
		\caption{Sensitivity of the inboard leg thicknesses to $f_{z,\mathrm{WP}}$ on an ITER-like wedging/316L case.}
		\label{fig:omega_scan}
	\end{figure}

	Figure~\ref{fig:omega_scan} shows the result of an $f_{z,\mathrm{WP}}$ scan on the ITER reference parameter set of Table~\ref{tab:tf_coil_benchmark}. Over the realistic range $f_{z,\mathrm{WP}} \in [0.2, 0.8]$, the total thickness varies by less than 10\%. This weak dependence has a simple origin: at fixed $F_z$ and $\sigma_\mathrm{lim}$, the total steel cross-section needed to react the axial tension is set by $\sigma_z = F_z / S_\mathrm{steel} \leq \sigma_\mathrm{lim}$. Varying $f_{z,\mathrm{WP}}$ only transfers steel from one region to the other.
	
	Two conclusions follow. First, the weak sensitivity of $c_\mathrm{total}$ on $f_{z,\mathrm{WP}}$ justifies treating it as a fixed input rather than introducing a dedicated convergence loop on the steel area distribution. Second, the default $f_{z,\mathrm{WP}} = 1/2$ is consistent with the ITER TF design (Fig.~\ref{fig:ITER}, Ref.~\cite{federici2026iter,sborchia2008design}), which shows comparable amounts of structural steel in the WP and the nose.
	
	\subsection{Radially graded conductor fraction}
	\label{appendix_grading}
	
	In the Refined model (Section~\ref{D0FUS}), the conductor fraction $f_c$ is uniform across the winding pack. The Tresca criterion is then saturated only at the most loaded radius $R_\mathrm{TF}^\mathrm{int}$, while regions closer to the outer surface operate below the stress limit. Steel is therefore over-provisioned everywhere except at $R_\mathrm{TF}^\mathrm{int}$.
	
	A graded model relaxes this assumption by allowing $f_c$ to vary with radius. Recalling from Eq.~\ref{eq:sigma_z_uniform} that $\sigma_z$ already denotes the axial stress in the \emph{steel} of the WP, the graded extension simply replaces $(1-f_c)$ by the area-weighted average steel fraction $\langle 1-f_c \rangle$ (defined in Eq.~\ref{eq:f_steel_graded}):
	\begin{equation}
		\sigma_z = \frac{f_{z,\mathrm{WP}}\,F_z}{\langle 1-f_c \rangle\,S_\mathrm{tot}}
		\label{eq:sigma_z_graded}
	\end{equation}
	With this approximation of a single value of $\sigma_z$ shared by all radii, the local Tresca criterion in the WP reads
	\begin{equation}
		\frac{\sigma_r(R)}{f_u} + \sigma_z = \sigma_\mathrm{lim}
		\label{eq:tresca_graded}
	\end{equation}
	and can be saturated everywhere in the winding pack by tuning $f_c(R)$ accordingly. At each radius $R$, the local conductor fraction $f_c(R)$ is determined by imposing
	\begin{equation}
		f_u = \frac{\sigma_r(R)}{\sigma_\mathrm{lim} - \sigma_z}
		\label{eq:gamma_graded}
	\end{equation}
	which is inverted numerically using the $f_u(f_c,n)$ relation derived in Appendix~\ref{Annexe_gamma}. Since $f_u$ is a decreasing function of $f_c$ (a higher conductor fraction means more stress concentration), and $\sigma_r$ increases monotonically from zero at $R = R_\mathrm{TF}^\mathrm{ext}$ to its peak at $R_\mathrm{TF}^\mathrm{int}$, the model naturally assigns more steel (lower~$f_c$) to inner regions where stresses are highest, and more conductor (higher~$f_c$) to outer regions.
	
	The radial stress $\sigma_r(R)$ and the toroidal field $B(R)$ within the winding pack are coupled through the electromagnetic body force. We integrate from the outer surface $R_\mathrm{TF}^\mathrm{ext}$ inward. At radius $R$, the enclosed ampere-turns $NI(R)$ and the field $B(R)$ follow from Amp\`ere's law:
	\begin{equation}
		B(R) = \frac{\mu_0\, NI(R)}{2\pi R}, \qquad \frac{dNI}{dR} = -f_c(R)\,J_\mathrm{TF}^\mathrm{wost}\,2\pi R
		\label{eq:ampere_graded}
	\end{equation}
	with $NI(R_\mathrm{TF}^\mathrm{ext}) = B_\mathrm{max}\,2\pi\,R_\mathrm{TF}^\mathrm{ext}/\mu_0$. The smeared radial stress accumulates as
	\begin{equation}
		\frac{d\sigma_r}{dR} = -f_c(R)\,J_\mathrm{TF}^\mathrm{wost}\,B(R)
		\label{eq:dsigma_r_graded}
	\end{equation}
	with $\sigma_r(R_\mathrm{TF}^\mathrm{ext}) = 0$ (free surface). Integration stops at $NI = 0$, which defines $R_\mathrm{TF}^\mathrm{sep}$ and thus $\Delta_\mathrm{WP} = R_\mathrm{TF}^\mathrm{ext} - R_\mathrm{TF}^\mathrm{sep}$.
	
	The vertical stress $\sigma_z$ depends on the geometry through the same expression as in the uniform model (Section~\ref{gamma_usefull}), but with $(1-f_c)$ replaced by the area-weighted average steel fraction:
	\begin{equation}
		\langle 1 - f_c \rangle = \frac{\displaystyle\int_{R_\mathrm{TF}^\mathrm{sep}}^{R_\mathrm{TF}^\mathrm{ext}} \bigl(1-f_c(R)\bigr)\,R\,dR}{\displaystyle\int_{R_\mathrm{TF}^\mathrm{sep}}^{R_\mathrm{TF}^\mathrm{ext}} R\,dR}
		\label{eq:f_steel_graded}
	\end{equation}
	Since $\sigma_z$ itself enters the Tresca budget (Eq.~\ref{eq:gamma_graded}) and therefore affects $f_c(R)$, the system is solved iteratively: $\sigma_z$ is initialized from a first guess, the inward integration is performed, the resulting $\langle 1-f_c \rangle$ and $R_\mathrm{TF}^\mathrm{sep}$ are used to update $\sigma_z$, and the process is repeated until convergence (Picard iteration, typically fewer than 10 iterations, with under-relaxation factor 0.5).
	
	The graded model removes structural steel where it is not needed, packing more current density in the outer portion of the winding pack and reducing the total thickness required to carry the same ampere-turns.
	
\begin{figure}[ht]
		\centering
		\includegraphics[width=0.5\textwidth]{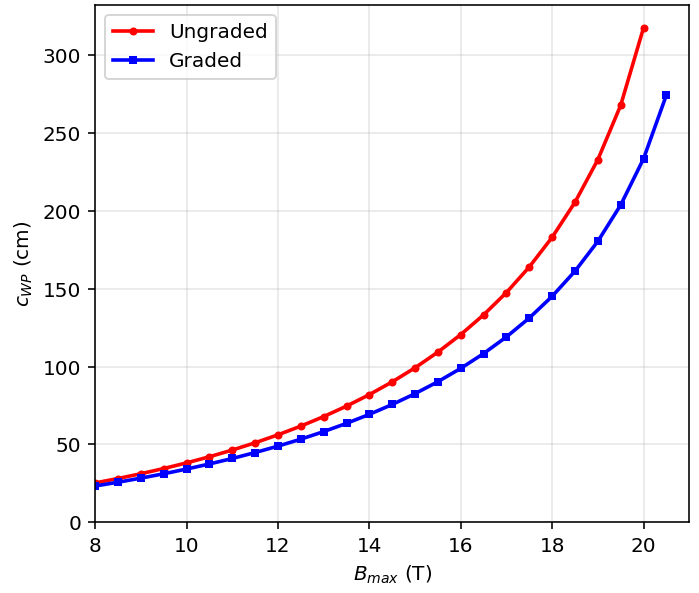}
		\caption{TF winding pack thickness $c_\mathrm{WP}$ as a function of $B_\mathrm{max}$ for the ungraded and radially graded conductor fraction models ($R_0 = 9$~m, $a = 3$~m, $\Delta_B = 1.7$~m, $\sigma_\mathrm{lim} = 660$~MPa, $J_\mathrm{TF}^\mathrm{wost} = 50$~A/mm$^2$).}
		\label{fig:grading}
	\end{figure}

	Figure~\ref{fig:grading} compares the winding pack thickness $c_\mathrm{WP}$ obtained with and without radial grading as a function of $B_\mathrm{max}$. The relative gain is visible across the entire field range and is largest at moderate fields where $\sigma_r$ dominates the Tresca budget.
	
	This grading model is compatible with all three mechanical configurations (wedging, bucking, plug) and all superconductor scalings. It does not account for manufacturing constraints that would limit the number of distinct conductor grades in practice (multiple conductor production lines, and local structural inhomogeneity of the winding pack that can introduce 3D stress concentrations). A concrete example is the CFETR TF coil, designed with three graded sub-winding-packs (high-Jc Nb$_3$Sn, ITER-like Nb$_3$Sn, NbTi)~\cite{hao2022conductor} and currently prototyped under the CRAFT project~\cite{wu2021preliminary}, which will provide a first fusion-scale demonstration that even a small number of grades captures most of the theoretical gain.
	
	\bibliographystyle{unsrt}
	\bibliography{bib}

@article{auclair2025tokamak,
  author    = {Auclair, Timoth{\'e} and Nardon, Eric and Sarazin, Yanick and Artaud, Jean-Fran{\c{c}}ois and Boudes, Baptiste and Bourdelle, Clarisse and Duchateau, Jean-Luc and Torre, Alexandre},
  title     = {The tokamak system code D0FUS and its first applications: Impact of B and confinement scaling law on power plant design},
  journal   = {Fusion Engineering and Design},
  publisher = {Elsevier},
  volume    = {219},
  pages     = {115270},
  year      = {2025}
}

@article{bachmann2023influence,
  author    = {Bachmann, C and Siccinio, M and Albino, Martin and Chiappa, A and Falcitelli, G and Federici, Gianfranco and Giannini, L and Luongo, C},
  title     = {Influence of a high magnetic field to the design of EU DEMO},
  journal   = {Fusion Engineering and Design},
  publisher = {Elsevier},
  volume    = {197},
  pages     = {114050},
  year      = {2023}
}

@techreport{bajas2022ship,
  author      = {Bajas, H. and Tommasini, D.},
  title       = {The {SHiP} spectrometer magnet -- Superconducting options},
  institution = {CERN},
  number      = {CERN-SHiP-NOTE-2022-001, EDMS 2440157},
  year        = {2022},
  url         = {https://edms.cern.ch/document/2440157}
}

@article{barth2015electro,
  author    = {Barth, Christian and Mondonico, Giorgio and Senatore, Carmine},
  title     = {Electro-mechanical properties of REBCO coated conductors from various industrial manufacturers at 77 K, self-field and 4.2 K, 19 T},
  journal   = {Superconductor Science and Technology},
  publisher = {IOP publishing},
  volume    = {28},
  number    = {4},
  pages     = {045011},
  year      = {2015}
}

@incollection{benzinger1980manufacturing,
  author    = {Benzinger, JR},
  title     = {Manufacturing capabilities of CR-grade laminates},
  booktitle = {Advances in Cryogenic Engineering Materials: Volume 26},
  publisher = {Springer},
  pages     = {252--258},
  year      = {1980}
}

@article{bottura2009jc,
  author  = {Bottura, L. and Bordini, B.},
  title   = {{$J_c(B,T,\varepsilon)$} parameterization for the {ITER} {Nb$_3$Sn} production},
  journal = {IEEE Transactions on Applied Superconductivity},
  volume  = {19},
  number  = {3},
  pages   = {1521--1524},
  year    = {2009},
  doi     = {10.1109/TASC.2009.2018278}
}

@article{boudes2025circe,
  author    = {Boudes, B and Torre, A and Nunio, F and Dao, C Nguyen Thanh},
  title     = {The CIRCE Code: Cast3M Investigation and Research for Coils Mechanical Evaluation},
  journal   = {IEEE Transactions on Applied Superconductivity},
  publisher = {IEEE},
  year      = {2025}
}

@article{burkhard1975magnetic,
  author    = {Burkhard, H},
  title     = {Magnetic-and stress-fields of cylindrical non-uniform current density superconducting coils},
  journal   = {Applied physics},
  publisher = {Springer},
  volume    = {6},
  number    = {3},
  pages     = {357--362},
  year      = {1975}
}

@article{chen2008design,
  author    = {Chen, Wenge and Pan, Yannian and Chen, Zuoming and Wei, Jin},
  title     = {The design and the manufacturing process of the superconducting toroidal field magnet system for EAST device},
  journal   = {Fusion engineering and design},
  publisher = {Elsevier},
  volume    = {83},
  number    = {1},
  pages     = {45--49},
  year      = {2008}
}

@article{chen20163d,
  author    = {Chen, SL and Villone, F and Xiao, BJ and Barbato, L and Luo, ZP and Liu, L and Mastrostefano, S and Xing, Z},
  title     = {3D passive stabilization of n= 0 MHD modes in EAST tokamak},
  journal   = {Scientific Reports},
  publisher = {Nature Publishing Group UK London},
  volume    = {6},
  number    = {1},
  pages     = {32440},
  year      = {2016}
}

@article{Clapeyron1829,
  author  = {Clapeyron, Benoit Paul Emile},
  title   = {Mémoire sur l'équilibre intérieur des corps solides homogènes},
  journal = {Journal de l'Ecole Polytechnique},
  volume  = {13},
  number  = {23},
  pages   = {153--190},
  year    = {1829}
}

@article{coleman2025definition,
  author    = {Coleman, M and Zohm, H and Bourdelle, C and Maviglia, F and Pearce, AJ and Siccinio, M and Spagnuolo, A and Wiesen, S},
  title     = {Definition of an EU-DEMO design point robust to epistemic plasma physics uncertainties},
  journal   = {Nuclear Fusion},
  publisher = {IOP Publishing},
  volume    = {65},
  number    = {3},
  pages     = {036039},
  year      = {2025}
}

@article{corato2016common,
  author  = {Corato, V and Bonifetto, R and Bruzzone, P and Ciazynski, D and Coleman, M and Gaio, E and Heller, R and Lacroix, B and Lewandowska, M and Maistrello, A and others},
  title   = {Common operating values for DEMO magnets design for 2016},
  journal = {EUROFusion Programme Management Unit, Oxon, United Kingdom, IDM Ref. No. EFDA\_D\_2MMDTG},
  year    = {2016}
}

@article{creely2020overview,
  author    = {Creely, AJ and Greenwald, Martin J and Ballinger, Sean B and Brunner, D and Canik, J and Doody, Jeffrey and F{\"u}l{\"o}p, T and Garnier, DT and Granetz, R and Gray, TK and others},
  title     = {Overview of the SPARC tokamak},
  journal   = {Journal of Plasma Physics},
  publisher = {Cambridge University Press},
  volume    = {86},
  number    = {5},
  pages     = {865860502},
  year      = {2020}
}

@misc{creely2024aps,
  author       = {Creely, Alexander J. and {SPARC, CFS}},
  title        = {Advances in the {SPARC} and {ARC} projects at {Commonwealth Fusion Systems}},
  address      = {Atlanta, GA, USA},
  year         = {2024},
  note         = {Session TO06, 10 October 2024. \url{https://meetings.aps.org/Meeting/DPP24/Session/TO06.1}},
  howpublished = {Bull.\ Am.\ Phys.\ Soc., 66th Annual Meeting of the APS Division of Plasma Physics, Abstract TO06.00001}
}

@article{creely2024comment,
  author    = {Creely, A. J. and Brunner, D. and Eich, T. and Greenwald, M. J. and LaBombard, B. and Mumgaard, R. T. and Segal, M. and Sorbom, B. N. and Whyte, D. G.},
  title     = {Comment on `{Relationship between magnetic field and tokamak size --- a system engineering perspective and implications to fusion development}'},
  journal   = {Nuclear Fusion},
  publisher = {IOP Publishing},
  volume    = {64},
  number    = {10},
  pages     = {108001},
  year      = {2024},
  doi       = {10.1088/1741-4326/ad6c5f}
}

@misc{creely2025eps,
  author       = {Creely, Alexander J.},
  title        = {The {SPARC} and {ARC} tokamaks: Predictive results and open physics questions on the path to a commercial fusion power plant},
  address      = {Vilnius, Lithuania},
  year         = {2025},
  note         = {Session MCF~2, 8 July 2025. \url{https://lac913.epfl.ch/epsppd3/2025/html/EPS_2025_Abstract_Book_Final.pdf}},
  howpublished = {51st {EPS} Conference on Plasma Physics, Abstract O~031, Europhysics Conference Abstracts Vol.~51A, ISBN 978-80-909294-1-8}
}

@article{devred2012nb3sn,
  author  = {Devred, A. and others},
  title   = {Status of {ITER} conductor development and production},
  journal = {IEEE Transactions on Applied Superconductivity},
  volume  = {22},
  number  = {3},
  pages   = {4804909},
  year    = {2012},
  note    = {Nb$_3$Sn specification: $J_c > 800$~A/mm$^2$ at 12~T, 4.2~K},
  doi     = {10.1109/TASC.2012.2182980}
}

@article{devred2012nbti,
  author  = {Devred, A. and others},
  title   = {Challenges and status of {ITER} conductor production},
  journal = {Superconductor Science and Technology},
  volume  = {27},
  number  = {4},
  pages   = {044001},
  year    = {2014},
  note    = {NbTi specification: $J_c > 2900$~A/mm$^2$ at 5~T, 4.2~K},
  doi     = {10.1088/0953-2048/27/4/044001}
}

@article{devries2019breakdown,
  author  = {de Vries, P. C. and Gribov, Y.},
  title   = {{ITER} breakdown and plasma initiation revisited},
  journal = {Nuclear Fusion},
  volume  = {59},
  number  = {9},
  pages   = {096043},
  year    = {2019},
  doi     = {10.1088/1741-4326/ab2ef4}
}

@article{di2014overview,
  author    = {Di Pietro, Enrico and Barabaschi, Pietro and Kamada, Yutaka and Ishida, Shinichi and others},
  title     = {Overview of engineering design, manufacturing and assembly of JT-60SA machine},
  journal   = {Fusion Engineering and Design},
  publisher = {Elsevier},
  volume    = {89},
  number    = {9-10},
  pages     = {2128--2135},
  year      = {2014}
}

@misc{diazpacheco2025electromechanical,
  author       = {Diaz-Pacheco, R. and Greenberg, A. and Armiri, B. and Sanabria, C. and Chavarria, D. and Rajan, G. and Colque, J. and Metcalfe, K. and Joseph, L. and Litchfield, M. and Duke, O. and Padukone, R.},
  title        = {Electromechanical properties of {SPARC} {CS} and {PF} superconductor cables under relevant transverse and axial compression},
  month        = {July},
  year         = {2025},
  url          = {https://indico.cern.ch/event/1431972/contributions/6420189/},
  howpublished = {Invited talk, 29th International Conference on Magnet Technology (MT29), Boston, MA, USA}
}

@article{duchateau2014conceptual,
  author    = {Duchateau, J-L and Hertout, P and Saoutic, B and Artaud, J-F and Zani, L and Reux, C},
  title     = {Conceptual integrated approach for the magnet system of a tokamak reactor},
  journal   = {Fusion Engineering and Design},
  publisher = {Elsevier},
  volume    = {89},
  number    = {11},
  pages     = {2606--2620},
  year      = {2014}
}

@incollection{elber1971significance,
  author    = {Elber, Wolf},
  title     = {The significance of fatigue crack closure},
  booktitle = {Damage Tolerance in Aircraft Structures, ASTM STP 486},
  publisher = {ASTM International},
  pages     = {230--242},
  year      = {1971}
}

@techreport{eurodemo2017process,
  author      = {{EUROfusion Consortium}},
  title       = {{EU-DEMO1} 2017 Reference Design: {PROCESS} v1.0.10 Input File ({EU~2NDSKT~v1.0})},
  institution = {EUROfusion},
  month       = {March},
  year        = {2017}
}

@techreport{fable_astra,
  author      = {Fable, E.},
  title       = {{Trapped particle fraction formula implemented in ASTRA}},
  institution = {IPP Garching},
  year        = {2015},
  note        = {Formula documented in the PROCESS systems code: \url{https://ukaea.github.io/PROCESS/physics-models/plasma_current/bootstrap_current/}},
  type        = {Private communication}
}

@article{federici2024relationship,
  author    = {Federici, G and Siccinio, M and Bachmann, C and Giannini, L and Luongo, C and Lungaroni, M},
  title     = {Relationship between magnetic field and tokamak size—a system engineering perspective and implications to fusion development},
  journal   = {Nuclear Fusion},
  publisher = {IOP Publishing},
  volume    = {64},
  number    = {3},
  pages     = {036025},
  year      = {2024}
}

@misc{federici2026iter,
  author       = {Federici, Gianfranco},
  title        = {ITER Engineering Basis Handbook},
  year         = {2026},
  howpublished = {ITER Organization}
}

@techreport{fleiter2014rebco,
  author      = {Fleiter, J. and Ballarino, A.},
  title       = {Parameterization of the critical surface of {REBCO} conductors from {Fujikura}},
  institution = {CERN},
  number      = {EDMS 1426239},
  year        = {2014},
  url         = {https://edms.cern.ch/document/1426239}
}

@article{formisano2017analysis,
  author  = {Formisano, A. and Albanese, R. and Ambrosino, G. and de Magistris, M. and De Vries, P. and Gribov, Y. and Ledda, F. and Martone, R. and Mattei, M. and Minucci, S. and Pironti, A. and Pizzo, F. and Snipes, J. and Villone, F. and Zabeo, L.},
  title   = {{3D} Analysis of magnetic field lines to assess the impact of stray fields at breakdown in {ITER}},
  journal = {Fusion Engineering and Design},
  volume  = {123},
  year    = {2017},
  doi     = {10.1016/j.fusengdes.2017.03.015}
}

@article{freethy2024optimisation,
  author  = {Freethy, S. and others},
  title   = {The optimisation of the {STEP} electron cyclotron current drive concept},
  journal = {Nuclear Fusion},
  volume  = {64},
  pages   = {126035},
  year    = {2024},
  doi     = {10.1088/1741-4326/ad7a8a}
}

@article{freidberg2015designing,
  author    = {Freidberg, JP and Mangiarotti, FJ and Minervini, J},
  title     = {Designing a tokamak fusion reactor—How does plasma physics fit in?},
  journal   = {Physics of Plasmas},
  publisher = {AIP Publishing},
  volume    = {22},
  number    = {7},
  year      = {2015}
}

@article{Fu2006,
  author  = {Fu, Y. and Jong, C. and Michael, P. and Mitchell, N.},
  title   = {Pre-Compression Requirements for the {ITER} Central Solenoid},
  journal = {IEEE Transactions on Applied Superconductivity},
  volume  = {16},
  number  = {2},
  pages   = {791--794},
  year    = {2006},
  doi     = {10.1109/TASC.2006.873257}
}

@article{giannini2023magnet,
  author    = {Giannini, Lorenzo and Muzzi, Luigi and Portone, Alfredo and Romanelli, Gherardo and Boso, Daniela P and Zoboli, Lorenzo and Sarasola, Xabier and Zignani, Chiarasole Fiamozzi and Luongo, Cesar and Corato, Valentina and others},
  title     = {The MAgnet Design Explorer algorithm (MADE) for LTS, Hybrid or HTS toroidal and poloidal systems of a tokamak with a view to DEMO},
  journal   = {Fusion Engineering and Design},
  publisher = {Elsevier},
  volume    = {193},
  pages     = {113659},
  year      = {2023}
}

@article{Godeke2006_Nb3Sn_review,
  author  = {A. Godeke},
  title   = {A review of the properties of Nb3Sn and their variation with A15 composition, morphology and strain state},
  journal = {Superconductor Science and Technology},
  volume  = {19},
  number  = {8},
  pages   = {R68--R80},
  year    = {2006},
  doi     = {10.1088/0953-2048/19/8/R02}
}

@article{gray1977electromechanical,
  author    = {Gray, WH and Ballou, JK},
  title     = {Electromechanical stress analysis of transversely isotropic solenoids},
  journal   = {Journal of Applied Physics},
  publisher = {American Institute of Physics},
  volume    = {48},
  number    = {7},
  pages     = {3100--3109},
  year      = {1977}
}

@article{grigorenko2006solving,
  author    = {Grigorenko, Ya M and Grigorenko, A Ya and Rozhok, LS},
  title     = {Solving the stress problem for solid cylinders with different end conditions},
  journal   = {International Applied Mechanics},
  publisher = {Springer},
  volume    = {42},
  number    = {6},
  pages     = {629--635},
  year      = {2006}
}

@article{hahn2010hts,
  author    = {Hahn, Seungyong and Park, Dong Keun and Bascunan, Juan and Iwasa, Yukikazu},
  title     = {HTS pancake coils without turn-to-turn insulation},
  journal   = {IEEE transactions on applied superconductivity},
  publisher = {IEEE},
  volume    = {21},
  number    = {3},
  pages     = {1592--1595},
  year      = {2010}
}

@article{hao2022conductor,
  author  = {Hao, Qiangwang and Hussain, Muhammad Talib and Dai, Chao and Wu, Yu and Shi, Yi and Hu, Libiao and Liu, Xiaochuan},
  title   = {Conductor design and performance analysis for {CFETR} magnet},
  journal = {Fusion Engineering and Design},
  volume  = {182},
  pages   = {113224},
  year    = {2022},
  doi     = {10.1016/j.fusengdes.2022.113224}
}

@article{hartwig2023sparc,
  author    = {Hartwig, Zachary S and Vieira, Rui F and Dunn, Darby and Golfinopoulos, Theodore and LaBombard, Brian and Lammi, Christopher J and Michael, Philip C and Agabian, Susan and Arsenault, David and Barnett, Raheem and others},
  title     = {The SPARC toroidal field model coil program},
  journal   = {IEEE Transactions on Applied Superconductivity},
  publisher = {IEEE},
  volume    = {34},
  number    = {2},
  pages     = {1--16},
  year      = {2023}
}

@article{hirshman1986external,
  author  = {Hirshman, S. P. and Neilson, G. H.},
  title   = {External inductance of an axisymmetric plasma},
  journal = {Physics of Fluids},
  volume  = {29},
  number  = {3},
  pages   = {790--793},
  year    = {1986},
  doi     = {10.1063/1.865934}
}

@article{hong2019optimal,
  author  = {Hong, B.G. and Kim, T.-H.},
  title   = {On the optimal radial build of a normal aspect ratio tokamak fusion system},
  journal = {Fusion Engineering and Design},
  volume  = {139},
  pages   = {148--154},
  year    = {2019},
  doi     = {10.1016/j.fusengdes.2019.01.042}
}

@book{humphries1990charged,
  author    = {Humphries, Stanley, Jr.},
  title     = {Charged Particle Beams},
  publisher = {John Wiley \& Sons},
  address   = {New York},
  year      = {1990},
  note      = {Republished by Dover, 2013, ISBN 978-0-486-49868-3},
  isbn      = {978-0-471-60014-5}
}

@article{imbeaux2011current,
  author  = {Imbeaux, F. and Citrin, J. and Hobirk, J. and Hogeweij, G. M. D. and K{\"o}chl, F. and Leonov, V. M. and Miyamoto, S. and Nakamura, Y. and Parail, V. and Pereverzev, G. and Polevoi, A. and Voitsekhovitch, I. and others},
  title   = {Current ramps in tokamaks: from present experiments to {ITER} scenarios},
  journal = {Nuclear Fusion},
  volume  = {51},
  number  = {8},
  pages   = {083026},
  year    = {2011},
  doi     = {10.1088/0029-5515/51/8/083026}
}

@article{ipb1999,
  author  = {{ITER Physics Expert Groups on Confinement and Transport and Confinement Modelling and Database}},
  title   = {{Chapter 2: Plasma confinement and transport}},
  journal = {Nucl. Fusion},
  volume  = {39},
  pages   = {2175},
  year    = {1999}
}

@book{iwasa2009casestudies,
  author    = {Iwasa, Yukikazu},
  title     = {Case Studies in Superconducting Magnets: Design and Operational Issues},
  publisher = {Springer},
  address   = {New York},
  edition   = {2nd},
  year      = {2009},
  doi       = {10.1007/b112047},
  isbn      = {978-0-387-09799-2}
}

@book{jackson1998classical,
  author    = {Jackson, John David},
  title     = {Classical Electrodynamics},
  publisher = {John Wiley \& Sons},
  address   = {New York},
  edition   = {3rd},
  year      = {1998},
  isbn      = {978-0-471-30932-1}
}

@article{jean2011helios,
  author    = {Jean, Johner},
  title     = {HELIOS: a zero-dimensional tool for next step and reactor studies},
  journal   = {Fusion Science and Technology},
  publisher = {Taylor \& Francis},
  volume    = {59},
  number    = {2},
  pages     = {308--349},
  year      = {2011}
}

@incollection{johnson1977stress,
  author    = {Johnson, Neil E},
  title     = {Stress analysis of nonhomogeneous superconducting solenoids},
  booktitle = {Advances in Cryogenic Engineering: Volume 22},
  publisher = {Springer},
  pages     = {490--499},
  year      = {1977}
}

@inproceedings{jong2007iter,
  author       = {Jong, CTJ and Mitchell, N and Knaster, J},
  title        = {ITER magnet design criteria and their impact on manufacturing and assembly},
  booktitle    = {2007 IEEE 22nd Symposium on Fusion Engineering},
  organization = {IEEE},
  pages        = {1--4},
  year         = {2007}
}

@inproceedings{jong2009mechanical,
  author       = {Jong, CTJ and Beemsterboer, CJJ and Libeyre, P and Lyraud, C and Mitchell, N and Vollmann, Th},
  title        = {Mechanical behaviour of ITER central solenoid},
  booktitle    = {2009 23rd IEEE/NPSS Symposium on Fusion Engineering},
  organization = {IEEE},
  pages        = {1--4},
  year         = {2009}
}

@incollection{kasen1980mechanical,
  author    = {Kasen, MB and MacDonald, GR and Beekman Jr, DH and Schramm, RE},
  title     = {Mechanical, electrical, and thermal characterization of G-10CR and G-11CR glass-cloth/epoxy laminates between room temperature and 4 K},
  booktitle = {Advances in Cryogenic Engineering Materials: Volume 26},
  publisher = {Springer},
  pages     = {235--244},
  year      = {1980}
}

@article{kovari2016process,
  author    = {Kovari, M and Fox, F and Harrington, C and Kembleton, R and Knight, P and Lux, H and Morris, J},
  title     = {“PROCESS”: A systems code for fusion power plants--Part 2: Engineering},
  journal   = {Fusion Engineering and Design},
  publisher = {Elsevier},
  volume    = {104},
  pages     = {9--20},
  year      = {2016}
}

@article{LameClapeyron1833,
  author    = {Lam{\'e}, Gabriel and Clapeyron, {\'E}mile},
  title     = {M{\'e}moire sur l'{\'e}quilibre int{\'e}rieur des corps solides homog{\`e}nes},
  journal   = {M{\'e}moires pr{\'e}sent{\'e}s par divers savants {\`a} l'Acad{\'e}mie royale des sciences de l'Institut de France},
  publisher = {Bachelier},
  address   = {Paris},
  volume    = {4},
  pages     = {465--562},
  year      = {1833}
}

@inproceedings{li2019flashover,
  author       = {Li, J and Luo, LM and Huang, RJ and Ma, JH and Liang, YQ and Huang, CJ and Wu, ZX and Shen, FZ and Wang, YG and Xu, D and others},
  title        = {Flashover characteristics of G-11 at different cryogenic temperatures},
  booktitle    = {IOP Conference Series: Materials Science and Engineering},
  organization = {IOP Publishing},
  volume       = {502},
  pages        = {012186},
  year         = {2019}
}

@article{libeyre2009detailed,
  author    = {Libeyre, Paul and Mitchell, Neil and Bessette, Denis and Gribov, Yuri and Jong, Cornelis and Lyraud, Charles},
  title     = {Detailed design of the ITER central solenoid},
  journal   = {Fusion Engineering and Design},
  publisher = {Elsevier},
  volume    = {84},
  number    = {7-11},
  pages     = {1188--1191},
  year      = {2009}
}

@article{lloyd1991plasma,
  author  = {Lloyd, B. and Jackson, G. L. and Taylor, T. S. and Lazarus, E. A. and Luce, T. C. and Prater, R.},
  title   = {Low voltage ohmic and electron cyclotron heating assisted startup in {ITER}},
  journal = {Nuclear Fusion},
  volume  = {31},
  number  = {11},
  pages   = {2031--2053},
  year    = {1991},
  doi     = {10.1088/0029-5515/31/11/001}
}

@article{maddock1969,
  author  = {Maddock, B. J. and James, G. B. and Norris, W. T.},
  title   = {Superconductive composites: heat transfer and steady state stabilization},
  journal = {Cryogenics},
  volume  = {9},
  number  = {4},
  pages   = {261--273},
  year    = {1969},
  doi     = {10.1016/0011-2275(69)90231-X}
}

@misc{maglab2024,
  author       = {{National High Magnetic Field Laboratory}},
  title        = {Superconducting Properties Database},
  year         = {2025},
  howpublished = {\url{https://nationalmaglab.org/magnet-development/applied-superconductivity-center}}
}

@article{manta2024negative,
  author  = {{The MANTA Collaboration}},
  title   = {{MANTA}: a negative-triangularity {NASEM}-compliant fusion pilot plant},
  journal = {Plasma Physics and Controlled Fusion},
  volume  = {66},
  number  = {10},
  pages   = {105006},
  year    = {2024},
  doi     = {10.1088/1361-6587/ad6708}
}

@article{meyer2024plasma,
  author  = {Meyer, H. and {the STEP Team}},
  title   = {Plasma burn---mind the gap},
  journal = {Philosophical Transactions of the Royal Society A},
  volume  = {382},
  pages   = {20230406},
  year    = {2024},
  doi     = {10.1098/rsta.2023.0406}
}

@article{mitchell1999iter,
  author    = {Mitchell, N},
  title     = {ITER magnet design and R\&D},
  journal   = {Fusion engineering and design},
  publisher = {Elsevier},
  volume    = {46},
  number    = {2-4},
  pages     = {129--150},
  year      = {1999}
}

@article{mitchell2002stress,
  author    = {Mitchell, N and Mszanowski, U},
  title     = {Stress analysis of structurally graded long solenoid coils},
  journal   = {IEEE transactions on magnetics},
  publisher = {IEEE},
  volume    = {28},
  number    = {1},
  pages     = {226--229},
  year      = {2002}
}

@article{mitchell2011iter,
  author    = {Mitchell, N and Devred, A and Libeyre, P and Lim, B and Savary, F},
  title     = {The ITER magnets: Design and construction status},
  journal   = {IEEE transactions on applied superconductivity},
  publisher = {IEEE},
  volume    = {22},
  number    = {3},
  pages     = {4200809--4200809},
  year      = {2011}
}

@article{mitchell2021superconductors,
  author  = {Mitchell, Neil and Zheng, Jinxing and Vorpahl, Christian and Corato, Valentina and Sanabria, Charlie and Segal, Michael and Sorbom, Brandon and Slade, Robert and Brittles, Greg and Bateman, Rod and others},
  title   = {Superconductors for fusion: a roadmap},
  journal = {Superconductor Science and Technology},
  volume  = {34},
  number  = {10},
  pages   = {103001},
  year    = {2021},
  doi     = {10.1088/1361-6668/ac0992}
}

@book{montgomery1969solenoid,
  author    = {Montgomery, D. Bruce},
  title     = {Solenoid Magnet Design: The Magnetic and Mechanical Aspects of Resistive and Superconducting Systems},
  publisher = {Wiley-Interscience},
  address   = {New York},
  year      = {1969}
}

@article{morris2015implications,
  author    = {Morris, J and Kemp, R and Kovari, M and Last, J and Knight, P},
  title     = {Implications of toroidal field coil stress limits on power plant design using PROCESS},
  journal   = {Fusion Engineering and Design},
  publisher = {Elsevier},
  volume    = {98},
  pages     = {1118--1121},
  year      = {2015}
}

@article{morris2021preparing,
  author    = {Morris, J and Coleman, M and Kahn, S and Muldrew, SI and Pearce, AJ and Short, D and Cook, JE and Desai, S and Humphrey, L and Kovari, M and others},
  title     = {Preparing systems codes for power plant conceptual design},
  journal   = {Nuclear Fusion},
  publisher = {IOP Publishing},
  volume    = {61},
  number    = {11},
  pages     = {116020},
  year      = {2021}
}

@article{nasr2024magnetic,
  author  = {Nasr, E. and Wimbush, S. C. and Noonan, P. and Harris, P. and Gowland, R. and Petrov, A.},
  title   = {The magnetic cage},
  journal = {Philosophical Transactions of the Royal Society A},
  volume  = {382},
  number  = {2280},
  pages   = {20230407},
  year    = {2024},
  doi     = {10.1098/rsta.2023.0407}
}

@techreport{newman1981crack,
  author      = {Newman Jr, Jo C},
  title       = {A crack-closure model for predicting fatigue-crack growth under aircraft spectrum loading},
  institution = {NASA Langley Research Center},
  number      = {NASA-TM-81941},
  year        = {1981}
}

@techreport{no1999final,
  author      = {{ITER EDA Documentation Series}},
  title       = {Final Design Report, Cost Review and Safety Analysis (FDR) and Relevant Documents},
  institution = {International Atomic Energy Agency},
  year        = {1999}
}

@article{nunio2019mechanical,
  author    = {Nunio, Francois and Torre, Alexandre and Zani, Louis},
  title     = {Mechanical analysis of the European DEMO central solenoid pre-load structure and coils},
  journal   = {Fusion Engineering and Design},
  publisher = {Elsevier},
  volume    = {146},
  pages     = {168--172},
  year      = {2019}
}

@article{panin2017mechanical,
  author    = {Panin, Anatoly and Biel, Wolfgang and Mertens, Philippe and Nunio, Francois and Zani, Louis},
  title     = {Mechanical pre-dimensioning and pre-optimization of the tokamaks’ toroidal coils featuring the winding pack layout},
  journal   = {Fusion Engineering and Design},
  publisher = {Elsevier},
  volume    = {124},
  pages     = {77--81},
  year      = {2017}
}

@article{pippan2017fatigue,
  author    = {Pippan, Reinhard and Hohenwarter, Anton},
  title     = {Fatigue crack closure: a review of the physical phenomena},
  journal   = {Fatigue \& fracture of engineering materials \& structures},
  publisher = {Wiley Online Library},
  volume    = {40},
  number    = {4},
  pages     = {471--495},
  year      = {2017}
}

@techreport{puthoff1969digital,
  author      = {Puthoff, Richard L},
  title       = {A digital computer program for determining the elastic-plastic deformation and creep strains in cylindrical rods, tubes and vessels},
  institution = {NASA Lewis Research Center},
  year        = {1969}
}

@article{rebut1976jet,
  author  = {Rebut, PH and Bertolini, E},
  title   = {The JET project: design proposal for the Joint European Torus},
  journal = {Nuclear Science and Technology, Commission of the European Communities},
  year    = {1976}
}

@techreport{rebut1981jet,
  author      = {Rebut, Paul-Henri and Green, B. J.},
  title       = {The JET Project},
  institution = {Commission of the European Communities, JET Joint Undertaking},
  number      = {EUR-JET-R7},
  year        = {1981}
}

@article{reccia2023iter,
  author    = {Reccia, L and Portone, A and Sawa, N and Mitchell, N and Koczorowski, S},
  title     = {ITER pre-compression ring tightening analysis},
  journal   = {IEEE Transactions on Applied Superconductivity},
  publisher = {IEEE},
  volume    = {34},
  number    = {5},
  pages     = {1--5},
  year      = {2023}
}

@article{redl2021new,
  author  = {Redl, A. and Angioni, C. and Belli, E. and Sauter, O.},
  title   = {A new set of analytical formulae for the computation of the bootstrap current and the neoclassical conductivity in tokamaks},
  journal = {Physics of Plasmas},
  volume  = {28},
  number  = {2},
  pages   = {022502},
  year    = {2021},
  doi     = {10.1063/5.0012664}
}

@article{reux2018demo,
  author    = {Reux, C{\'e}dric and Kahn, S{\'e}bastien and Zani, L and P{\'e}gouri{\'e}, Bernard and Piot, N and Owsiak, Michal and Aiello, Giacomo and Artaud, J-F and Boutry, Arthur and Dardour, Saied and others},
  title     = {DEMO design using the SYCOMORE system code: Influence of technological constraints on the reactor performances},
  journal   = {Fusion Engineering and Design},
  publisher = {Elsevier},
  volume    = {136},
  pages     = {1572--1576},
  year      = {2018}
}

@techreport{sackett1978effi,
  author      = {Sackett, S. J.},
  title       = {{EFFI}: A code for calculating the electromagnetic field, force, and inductance in coil systems of arbitrary geometry},
  institution = {Lawrence Livermore Laboratory},
  number      = {UCRL-52402},
  year        = {1978}
}

@article{salpietro2007precompression,
  author  = {Salpietro, E. and others},
  title   = {The pre-compression system of the toroidal field coils in {ITER}},
  journal = {Fusion Engineering and Design},
  volume  = {82},
  pages   = {1538--1543},
  year    = {2007},
  doi     = {10.1016/j.fusengdes.2007.06.027}
}

@article{Sanabria2024PITVIPER,
  author  = {Charlie Sanabria and Alexey Radovinsky and Christopher L. Craighill and others},
  title   = {Development of a high current density, high temperature superconducting cable for pulsed magnets},
  journal = {Superconductor Science and Technology},
  year    = {2024},
  doi     = {10.1088/1361-6668/ad7efc}
}

@article{sarasola2020progress,
  author    = {Sarasola, Xabier and Wesche, Rainer and Ivashov, Ilia and Sedlak, Kamil and Uglietti, Davide and Bruzzone, Pierluigi},
  title     = {Progress in the design of a hybrid HTS-Nb 3 Sn-NbTi central solenoid for the EU DEMO},
  journal   = {IEEE Transactions on Applied Superconductivity},
  publisher = {IEEE},
  volume    = {30},
  number    = {4},
  pages     = {1--5},
  year      = {2020}
}

@article{sarasola2023parametric,
  author    = {Sarasola, Xabier and Bruzzone, Pierluigi and Sedlak, Kamil and Corato, Valentina and Giannini, Lorenzo and Bachmann, Christian and Luongo, Cesar and Siccinio, Mattia},
  title     = {Parametric studies of the EU DEMO central solenoid},
  journal   = {Ieee Transactions On Applied Superconductivity},
  publisher = {IEEE},
  volume    = {33},
  number    = {5},
  pages     = {1--5},
  year      = {2023}
}

@article{sauter1999neoclassical,
  author  = {Sauter, O. and Angioni, C. and Lin-Liu, Y. R.},
  title   = {Neoclassical conductivity and bootstrap current formulas for general axisymmetric equilibria and arbitrary collisionality regime},
  journal = {Physics of Plasmas},
  volume  = {6},
  number  = {7},
  pages   = {2834--2839},
  year    = {1999},
  doi     = {10.1063/1.873240}
}

@article{sborchia2008design,
  author    = {Sborchia, C and Fu, Y and Gallix, R and Jong, C and Knaster, J and Mitchell, N},
  title     = {Design and specifications of the ITER TF coils},
  journal   = {IEEE transactions on applied superconductivity},
  publisher = {IEEE},
  volume    = {18},
  number    = {2},
  pages     = {463--466},
  year      = {2008}
}

@incollection{scanlan1980mechanical,
  author    = {Scanlan, RM and Hoard, RW and Cornish, DN and Zbasnik, JP},
  title     = {Mechanical properties of high-current multifilamentary Nb3Sn conductors},
  booktitle = {Filamentary A15 Superconductors},
  publisher = {Springer},
  pages     = {221--232},
  year      = {1980}
}

@article{senatore2024rebco,
  author  = {Senatore, C. and others},
  title   = {{REBCO} tapes for applications in ultra-high fields: critical current surface and scaling relations},
  journal = {Superconductor Science and Technology},
  volume  = {37},
  number  = {11},
  pages   = {115013},
  year    = {2024},
  doi     = {10.1088/1361-6668/ad7f95}
}

@article{shih1974study,
  author    = {Shih, True-Tsai and Wei, Robert Peh-ying},
  title     = {A study of crack closure in fatigue},
  journal   = {Engineering Fracture Mechanics},
  publisher = {Elsevier},
  volume    = {6},
  number    = {1},
  pages     = {19--32},
  year      = {1974}
}

@article{shimada2007chapter,
  author  = {Shimada, M. and Campbell, D. J. and Mukhovatov, V. and Fujiwara, M. and Kirneva, N. and Lackner, K. and Nagami, M. and Pustovitov, V. D. and Uckan, N. and Wesley, J. and Asakura, N. and Costley, A. E. and Donn{\'e}, A. J. H. and Doyle, E. J. and Fasoli, A. and Gormezano, C. and Gribov, Y. and Gruber, O. and Hender, T. C. and Houlberg, W. and Ide, S. and Kamada, Y. and Leonard, A. and Lipschultz, B. and Loarte, A. and Miyamoto, K. and Mukhovatov, V. and Osborne, T. H. and Polevoi, A. and Sips, A. C. C.},
  title   = {Chapter 1: {Overview} and summary},
  journal = {Nuclear Fusion},
  volume  = {47},
  number  = {6},
  pages   = {S1--S17},
  year    = {2007},
  doi     = {10.1088/0029-5515/47/6/S01}
}

@article{siccinio2019figure,
  author  = {Siccinio, M. and Federici, G. and Kembleton, R. and Lux, H. and Maviglia, F. and Morris, J.},
  title   = {Figure of merit for divertor protection in the preliminary design of the {EU-DEMO} reactor},
  journal = {Nuclear Fusion},
  volume  = {59},
  number  = {10},
  pages   = {106026},
  year    = {2019},
  doi     = {10.1088/1741-4326/ab3153}
}

@article{sorbom2015arc,
  author    = {Sorbom, BN and Ball, J and Palmer, TR and Mangiarotti, FJ and Sierchio, JM and Bonoli, P and Kasten, C and Sutherland, DA and Barnard, HS and Haakonsen, CB and others},
  title     = {ARC: A compact, high-field, fusion nuclear science facility and demonstration power plant with demountable magnets},
  journal   = {Fusion Engineering and Design},
  publisher = {Elsevier},
  volume    = {100},
  pages     = {378--405},
  year      = {2015}
}

@article{stambaugh2011fusion,
  author    = {Stambaugh, RD and Chan, VS and Garofalo, AM and Sawan, M and Humphreys, DA and Lao, LL and Leuer, JA and Petrie, TW and Prater, R and Snyder, PB and others},
  title     = {Fusion nuclear science facility candidates},
  journal   = {Fusion Science and Technology},
  publisher = {Taylor \& Francis},
  volume    = {59},
  number    = {2},
  pages     = {279--307},
  year      = {2011}
}

@article{sutcliffe2025magnet,
  author    = {Sutcliffe, MF and Torre, A and Boudes, B and Lacroix, B and Le Coz, Q and Dao, C Nguyen Thanh and Nicollet, S and Corato, V},
  title     = {Magnet Design Activities at CEA for EU-DEMO LAR Baseline 2024},
  journal   = {IEEE Transactions on Applied Superconductivity},
  publisher = {IEEE},
  volume    = {36},
  number    = {3},
  pages     = {1--6},
  year      = {2025},
  doi       = {10.1109/TASC.2025.3637189}
}

@article{swanson2022validation,
  author  = {Swanson, CPS and Kahn, S and Rana, C and Titus, PH and Brooks, AW and Guttenfelder, W and Zhai, Y and Brown, TG and Menard, JE},
  title   = {Validation and results of an approximate model for the stress of a Tokamak toroidal field coil at the inboard midplane},
  journal = {arXiv preprint arXiv:2206.14699},
  year    = {2022}
}

@book{thome1982mhd,
  author    = {Thome, Richard J and Tarrh, John M},
  title     = {MHD and fusion magnets: field and force design concepts},
  publisher = {John Wiley and Sons, Inc., New York, NY},
  year      = {1982}
}

@book{timoshenko1970elasticity,
  author    = {Timoshenko, Stephen P. and Goodier, J. N.},
  title     = {Theory of Elasticity},
  publisher = {McGraw-Hill},
  address   = {New York},
  edition   = {3rd},
  year      = {1970},
  isbn      = {978-0-07-064720-9}
}

@inproceedings{titus1995structural,
  author       = {Titus, Peter H},
  title        = {Structural analysis of the ITER EDA magnet system},
  booktitle    = {Proceedings of 16th International Symposium on Fusion Engineering},
  organization = {IEEE},
  volume       = {2},
  pages        = {1522--1525},
  year         = {1995}
}

@article{titus1998analysis,
  author    = {Titus, Peter H},
  title     = {Analysis of selected mechanical details of the ITER magnet system},
  journal   = {Fusion technology},
  publisher = {Taylor \& Francis},
  volume    = {34},
  number    = {3P2},
  pages     = {675--679},
  year      = {1998}
}

@article{titus2002provisions,
  author    = {Titus, Peter H},
  title     = {Provisions for out-of-plane support of the TF coils in recent Tokamaks},
  journal   = {IEEE transactions on applied superconductivity},
  publisher = {IEEE},
  volume    = {10},
  number    = {1},
  pages     = {636--640},
  year      = {2002}
}

@article{titus2013tf,
  author    = {Titus, Peter H and Zolfaghari, Ali},
  title     = {TF Inner Leg Space Allocation for Pilot Plant Design Studies},
  journal   = {Fusion Science and Technology},
  publisher = {Taylor \& Francis},
  volume    = {64},
  number    = {3},
  pages     = {680--686},
  year      = {2013}
}

@article{torre2016tools,
  author    = {Torre, A and Ciazynski, D and Hertout, P and Zani, L},
  title     = {Tools used at CEA for designing the DEMO toroidal field coils winding pack},
  journal   = {IEEE Transactions on Applied Superconductivity},
  publisher = {IEEE},
  volume    = {26},
  number    = {4},
  pages     = {1--5},
  year      = {2016}
}

@article{tsuchiya2008design,
  author  = {Tsuchiya, K. and others},
  title   = {Superconducting properties of {Nb$_3$Sn} strands for the {JT-60SA} {CS} conductor},
  journal = {IEEE Trans. Appl. Supercond.},
  volume  = {18},
  number  = {2},
  pages   = {208--211},
  year    = {2008}
}

@techreport{uckan1990,
  author      = {Uckan, N. A. and {ITER Physics Group}},
  title       = {{ITER Physics Design Guidelines: 1989}},
  institution = {IAEA},
  number      = {IAEA/ITER/DS/10},
  year        = {1990},
  note        = {See also Fusion Sci. Technol. 19 (1991) 1493}
}

@article{wade2021cost,
  author    = {Wade, MR and Leuer, JA},
  title     = {Cost drivers for a tokamak-based compact pilot plant},
  journal   = {Fusion Science and Technology},
  publisher = {Taylor \& Francis},
  volume    = {77},
  number    = {2},
  pages     = {119--143},
  year      = {2021}
}

@article{wakatsuki2019safety,
  author  = {Wakatsuki, T. and Suzuki, T. and Hayashi, N. and Oyama, N. and Ide, S.},
  title   = {Safety factor profile control with reduced central solenoid flux consumption during plasma current ramp-up phase using a reinforcement learning technique},
  journal = {Nuclear Fusion},
  volume  = {59},
  pages   = {066022},
  year    = {2019},
  doi     = {10.1088/1741-4326/ab1571}
}

@article{wang2023study,
  author    = {Wang, Weijun and Jin, Jing and Wu, Lei and Deng, Ming and Shi, Jinhao and Jin, Huan and Huang, Chuanjun and Yuan, Yuan and Liu, Kun and Wang, Songtao and others},
  title     = {Study on the welding properties of modified N50 CICC jacket for future fusion applications},
  journal   = {Journal of Materials Research and Technology},
  publisher = {Elsevier},
  volume    = {27},
  pages     = {6094--6103},
  year      = {2023}
}

@article{wang2024structure,
  author    = {Wang, Weijun and Wu, Yongsheng and Jin, Jing and Shi, Jinhao and Jin, Huan and Wang, Changjun and Liu, Yu and Huang, Chuanjun and Li, Laifeng and Qin, Jinggang},
  title     = {Structure optimization and performance evaluation of CS CICC jacket based on N50H austenitic steel for future fusion reactor},
  journal   = {Cryogenics},
  publisher = {Elsevier},
  volume    = {139},
  pages     = {103836},
  year      = {2024}
}

@article{wang2026mass,
  author    = {Wang, Wei-Jun and Qin, Jing-Gang and Wu, Yong-Sheng and Jin, Jing and Shi, Jin-Hao and Wu, Yi-Fei and Tu, Zheng-Ping and Chen, Xiao-Wei and Li, Jian-Gang and Jin, Huan},
  title     = {Mass production and performance evaluation of CHSN01 jacket for future fusion applications: W.-J. Wang et al.},
  journal   = {Nuclear Science and Techniques},
  publisher = {Springer},
  volume    = {37},
  number    = {2},
  pages     = {32},
  year      = {2026}
}

@article{wang2026mechanical,
  author    = {Wang, Weijun and Ye, Xinlan and Zhang, Baozhu and Wu, Yifei and Zhi, Huihui and Tu, Zhengping and Shen, Xiaogang and Qin, Jinggang and Li, Jiangang},
  title     = {Mechanical properties and microstructure of CHSN01 conductor jacket under long-term cryogenic service in future fusion reactors},
  journal   = {Journal of Nuclear Materials},
  publisher = {Elsevier},
  pages     = {156463},
  year      = {2026}
}

@article{wenninger2015,
  author  = {Wenninger, R. and others},
  title   = {{The physics and technology basis entering European system code studies for DEMO}},
  journal = {Nucl. Fusion},
  volume  = {55},
  pages   = {063003},
  year    = {2015}
}

@book{wesson2011tokamaks,
  author    = {Wesson, John},
  title     = {Tokamaks},
  publisher = {Oxford University Press},
  address   = {Oxford},
  series    = {International Series of Monographs on Physics},
  edition   = {4},
  year      = {2011}
}

@book{wilson1983superconducting,
  author    = {Wilson, M. N.},
  title     = {Superconducting Magnets},
  publisher = {Oxford University Press},
  series    = {Monographs on Cryogenics},
  year      = {1983},
  isbn      = {978-0-19-854810-2}
}

@article{wu2003east,
  author  = {Wu, Weiyue and others},
  title   = {Design of the {EAST} {PF} system},
  journal = {Fusion Engineering and Design},
  volume  = {65},
  pages   = {331--337},
  year    = {2003}
}

@article{wu2021preliminary,
  author  = {Wu, Yu and Li, Jiangang and Shen, Guang and Zheng, Jinxing and Liu, Xiaogang and Long, Feng and Dai, Chao and Liu, Xufeng and Shi, Yi and Li, Junjun and Hao, Qiangwang and Hu, Yanlan and Xiao, Yezheng and Wen, Wei and Yu, Xiaowu and Fang, Chao and Wei, Jing and Zhu, Lina and Han, Houxiang},
  title   = {Preliminary Design of {CFETR} {TF} Prototype Coil},
  journal = {Journal of Fusion Energy},
  volume  = {40},
  number  = {1},
  pages   = {5},
  year    = {2021},
  doi     = {10.1007/s10894-021-00291-8}
}

@article{wu2025development,
  author    = {Wu, Yongsheng and Wang, Weijun and Jin, Jing and Shi, Jinhao and Deng, Ming and Qin, Jinggang},
  title     = {The Development and Challenge of the CHSN01 Jacket for the CS Magnet in China’s Future Fusion Device},
  journal   = {Applied Sciences},
  publisher = {MDPI},
  volume    = {15},
  number    = {9},
  pages     = {5201},
  year      = {2025}
}

@article{yiannopoulos5stress,
  author    = {Yiannopoulos, Andrew Ch},
  title     = {Stress analysis and design of cylinder bearings supporting metal bridges},
  journal   = {International Journal of Engineering and Applied Sciences},
  publisher = {Engineering Research Publication},
  volume    = {5},
  number    = {5},
  pages     = {257225},
  year      = {n.d.}
}

@article{yoshida2010design,
  author  = {Yoshida, Kiyoshi and Tsuchiya, Katsuhiko and Kizu, Kaname and Murakami, Haruyuki and Kamiya, Koji and Peyrot, Marc and Barabaschi, Pietro},
  title   = {Design and construction of JT-60SA superconducting magnet system},
  journal = {J. Plasma Fusion Res. SERIES},
  volume  = {9},
  pages   = {214--219},
  year    = {2010}
}

@article{zani2019parametric,
  author    = {Zani, Louis and Ciazynski, Daniel and Corato, Valentina and Lacroix, Benoit and Le Coz, Quentin and Misiara, Nicolas and Nicollet, Sylvie and Nunio, Francois and Sedlak, Kamil and Torre, Alexandre and others},
  title     = {Parametric optimization of the CEA TF magnet design of the EU DEMO updated configuration},
  journal   = {IEEE Transactions on Applied Superconductivity},
  publisher = {IEEE},
  volume    = {29},
  number    = {5},
  pages     = {1--5},
  year      = {2019}
}

@article{zhao2022structural,
  author  = {Zhao, Z. and Moore, P. and Chiesa, L.},
  title   = {Structural Modeling of {REBCO} {VIPER} Cable for High-Field Magnet Applications},
  journal = {IEEE Transactions on Applied Superconductivity},
  volume  = {32},
  number  = {6},
  pages   = {1--5},
  year    = {2022},
  doi     = {10.1109/TASC.2022.3146813}
}

@article{Zhou2023_REBCO_mech,
  author  = {Y. H. Zhou and others},
  title   = {Review of progress and challenges of key mechanical issues in high-field superconducting magnets},
  journal = {National Science Review},
  year    = {2023},
  note    = {Review; REBCO coated conductors mechanics}
}

@article{zhu2025electromagnetic,
  author    = {Zhu, Jiandong and Liu, Xiaogang and Gao, Xiang and Zhang, Jie and Yu, Lijuan and Ding, Fanping and Li, Guoqiang},
  title     = {Electromagnetic design of a hybrid central solenoid for a medium-sized tokamak},
  journal   = {AIP Advances},
  publisher = {AIP Publishing},
  volume    = {15},
  number    = {3},
  year      = {2025}
}

@article{huguet2001iter,
	author  = {Huguet, M.},
	title   = {Key engineering features of the {ITER-FEAT} magnet system and implications for the {R\&D} programme},
	journal = {Nuclear Fusion},
	volume  = {41},
	number  = {10},
	pages   = {1503},
	year    = {2001},
	doi     = {10.1088/0029-5515/41/10/317},
}
	
\end{document}